\documentclass[usenatbib]{mn2e}
\pdfoutput=1
\usepackage{amsmath}
\usepackage{amssymb}
\usepackage{graphicx}
\usepackage{txfonts}
\usepackage{natbib}
\usepackage{wrapfig}
\usepackage{longtable}
\usepackage[usenames]{color}

\def\clusterfields{51}
\def\cosmosample{28~}
\def\relaxedsample{18~}

\bibpunct{(}{)}{;}{a}{}{,}

\DeclareMathAlphabet{\mymath}{U}{eus}{m}{n}

\def\la{\mathrel{\hbox{\rlap{\hbox{\lower4pt\hbox{$\sim$}}}\hbox{$<$}}}}
\def\ga{\mathrel{\hbox{\rlap{\hbox{\lower4pt\hbox{$\sim$}}}\hbox{$>$}}}}

\newcommand{\be}{\begin{equation}}
\newcommand{\ee}{\end{equation}}

\newcommand{\bi}{\begin{itemize}}
\newcommand{\ei}{\end{itemize}}

\newcommand{\ben}{\begin{enumerate}}
\newcommand{\een}{\end{enumerate}}

\newcommand{\bfig}{\begin{figure}\begin{minipage}{140mm}}
\newcommand{\efig}{\end{minipage}\end{figure}}

\newcommand{\btab}{\begin{table}\begin{minipage}{140mm}}
\newcommand{\etab}{\end{minipage}\end{table}}

\newcommand{\bfigMore}{\begin{figure}\begin{minipage}{160mm}}
\newcommand{\efigMore}{\end{minipage}\end{figure}}

\newcommand{\btabMore}{\begin{table}\begin{minipage}{160mm}}
\newcommand{\etabMore}{\end{minipage}\end{table}}

\newcommand{\bea}{\begin{eqnarray}}
\newcommand{\eea}{\end{eqnarray}}

\newcommand{\bega}{\begin{gather}}
\newcommand{\eega}{\end{gather}}

\newcommand{\bc}{\begin{center}}
\newcommand{\ec}{\end{center}}

\renewcommand{\vec}[1]{\bm #1}

\def\arcsecf {\hbox{$.\!\!^{\prime\prime}$}}

\newcommand{\dif}{{\rm d}}

\newcommand{\ave}[1]{\langle #1 \rangle}

\newcommand{\bfmath}[1]
{\mbox{\boldmath${\rm #1}$\unboldmath}}
\newcommand{\bm}{\boldsymbol}

\def\bbullet{MACSJ0025.4$-$1222}

\def\rxj{RX~J1347$-$1145}
\def\mtt{MACSJ2243.3$-$0935}
\def\mof{MACSJ0417.5$-$1154}
\def\mos{MACSJ0717.5$+$3745}
\def\mosf{MACSJ0744.8$+$3927}
\def\mnt{MACSJ1931.8-2634}
\def\mef{MACSJ1149.5$+$2223}

\title
[Weighing the Giants I]
{Weighing the Giants – I. Weak-lensing masses for \clusterfields\, massive galaxy
clusters: project overview, data analysis methods and cluster images}
\author[Anja von der Linden et al.]
{
\parbox[t]{\textwidth}
{\vspace{-1.3cm}
\begin{flushleft}
Anja von der Linden$^{1,2,3}$
\thanks{\vspace{-1cm}E-mail:anja@slac.stanford.edu}, 
Mark T. Allen$^{1,2}$,
Douglas E. Applegate$^{1,2,4,5}$,
Patrick L. Kelly$^{1,2,4,6}$,
Steven W. Allen$^{1,2,4}$,
Harald Ebeling$^{7}$,
Patricia R. Burchat$^{1,2}$,
David L. Burke$^{1,4}$,
David Donovan$^{7}$,
R. Glenn Morris$^{1,4}$,
Roger Blandford$^{1,2,4}$,
Thomas Erben$^{5}$,
Adam Mantz$^{8,9}$
\end{flushleft}
}
\\
        \vspace*{2pt}
\\
$^{1}$Kavli Institute for Particle Astrophysics and Cosmology,
Stanford University,
452 Lomita Mall,
Stanford, CA  94305-4085, USA\\
$^{2}$Department of Physics,
Stanford University,
382 Via Pueblo Mall, 
Stanford, CA  94305-4060, USA\\
$^{3}$Dark Cosmology Centre , 
Niels Bohr Institute, University of Copenhagen,
Juliane Maries Vej 30,
2100 Copenhagen {\O},
Denmark\\
$^{4}$SLAC National Accelerator Laboratory, 
2575 Sand Hill Road, 
Menlo Park, CA 94025, USA\\
$^{5}$Argelander-Institut f\"ur Astronomie,
Universit\"at Bonn, 
Auf dem H\"ugel 71, 
53121 Bonn, Germany\\
$^{6}$Department of Astronomy, 
University of California, 
B-20 Hearst Field Annex \# 3411\\
Berkeley, CA 94720-3411, 
USA\\
$^{7}$Institute for Astronomy, 
2680 Woodlawn Drive, 
Honolulu, HI 96822, USA\\
$^{8}$Kavli Institute for Cosmological Physics,
University of Chicago,
5640 South Ellis Avenue,
Chicago, IL 60637-1433, USA\\
$^{9}$Department of Astronomy and Astrophysics, 
University of Chicago,
5640 South Ellis Avenue,
Chicago, IL 60637-1433, USA\\
}

\begin{document}

\date{Accepted 2013 October 9. Received 2013 October 2; in original form 2012 August 1}

\pagerange{\pageref{firstpage}--\pageref{lastpage}} \pubyear{2014}

\maketitle

\label{firstpage}
\vspace{-2cm}
\begin{abstract}
\noindent
  This is the first in a series of papers in which we measure accurate
  weak-lensing masses for \clusterfields\, of the most X-ray luminous
  galaxy clusters known at redshifts $0.15\lesssim z_{\rm Cl}\lesssim 0.7$, in
  order to calibrate X-ray and other mass proxies for cosmological
  cluster experiments. The primary aim is to improve the absolute mass
  calibration of cluster observables, currently the dominant
  systematic uncertainty for cluster count experiments. 
  Key elements of this work are the rigorous quantification of systematic
  uncertainties, high quality data reduction and photometric calibration, 
  and the ``blind'' nature of the analysis to avoid confirmation
  bias. Our target
  clusters are drawn from X-ray catalogs based on the
  ROSAT All-Sky Survey, and provide a versatile calibration sample for
  many aspects of cluster cosmology. We have acquired wide-field,
  high-quality imaging using the Subaru and CFHT telescopes for all
  \clusterfields\, clusters, in at least three bands per cluster. For a
  subset of 27 clusters, we have data in at least five bands, allowing
  accurate photometric redshift estimates of lensed galaxies. In this paper, we describe
  the cluster sample and observations, and detail the processing of
  the SuprimeCam data to yield high-quality images suitable for robust
  weak-lensing shape measurements and precision photometry. For each
  cluster, we present wide-field three-color optical images and maps
  of the weak-lensing mass distribution, the optical light
  distribution, and the X-ray emission. These provide insights into
  the large-scale structure in which the clusters are embedded. We
  measure the offsets between X-ray flux centroids and the Brightest
  Cluster Galaxies in the clusters, finding these to be small in
  general, with a median of 20 kpc. For offsets $\lesssim 100$ kpc,
  weak-lensing mass measurements centered on the Brightest Cluster
  Galaxies agree well with values determined relative to the X-ray
  centroids; miscentering is therefore not a significant source of
  systematic uncertainty for our weak-lensing mass measurements. In
  accompanying papers we discuss the key aspects of our photometric
  calibration and photometric redshift measurements (Kelly et al.),
  and measure cluster masses using two methods, including a novel
  Bayesian weak-lensing approach that makes full use of the
  photometric redshift probability distributions for individual
  background galaxies (Applegate et al.). In subsequent papers, we
  will incorporate these weak-lensing mass measurements into a
  self-consistent framework to simultaneously determine cluster
  scaling relations and cosmological parameters.
\end{abstract}


\section{Introduction}
\label{sect:intro}

The formation of cosmic structure depends sensitively on the mass and
energy content of the Universe, and the physical nature of dark matter
and dark energy.  Galaxy clusters are the most massive gravitationally
bound structures, sitting at the largest nodes of the cosmic web.  As
such, their number density, baryon content, and evolution are
sensitive probes of cosmological parameters, in particular the
amplitude of matter fluctuations ($\sigma_8$), the mean matter and
dark energy densities ($\Omega_{\rm m}$ and $\Omega_{\rm DE}$), and
the dark energy equation of state parameter ($w$) \citep[for a recent
review, see][]{aem11}.

The idea of ``counting clusters'' as a way to test cosmology has
existed for decades \citep[e.g.,][]{kai84,hea91}.  The discovery of
massive clusters at high redshifts \citep{dvg98,baf98} provided
supporting evidence for a low matter density Universe, and presaged
the discovery of dark energy from Type Ia supernovae studies
\citep{rfc98,pag99}.  Cluster counts paved the way in determining the
now accepted value of $\sigma_8\sim0.8$ \citep[e.g.][]{brt01,gbc03}.
Recently, measurements of the evolution of the cluster number density
have provided some of the most precise and robust constraints on dark
energy \citep{vkb09,mar10}, as well as departures from General
Relativity on cosmological scales \citep{ram10,rba12,svh09}, and the
species-summed neutrino mass \citep{mar10c,rvj10}.

A fundamental challenge for cluster count experiments is that the
survey observations do not measure cluster masses directly,
but rather a property that correlates with cluster mass, typically
with significant associated scatter. For X-ray surveys, the standard
survey observable is the X-ray flux, which with the cluster redshift
gives the X-ray luminosity; for optical red-sequence finders, survey
measures are typically based on optical richness; and for millimeter
surveys the typical observable is the Sunyaev-Zel'dovich (SZ) flux. In order
to reconstruct the underlying mass function, the scaling relation
between the survey observable and cluster mass, as well as the
scatter in this relation as a function of mass and redshift, must be
measured.  This process can be improved if, for a representative
subsample of the survey clusters, one can also obtain deeper follow-up
measurements of precise mass proxies with lower systematic scatter
\citep{mar10,mae10,vbe09,vkb09}. The subsample re-observed need not be
large in order to bring a substantial boost in constraining power
\citep{mar10,wrw10}.

\subsection{The role of mass proxies}

X-ray observations provide a critical element of this work, offering
several observables that are straightforward to measure and which
correlate tightly with true cluster mass. For example, the temperature
of the intracluster medium, $T_{\rm X}$, traces cluster mass with a scatter
of 10--15\%, far better than the total X-ray luminosity (scatter
$\sim$40\%). Other X-ray proxies such as gas mass, $M_{\rm gas}$,
thermal energy, $Y_{\rm X} \,(=M_{\rm gas} T_{\rm X})$, and center-excised X-ray
luminosity provide comparable or possibly even lower scatter
\citep{ars08,kvn06,mau07,mae10}.

However, even for these low-scatter mass proxies, the absolute scaling
with true cluster mass must also be determined, accurately and
robustly. For X-ray data, under the assumptions of hydrostatic
equilibrium and spherical symmetry, one can relate the observed gas
density and temperature profiles to the underlying mass profile.  Yet
even for the most dynamically relaxed clusters, and at optimal
measurement radii ($r\sim r_{2500}$), hydrostatic X-ray mass estimates
are expected to be biased at the 5--10\% level due to non-thermal
pressure support from residual gas bulk motion and other processes
\citep{nvk07,rmm12}. For less relaxed systems, and for measurements at
larger radii ($r\gtrsim r_{500}$), the biases in hydrostatic
measurements can be significantly worse
\citep[20--30\%,][]{nvk07}. This uncertainty in the absolute mass
scaling is currently the dominant systematic uncertainty in the
constraints on $\sigma_8$ from cluster counts
\citep{mar10,vkb09,rwr10,sta11,bhd11}. In order for future surveys to
access their full constraining power, it is imperative to calibrate
these mass proxies to within 5\% and over the entire mass and redshift
range of interest \citep{wrw10}.

\subsection{Weak-lensing mass measurements as calibrators for cluster
  masses}

The most promising method currently capable of absolutely calibrating
mass measurements for statistical cluster samples is cluster weak
gravitational lensing. Weak-lensing mass measurements do not require a
baryonic tracer, but directly measure the total gravitating matter.
For individual clusters, weak lensing is inherently noisy since the
intrinsic ellipticity distribution of galaxies is broad and lensing
measurements are sensitive to all structure along the line of
sight. Utilizing cluster weak-lensing mass measurements for precision
cosmology requires a thorough understanding of the systematic biases
involved. Since the shear induced on a background galaxy depends on
the cluster mass, the ratios of angular diameter distances between the
observer, cluster and source, and cosmology, there are three
possible sources of systematic uncertainties. Observationally, biases
in the shear measurements and in the redshifts of background galaxies
translate to biased mass measurements.  Even in the absence of
observational biases, systematic uncertainties may arise from the
assumptions made to relate the measured lensing signal to an intrinsic
cluster mass.

Lensing inherently measures projected, 2D masses; however, to compare
these to the halo mass function, they need to be related to 3D
masses. The most common method to do so is to fit spherically
symmetric density models \citep[such as the NFW profile, ][]{NFW97} to
the measured shear profiles. Adopting a profile shape has the added
advantages that it breaks the mass-sheet degeneracy, and that
significantly fewer galaxies are required compared to non-parametric
mass reconstruction. (Note that the aperture mass method, which also
assumes spherical symmetry but does not directly fit a specific
profile, still requires a profile assumption at large radii to break
the mass-sheet degeneracy.) However, because clusters are generally
triaxial, the assumption of spherical symmetry leads to
over-/underestimates of the mass if the cluster major axis is aligned
along/perpendicular to the line of sight \citep{cok07,mrm10}.  Mass in
the infall region of clusters (e.g., filaments and infalling groups)
and/or unassociated structures along the line of sight can similarly
bias individual mass measurements \citep{hoe01,hoe03}.  Quantifying
the expected scatter due to these sources, as well as any expected
bias due to the profile assumption, can be achieved straightforwardly from
cosmological N-body simulations, by applying the same mass measurement
methods to the simulations as to the real data. For the NFW profile
(or closely related profiles), this has recently been done by a number
of groups \citep{bek11,ogh11,bmk11}. The intrinsic scatter due to
projection effects is found to be $\sim 25$\%
\citep{bek11,bmk11}, while the expected bias is dependent on the outer
fit radius -- if this is restricted to be close to the virial radius,
the average mass can be recovered with little bias
\citep{bek11,ogh11}.

The unbiased mean, yet considerable intrinsic scatter, for cluster weak
lensing measurements implies that relatively large samples of clusters
are necessary to meet the calibration needs of cluster cosmology.
For such work, the clusters used should ideally be drawn
representatively from the surveys in question, so as to have the same
selection function. This is a fundamental reason why, for example,
strong-lensing selected clusters should not be used for this purpose
-- the incidence of strong lensing is highly biased towards clusters
that are elongated and/or have additional structures along the
line of sight.

To date, only a handful of studies have measured individual
weak-lensing masses for more than a few clusters, and none have fully
incorporated the results into a robust cosmological work, which would
require solving simultaneously for the scaling relations and
cosmological parameters \citep{mar10,aem11}.  A number of early works
\citep{all98,hok98,csk04,sks05} compared lensing mass estimates of
massive clusters to X-ray mass proxies, but the weak-lensing mass
measurements were generally limited by the small fields of view of
existing cameras. The work of \citet[][see also
\citealt{dah06,ped07}]{dki02} provides the so-far largest compilation
of weak-lensing mass measurements of individual clusters (38
clusters).  With the increasing availability of high-quality,
wide-field mosaic cameras, the precision of weak-lensing mass
measurements at sufficiently large cluster radii has significantly
increased, providing the means to study cluster scaling relations with
total mass measurements.  \citet{hoe07} compared weak-lensing masses
of 20 clusters, derived from two-filter optical imaging, to
independently measured X-ray luminosities and temperatures, as well as
galaxy velocity dispersions. For 18 of these clusters, \citet{mhb08}
computed X-ray hydrostatic masses and compared these to the
weak-lensing mass estimates.  \citet{bsk07} compared weak-lensing
masses for 11 clusters measured from three-filter imaging to X-ray
luminosities and temperatures.  The LoCuSS project measured
weak-lensing masses with two-filter imaging for 30 clusters
\citep{otu10}.  For 12 of them, the lensing masses were compared to
$T_X$, $M_{\rm gas}$, and $Y_{\rm X}$ \citep{ozf10} as well as
hydrostatic mass estimates \citep{zof10}. Using 18 of these clusters,
\citet{mso11} present a first comparison of integrated Compton
parameters from SZ observations to weak-lensing mass determinations.
\citet{hhl12} present a second SZ--weak lensing comparison for 5
clusters. \citet{hdc11} used single-filter Hubble Space Telescope
observations to measure weak-lensing masses for 25 clusters of
moderate X-ray luminosity, and compared these to the cluster X-ray
luminosities and temperatures. For larger samples of less massive
systems, stacking analyses enable the determination of the mean
cluster mass in bins of survey observable \citep{jsw07,lfk10}. For
most studies listed here, the bulk of the cluster samples studied is
at $z_{\rm Cl} \sim 0.2 - 0.3$. A few studies have specifically
targeted higher-redshift clusters, both using space-based \citep[][22
clusters at $z\gtrsim1$]{jdh11} and ground-based imaging \citep[][7
clusters at $z \sim 0.4 - 0.8$]{ier11}.

A key assumption of these pathfinding lensing studies is to implicitly
place all background galaxies at the same effective redshift.  For
low-redshift clusters ($z_{\rm Cl}\sim0.2$, representing the bulk of
the clusters studied to date), this approximation should not severely
bias the mass measurements; the peak of the galaxy distribution is at
$z\sim0.8-1.0$ and, for clusters at low redshifts, the shear signal
varies only slowly over this range, causing errors in the effective
redshift to bias the mass only slightly.  For clusters at higher
redshifts ($z_{\rm Cl}\gtrsim0.4$), however, this is no longer the
case and one can significantly reduce systematic scatter and 
  potential bias (in case the assumed redshift distribution is not
  representative of the redshift distribution in cluster
  fields) by incorporating appropriate redshift information for
individual galaxies. Since weak lensing is based on shape measurements
of many faint galaxies, this is feasible only with photometric
redshifts.  Current and up-coming cluster surveys, such as the South
Pole Telescope survey \citep[SPT, ][]{vcd10}, the Atacama Cosmology
Telescope \citep[ACT, ][]{sta11}, Planck \citep{pla10}, the Dark
Energy Survey \citep[][DES]{des05}, and eROSITA \citep{pab10} will
find hundreds to many thousands of massive clusters in the redshift
range $0.5 \lesssim z \lesssim 1.5$. Lensing mass calibrations for
these surveys will be vital to maximizing their potential to constrain
cosmology. It is therefore essential to develop the strategies and
tools to measure unbiased cluster masses using photometric redshifts
in an optimal way.

\subsection{This study}

In this series of papers, we develop and apply techniques to enable
the determination of accurate weak-lensing masses for a total of
\clusterfields\, clusters from deep, high-quality multi-color Subaru
SuprimeCam and CFHT MegaPrime optical imaging.

In this first paper, we describe the cluster sample and the data
reduction methods: a careful data treatment is key to robust shear and
photometry measurements, and unbiased cluster mass determination. We
discuss the correspondence between the dark matter, gas, and optical
light distributions, and the relation of the positions of the
Brightest Cluster Galaxies (BCGs) and X-ray centroids. In Paper II
\citep{kla12}, we describe the details of our photometric calibration,
including a prescription to construct the ``star flat'', which
corrects flat-field errors due to varying pixel scale and scattered
light in wide-field cameras. In Paper II we also describe an
  improved and versatile technique to calibrate photometric zeropoints
  from stellar colors, whose implementation we have made publicly
  available. Using these methods, we show that we can estimate robust
  photometric redshifts even when calibration data are lacking, and
present an initial analysis of the source-redshift dependent shear
signal of the clusters. In Paper III \citep{alk12} we introduce a
novel Bayesian approach to weak-lensing mass estimation that makes
full use of photometric redshift probability distributions of lensed
galaxies. We compare the obtained masses to those derived from the
more common method of adopting a single effective redshift for the
background galaxies. Critically, we also include a detailed discussion
and quantification of the systematic uncertainties
involved. Additional papers will focus on the scaling relations
between weak-lensing masses and other observables, and present updated
cosmological constraints.

For a project such as this, where a central goal is the comparison of
measurements determined by independent techniques, and where the
measurements to be calibrated have already been used in cosmological
studies, there is a clear danger of ``observer bias'' or
``confirmation bias''. These biases are well-known in the wider
physics community, and can be avoided by implementing ``blind
analyses'' \citep{klr05}. While blind analyses have not yet been used
widely in astronomy to date, they will be essential for upcoming
precision cosmology measurements \citep[see also][]{aem11,crd11}. To
combat confirmation bias, we have chosen to explicitly avoid direct
comparison with X-ray mass proxies, or indeed any other mass
estimates, during the course of this study, revealing all comparisons
only at the end of a given part of the study.  To enforce this
restriction, in the few cases where intermediate results were
presented, all non-lensing mass estimates were multiplied by a random,
unrevealed number (all masses were multiplied by the same number),
thus removing the absolute scalings of the lensing vs. other mass
relations, the primary quantities of interest. Since the lensing data
are not altered, this procedure allows complete and accurate analyses
of statistical and systematic errors, while eliminating unintentional
bias towards the expected correlation with other mass proxies.

In the very early stages of this work blinding was not implemented and
preliminary comparisons of crude mass estimates for a small fraction
($\lesssim 20$ per cent) of the clusters were examined. We emphasize,
however, that the final, more sophisticated mass measurement methods
described in Paper III were developed independently of these early
analyses and that all lensing mass measurements presented in these
papers were determined blindly with respect to other mass proxies and
all results in the literature.  "Unblinding" with respect to lensing
mass estimates in the literature took place only after the lensing
analysis was completed, including internal review of Papers
I-III. "Unblinding" with respect to X-ray and other independent mass
proxies has not occurred at the time of completing papers I-III. Any
subsequent changes to the lensing analysis will be reported in Paper
III, or later work, if necessary.

This paper is structured as follows: In Sect.~\ref{sect:clusters} we
describe the cluster sample and the optical imaging
observations. Since the lensing analysis is performed mostly on
SuprimeCam data, we give a detailed description of the SuprimeCam
data reduction in Sect.~\ref{sect:optical_data} (with additional
details in App.~\ref{appendix:early_data}).
Sect.~\ref{sect:catalogs_photom} describes the object detection and
initial photometry measurements. In Sect.~\ref{sect:shearmeasurements}
we briefly summarize the shear measurement method based on
\citet{ksb95}, discuss our strategies to correct for the anisotropy
(also App.~\ref{appendix:psf}) and isotropic smearing of the point
spread function, and discuss the calibration using STEP2 simulations
\citep{mhb07}, including accounting for correlated noise. In
Sect.~\ref{sect:cluster_maps} and App.~\ref{appendix:clustermaps} we
present a gallery of cluster images and maps of the total mass
distribution as recovered from the weak-lensing data, the large-scale
structure around each cluster as traced by galaxies on the red
sequence, and the X-ray emission associated with the cluster.  In
Sect.~\ref{sect:cluster_centers}, we investigate the impact of
different choices for the cluster centers on the lensing results. We
summarize and provide an outlook on future work in
Sect.~\ref{sect:summary}.

The fiducial cosmology adopted in this paper is a flat $\Lambda$CDM
model with $\Omega_{\rm m} = 0.3$ and $H_0 = 100 \,h\,
\mbox{km/s/Mpc}$, where $h=0.7$.

\section{Cluster Sample}
\label{sect:clusters}

\begin{table*}
  \caption{Overview of the cluster sample. The columns are (1) cluster
    name; (2) redshift; (3) right ascension and (4) declination (both
    in J2000) of the X-ray centroid; (5) the filters in which the
    cluster was observed (see Table~\ref{tab:filters} for filter and
    instrument details); (6) the image(s) used as lensing band, with
    effective exposure time in seconds and seeing in arcsec. Column
    (7) indicates whether the cluster is part of the sample of
    \citet[][; M10]{mar10,mae10}; if so, the parent survey is also
    listed. Column (8) indicates whether the cluster is relaxed and in
    the \citet[][; A08]{ars08} sample. The table is sorted by
    increasing cluster redshifts, which are compiled from the MACS
    \citep{ebd07,eem10,mae12}, BCS \citep{eeb98}, and REFLEX
    \citep{bsg04} catalogs. }\vspace{-0.2cm}
\label{tab:clusters_table1}
\begin{tabular}{l c c c c c c c}
\hline
Cluster & $z_{\rm Cl}$ & R.A. & Dec. & Filter Bands & Lensing Band & M10 & A08 \\
 & & (J2000) & (J2000) & & (exp. time [s], seeing [\arcsec]) & Sample & Sample \\
 (1) & (2) & (3) & (4) & (5) & (6) & (7) & (8) \\
\hline
A2204 & 0.152 & 16:32:47.158 & 05:34:33.00 & {\it B}$_{\rm J}${\it V}$_{\rm J}${\it R}$_{\rm C}${\it g}$^{\star}${\it r}$^{\star}$ & {\it V}$_{\rm J}$ (1038, 0.58) & BCS & $\surd$ \\
A750 & 0.163 & 09:09:12.653 & 10:58:34.74 & {\it V}$_{\rm J}${\it R}$_{\rm C}${\it i}$^{+}$ & {\it V}$_{\rm J}$ (1728, 0.72) & BCS & - \\
RXJ1720.1+2638 & 0.164 & 17:20:09.996 & 26:37:28.70 & {\it B}$_{\rm J}${\it V}$_{\rm J}${\it R}$_{\rm C}${\it i}$^{+}$ & {\it V}$_{\rm J}$ (972, 0.51) & BCS & - \\
A383 & 0.188 & 02:48:03.268 & -03:31:46.43 & {\it B}$_{\rm J}${\it V}$_{\rm J}${\it R}$_{\rm C}${\it i}$^{+}${\it z}$^{+}${\it u}$^{\star}$ & {\it i}$^{+}$ (2160, 0.58) & REFLEX & $\surd$ \\
A209 & 0.206 & 01:31:53.139 & -13:36:48.35 & {\it V}$_{\rm J}${\it R}$_{\rm C}${\it i}$^{+}$ & {\it i}$^{+}$ (1188, 0.55) & REFLEX & - \\
A963 & 0.206 & 10:17:03.562 & 39:02:51.51 & {\it V}$_{\rm J}${\it R}$_{\rm C}${\it I}$_{\rm C}$ & {\it I}$_{\rm C}$ (2700, 0.61) & BCS & $\surd$ \\
A2261 & 0.224 & 17:22:26.986 & 32:07:57.89 & {\it B}$_{\rm J}${\it V}$_{\rm J}${\it R}$_{\rm C}${\it u}$^{\star}${\it g}$^{\star}${\it r}$^{\star}$ & {\it R}$_{\rm C}$ (1440, 0.55) & BCS & - \\
A2219 & 0.228 & 16:40:20.340 & 46:42:30.00 & {\it B}$_{\rm J}${\it V}$_{\rm J}${\it R}$_{\rm C}${\it I}$_{\rm 12}$ & {\it V}$_{\rm J}$ (924, 0.49) & BCS & - \\
A2390 & 0.233 & 21:53:37.070 & 17:41:45.39 & {\it B}$_{\rm J}${\it V}$_{\rm J}${\it R}$_{\rm C}${\it I}$_{\rm C}${\it i}$^{+}${\it z}$^{+}${\it u}$^{\star}$ & {\it R}$_{\rm C}$ (3420, 0.56) & BCS & $\surd$ \\
RXJ2129.6+0005 & 0.235 & 21:29:39.727 & 00:05:18.15 & {\it B}$_{\rm J}${\it V}$_{\rm J}${\it R}$_{\rm C}${\it i}$^{+}$ & {\it V}$_{\rm J}$ (1863, 0.58) & BCS & $\surd$ \\
A521 & 0.247 & 04:54:07.408 & -10:13:24.29 & {\it B}$_{\rm J}${\it V}$_{\rm J}${\it R}$_{\rm C}${\it i}$^{+}${\it z}$^{+}${\it g}$^{\star}${\it r}$^{\star}$ & {\it R}$_{\rm C}$ (1428, 0.61) & REFLEX & - \\
A1835 & 0.253 & 14:01:01.927 & 02:52:39.89 & {\it V}$_{\rm J}${\it I}$_{\rm C}${\it i}$^{+}${\it g}$^{\star}${\it r}$^{\star}$ & {\it i}$^{+}$ (1944, 0.91) & BCS & $\surd$ \\
A68 & 0.255 & 00:37:05.947 & 09:09:36.02 & {\it B}$_{\rm J}${\it R}$_{\rm C}${\it I}$_{\rm C}${\it i}$^{+}$ & {\it R}$_{\rm C}$ (2160, 0.55) & BCS & - \\
A2631 & 0.278 & 23:37:38.330 & 00:16:14.48 & {\it B}$_{\rm J}${\it V}$_{\rm J}${\it R}$_{\rm C}$ & {\it R}$_{\rm C}$ (1296, 0.60) & REFLEX & - \\
A1758N & 0.279 & 13:32:43.466 & 50:32:38.33 & {\it B}$_{\rm J}${\it R}$_{\rm C}${\it z}$^{+}${\it g}$^{\star}${\it r}$^{\star}$ & {\it R}$_{\rm C}$ (2880, 0.59) & - & - \\
RXJ0142.0+2131 & 0.280 & 01:42:03.311 & 21:31:22.64 & {\it B}$_{\rm J}${\it V}$_{\rm J}${\it i}$^{+}$ & {\it i}$^{+}$ (2136, 0.58) & - & - \\
A611 & 0.288 & 08:00:56.818 & 36:03:25.52 & {\it B}$_{\rm J}${\it V}$_{\rm J}${\it R}$_{\rm C}${\it I}$_{\rm C}${\it g}$^{\star}${\it r}$^{\star}$ & {\it I}$_{\rm C}$ (1896, 0.62) & - & $\surd$ \\
Zw7215 & 0.290 & 15:01:22.757 & 42:20:51.05 & {\it B}$_{\rm J}${\it V}$_{\rm J}${\it R}$_{\rm C}$ & {\it R}$_{\rm C}$ (1458, 0.50) & - & - \\
A2552 & 0.302 & 23:11:33.163 & 03:38:06.50 & {\it B}$_{\rm J}${\it V}$_{\rm J}${\it R}$_{\rm C}$ & {\it R}$_{\rm C}$ (1224, 0.59) & MACS & - \\
MS2137.3-2353 & 0.313 & 21:40:15.173 & -23:39:39.77 & {\it B}$_{\rm J}${\it V}$_{\rm J}${\it R}$_{\rm C}${\it I}$_{\rm C}${\it z}$^{+}$ & {\it R}$_{\rm C}$ (5562, 0.57) & MACS & $\surd$ \\
MACSJ1115.8+0129 & 0.355 & 11:15:51.881 & 01:29:54.98 & {\it V}$_{\rm J}${\it R}$_{\rm C}${\it I}$_{\rm C}$ & {\it R}$_{\rm C}$ (1944, 0.65) & MACS & $\surd$ \\
RXJ1532.8+3021 & 0.363 & 15:32:53.830 & 30:20:59.38 & {\it B}$_{\rm J}${\it V}$_{\rm J}${\it R}$_{\rm C}${\it I}$_{\rm C}${\it z}$^{+}${\it u}$^{\star}$ & {\it R}$_{\rm C}$ (2106, 0.55) & MACS & $\surd$ \\
A370 & 0.375$^{b}$ & 02:39:53.246 & -01:34:37.84 & {\it B}$_{\rm J}${\it R}$_{\rm C}${\it I}$_{\rm C}${\it i}$^{+}${\it z}$^{+}${\it u}$^{\star}${\it g}$^{\star}${\it r}$^{\star}${\it i}$^{\star}$ & {\it R}$_{\rm C}$ (3240, 0.52) & - & - \\
MACSJ0850.1+3604 & 0.378 & 08:50:06.986 & 36:04:20.45 & {\it B}$_{\rm J}${\it V}$_{\rm J}${\it R}$_{\rm C}${\it I}$_{\rm C}${\it i}$^{+}${\it z}$^{+}$ & {\it V}$_{\rm J}$ (1944, 0.81) & - & - \\
MACSJ0949.8+1708 & 0.384 & 09:49:51.785 & 17:07:08.31 & {\it B}$_{\rm J}${\it V}$_{\rm J}${\it R}$_{\rm C}${\it I}$_{\rm C}${\it i}$^{+}${\it z}$^{+}${\it u}$^{\star}$ & {\it V}$_{\rm J}$ (1692, 1.02) & MACS & - \\
MACSJ1720.2+3536 & 0.387 & 17:20:16.666 & 35:36:23.35 & {\it B}$_{\rm J}${\it V}$_{\rm J}${\it R}$_{\rm C}${\it I}$_{\rm C}${\it z}$^{+}$ & {\it V}$_{\rm J}$ (1944, 0.69) & MACS & $\surd$ \\
MACSJ1731.6+2252 & 0.389 & 17:31:39.192 & 22:51:49.96 & {\it B}$_{\rm J}${\it V}$_{\rm J}${\it R}$_{\rm C}${\it I}$_{\rm C}${\it z}$^{+}$ & {\it R}$_{\rm C}$ (864, 0.50) & MACS & - \\
MACSJ2211.7-0349 & 0.397 & 22:11:45.907 & -03:49:41.94 & {\it B}$_{\rm J}${\it V}$_{\rm J}${\it R}$_{\rm C}${\it I}$_{\rm C}${\it z}$^{+}${\it u}$^{\star}$ & {\it V}$_{\rm J}$ (1944, 0.63) & MACS & - \\
MACSJ0429.6-0253 & 0.399 & 04:29:36.001 & -02:53:05.63 & {\it V}$_{\rm J}${\it R}$_{\rm C}${\it I}$_{\rm C}$ & {\it R}$_{\rm C}$ (2592, 0.73) & MACS & $\surd$ \\
RXJ2228.6+2037 & 0.411 & 22:28:32.777 & 20:37:14.58 & {\it B}$_{\rm J}${\it V}$_{\rm J}${\it R}$_{\rm C}${\it I}$_{\rm C}${\it z}$^{+}$ & {\it R}$_{\rm C}$ (864, 0.47) & MACS & - \\
MACSJ0451.9+0006 & 0.429 & 04:51:54.684 & 00:06:18.52 & {\it B}$_{\rm J}${\it V}$_{\rm J}${\it R}$_{\rm C}${\it I}$_{\rm C}$ & {\it R}$_{\rm C}$ (2160, 0.62) & - & - \\
MACSJ1206.2-0847 & 0.439 & 12:06:12.293 & -08:48:06.22 & {\it V}$_{\rm J}${\it R}$_{\rm C}${\it I}$_{\rm C}${\it z}$^{+}${\it g}$^{\star}${\it r}$^{\star}${\it i}$^{\star}${\it z}$^{\star}$ & {\it R}$_{\rm C}$ (2520, 0.79) & MACS & - \\
MACSJ0417.5-1154 & 0.443 & 04:17:34.320 & -11:54:26.65 & {\it V}$_{\rm J}${\it R}$_{\rm C}${\it I}$_{\rm C}$ & {\it R}$_{\rm C}$ (2592, 0.74) & MACS & - \\
MACSJ2243.3-0935 & 0.447 & 22:43:21.437 & -09:35:42.76 & {\it V}$_{\rm J}${\it R}$_{\rm C}${\it I}$_{\rm C}${\it z}$^{+}${\it u}$^{\star}${\it g}$^{\star}${\it r}$^{\star}${\it i}$^{\star}${\it z}$^{\star}${\it B}$_{\rm 12}$ & {\it V}$_{\rm J}$ (972, 0.50) & MACS & - \\
MACSJ0329.6-0211 & 0.450 & 03:29:41.459 & -02:11:45.52 & {\it B}$_{\rm J}${\it V}$_{\rm J}${\it R}$_{\rm C}${\it I}$_{\rm C}${\it z}$^{+}${\it u}$^{\star}$ & {\it V}$_{\rm J}$ (1944, 0.55) & - & $\surd$ \\
RXJ1347.5-1144 & 0.451 & 13:47:30.778 & -11:45:09.43 & {\it B}$_{\rm J}${\it V}$_{\rm J}${\it R}$_{\rm C}${\it I}$_{\rm C}${\it z}$^{+}${\it u}$^{\star}${\it g}$^{\star}${\it r}$^{\star}${\it i}$^{\star}${\it z}$^{\star}$ & {\it R}$_{\rm C}$ (2592, 0.69) & MACS & $\surd$ \\
MACSJ1621.3+3810 & 0.463 & 16:21:24.749 & 38:10:09.31 & {\it B}$_{\rm J}${\it V}$_{\rm J}${\it R}$_{\rm C}${\it I}$_{\rm C}${\it z}$^{+}${\it u}$^{\star}$ & {\it I}$_{\rm C}$ (1568, 0.52) & - & $\surd$ \\
MACSJ1108.8+0906 & 0.466 & 11:08:55.154 & 09:06:02.79 & {\it B}$_{\rm J}${\it V}$_{\rm J}${\it R}$_{\rm C}${\it I}$_{\rm C}$ & {\it V}$_{\rm J}$ (1944, 0.86) & - & - \\
MACSJ1427.2+4407 & 0.487 & 14:27:16.025 & 44:07:30.51 & {\it V}$_{\rm J}${\it R}$_{\rm C}${\it z}$^{+}$ & {\it R}$_{\rm C}$ (2544, 0.59) & - & $\surd$ \\
MACSJ2214.9-1359 & 0.502 & 22:14:57.310 & -14:00:11.39 & {\it B}$_{\rm J}${\it V}$_{\rm J}${\it R}$_{\rm C}${\it I}$_{\rm C}${\it z}$^{+}${\it u}$^{\star}$ & {\it R}$_{\rm C}$ (2592, 0.52) & - & - \\
MACSJ0257.1-2325 & 0.505 & 02:57:09.089 & -23:26:03.90 & {\it B}$_{\rm J}${\it V}$_{\rm J}${\it R}$_{\rm C}${\it I}$_{\rm C}${\it z}$^{+}${\it u}$^{\star}$ & {\it V}$_{\rm J}$ (1080, 0.59) & - & - \\
MACSJ0911.2+1746 & 0.505 & 09:11:10.870 & 17:46:31.38 & {\it B}$_{\rm J}${\it V}$_{\rm J}${\it R}$_{\rm C}${\it I}$_{\rm C}${\it i}$^{+}${\it z}$^{+}$ & {\it V}$_{\rm J}$ (1908, 0.50) & - & - \\
MS0451.6-0305 & 0.538 & 04:54:11.444 & -03:00:50.76 & {\it B}$_{\rm J}${\it V}$_{\rm J}${\it R}$_{\rm C}${\it I}$_{\rm C}${\it i}$^{+}${\it z}$^{+}${\it u}$^{\star}${\it g}$^{\star}${\it r}$^{\star}${\it i}$^{\star}${\it z}$^{\star}$ & {\it R}$_{\rm C}$ (1944, 0.74) & - & - \\
MACSJ1423.8+2404 & 0.543 & 14:23:47.923 & 24:04:42.77 & {\it B}$_{\rm J}${\it V}$_{\rm J}${\it R}$_{\rm C}${\it I}$_{\rm C}${\it z}$^{+}${\it u}$^{\star}$ & {\it I}$_{\rm C}$ (1944, 0.73) & - & $\surd$ \\
MACSJ1149.5+2223 & 0.544 & 11:49:35.426 & 22:24:03.62 & {\it B}$_{\rm J}${\it V}$_{\rm J}${\it R}$_{\rm C}${\it I}$_{\rm C}${\it i}$^{+}${\it z}$^{+}${\it u}$^{\star}$ & {\it V}$_{\rm J}$ (1620, 0.54) & - & - \\
MACSJ0717.5+3745 & 0.546 & 07:17:32.088 & 37:45:20.94 & {\it B}$_{\rm J}${\it V}$_{\rm J}${\it R}$_{\rm C}${\it I}$_{\rm C}${\it i}$^{+}${\it z}$^{+}${\it u}$^{\star}${\it g}$^{\star}${\it r}$^{\star}$ & {\it V}$_{\rm J}$ (1728, 0.55) & - & - \\
CL0016+16 & 0.547 & 00:18:33.445 & 16:26:13.00 & {\it B}$_{\rm J}${\it V}$_{\rm J}${\it R}$_{\rm C}${\it I}$_{\rm C}${\it z}$^{+}${\it u}$^{\star}${\it g}$^{\star}${\it r}$^{\star}${\it i}$^{\star}${\it z}$^{\star}$ & {\it V}$_{\rm J}$ (5184, 0.62) & - & - \\
MACSJ0025.4-1222 & 0.585 & 00:25:29.907 & -12:22:44.64 & {\it B}$_{\rm J}${\it V}$_{\rm J}${\it R}$_{\rm C}${\it I}$_{\rm C}${\it z}$^{+}${\it u}$^{\star}$ & {\it V}$_{\rm J}$ (1944, 0.54) & - & - \\
MACSJ2129.4-0741 & 0.588 & 21:29:25.723 & -07:41:30.84 & {\it B}$_{\rm J}${\it V}$_{\rm J}${\it R}$_{\rm C}${\it z}$^{+}$ & {\it R}$_{\rm C}$ (3354, 0.59) & - & - \\
MACSJ0647.7+7015 & 0.592 & 06:47:49.682 & 70:14:56.05 & {\it B}$_{\rm J}${\it V}$_{\rm J}${\it R}$_{\rm C}${\it I}$_{\rm C}${\it i}$^{+}${\it z}$^{+}$ & {\it R}$_{\rm C}$ (2592, 0.62) & - & - \\
MACSJ0744.8+3927 & 0.698 & 07:44:52.310 & 39:27:26.80 & {\it B}$_{\rm J}${\it V}$_{\rm J}${\it R}$_{\rm C}${\it I}$_{\rm C}${\it i}$^{+}${\it z}$^{+}${\it u}$^{\star}$ & {\it R}$_{\rm C}$ (4869, 0.56) & - & $\surd$ \\
\hline
MACSJ1931.8-2634$^{a}$ & 0.352 & 19:31:49.608 & -26:34:33.60 & {\it B}$_{\rm J}${\it V}$_{\rm J}${\it R}$_{\rm C}${\it I}$_{\rm C}${\it z}$^{+}$ & {\it R}$_{\rm C}$ (2592, 0.73) & MACS & $\surd$ \\

\hline
\end{tabular}
\begin{flushleft}
  $^{a}$ High stellar density (Galactic bulge); only used for detailed
  PSF analysis and investigating the position of the BCG relative to
  the X-ray emission, not in the lensing analysis.\\
  $^{b}$ Redshift taken from the NASA/IPAC Extragalactic Database (NED).
\end{flushleft}
\end{table*}

\begin{figure}
\includegraphics[width=\hsize]{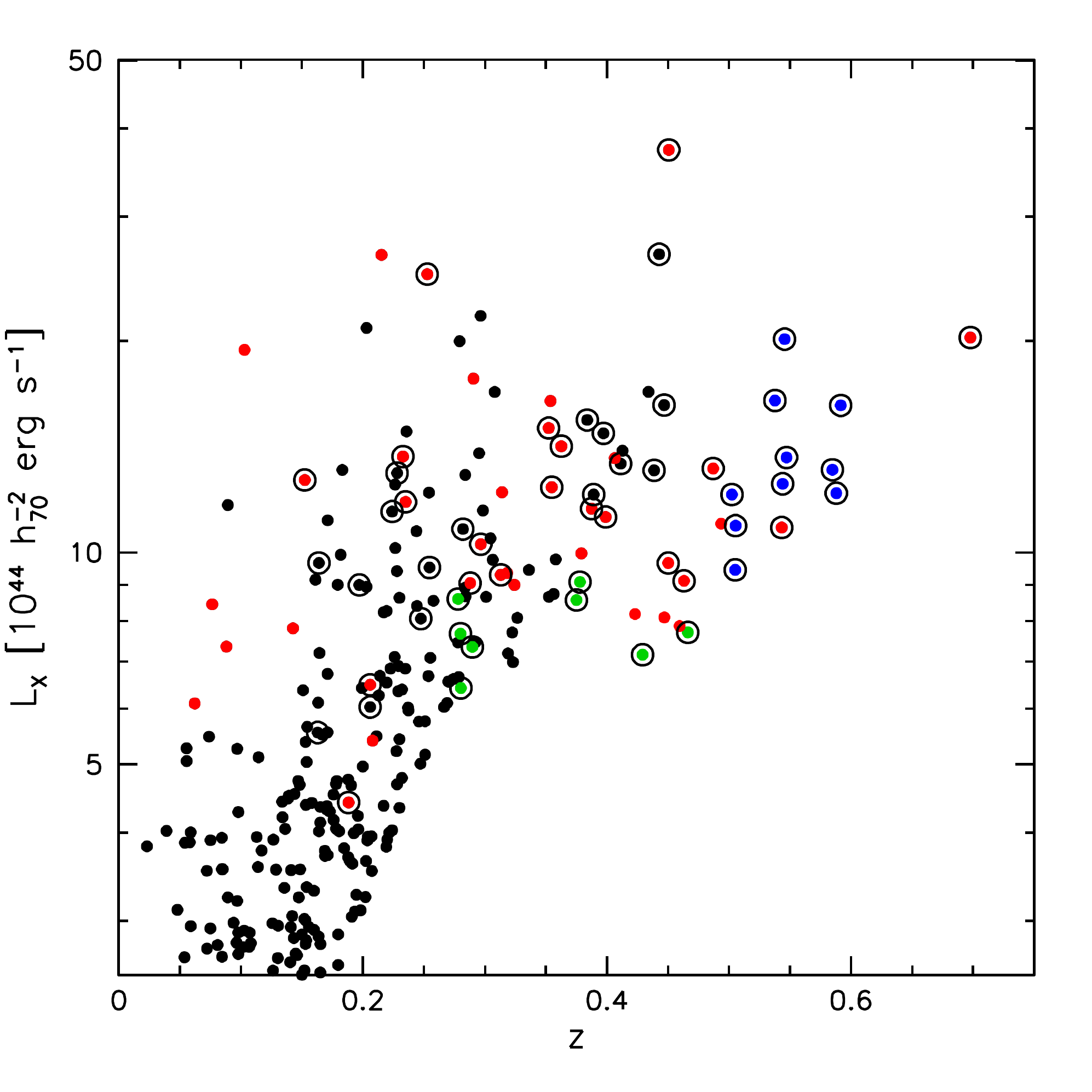}%
\caption{X-ray luminosities (in the fiducial cosmology)
  vs. redshifts of clusters in the M10 cosmology sample (black), the
  high-z MACS sample (blue), and the A08 sample of relaxed clusters
  (red). Clusters for which we derive weak-lensing mass measurements
  are marked with large open circles. Clusters that are in the weak
  lensing sample, but not in one of the three main sample are shown in
  green (see text for details). Clusters that belong to both the
  relaxed sample and the cosmology sample or the high-z sample are
  marked as relaxed clusters. (Note that the cosmology sample and
  high-z sample are disjoint.)}
\label{fig:surveys_lensing}
\end{figure}

We have acquired deep wide-field observations in at least three
filters for a total of \clusterfields\, clusters. Predominantly, these images were
taken as part of a dedicated program \citep{don07} to image clusters
selected from two cluster samples constructed from the Rosat All Sky
Survey \citep[RASS,][]{tru93}, namely the MAssive Cluster Survey
\citep[MACS;][]{eeh01,ebd07,eem10} and the Brightest Cluster Survey
\citep[BCS;][]{eeb98,eea00}. The clusters were observed with
SuprimeCam on the Subaru telescope, and with MegaPrime on the
Canada-France-Hawaii Telescope (CFHT) for {\it u}-band imaging. We
supplemented these data with further imaging from the SuprimeCam and
CFHT data archives, where available.  In addition, we searched the
archives for multi-color imaging of the sample of relaxed clusters
defined by \citet[][hereafter A08]{ars08}. For the SuprimeCam data
reduction, we made use of additional ``empty fields'' (extragalactic
fields without bright stars or large foreground galaxies) taken at
approximately the same epoch as our observations
(Sect.~\ref{sect:flatfields}); these data include an additional three
clusters (A1758, A370, and RXJ0142.0+2131).

The clusters included in this sample span a wide range in redshift
($0.15<z_{\rm Cl}<0.7$), as well as dynamical state. As such, several
(overlapping) subsamples can be identified which are of particular
interest to various aspects of cluster cosmology. For instance,
\cosmosample of the clusters are included in the cosmological analysis
of \citet[][hereafter jointly referred to as M10]{mar10,mae10}, which
used clusters selected from the BCS, the ROSAT-ESO Flux Limited X-ray
\citep[REFLEX,][]{bsg04}, and MACS.  In future work, we will
incorporate our lensing mass measurements for these clusters into our
framework to simultaneously determine cosmological parameters and
scaling relations between mass proxies and true mass, as measured on
average by the weak-lensing masses.  Relaxed clusters are of special
interest for investigating the bias in X-ray hydrostatic mass
estimates.  For this reason our sample includes \relaxedsample
dynamically relaxed clusters from the A08 sample, 13 of which are in
common with the M10 sample.  The sample also contains all 12 clusters
above $z>0.5$ in the Faint MACS sample \citep{ebd07}. Although these
are not actually part of the M10 sample, their calibration is
particularly interesting for future work with newer, larger cluster
surveys.  Our study is the first effort to calibrate mass measurements
with ground-based weak lensing at these redshifts.  An additional
seven clusters in the sample do not belong to any of these three
categories. Note that a significant fraction of the data taken
  for the {\it Weighing the Giants} project is currently being analyzed
  independently by the CLASH collaboration \citep{pcb12,umn12}.

Fig.~\ref{fig:surveys_lensing} illustrates these target clusters in a
plot of X-ray luminosity vs. redshift for the parent samples, marking
those clusters included in this study.
Table~\ref{tab:clusters_table1} summarizes the cluster sample and the
multi-color data used in this work. The lensing band is chosen as the
deepest image (i.e. the image with the highest number density of objects) with
the best seeing from the SuprimeCam {\it V}$_{\rm J}$, {\it R}$_{\rm
  C}$, {\it I}$_{\rm C}$, {\it i}$^{+}$, and MegaPrime {\it
  r}$^{\star}$ images (see Sect.~\ref{sect:set-processing}).

\begin{table}
  \caption{Overview of the filters used in this work. Note that the response 
    functions of the Subaru Johnson/Cousins 
    filters are considerably more top-hat-like than the original Johnson/Cousins 
    filter functions, making them well suited for photometric redshift 
    determination.
  }
\label{tab:filters}
\begin{tabular}{l c l}
\hline 
Instrument / Telescope & Short Filter Name & Long Filter Name \\
\hline
SuprimeCam & {\it B}$_{\rm J}$ & Johnson {\it B}-band \\
@ Subaru & {\it V}$_{\rm J}$ & Johnson {\it V}-band \\
 & {\it R}$_{\rm C}$ & Cousins {\it R}-band \\
 & {\it I}$_{\rm C}$ & Cousins {\it I}-band \\
 & {\it i}$^{+}$ & SDSS {\it i}-band \\
 & {\it z}$^{+}$ & SDSS {\it z}-band \\
\hline
MegaPrime  & {\it u}$^{\star}$ & SDSS {\it u}-band \\
@ CFHT & {\it g}$^{\star}$ & SDSS {\it g}-band \\
 & {\it r}$^{\star}$ & SDSS {\it r}-band \\
 & {\it i}$^{\star}$ & SDSS {\it i}-band \\
 & {\it z}$^{\star}$ & SDSS {\it z}-band \\
\hline
CFH12K & {\it B}$_{12}$ & Johnson {\it B}-band \\
@ CFHT & {\it I}$_{12}$ & Cousins {\it I}-band \\
\end{tabular}
\end{table}

\section{Data Reduction}
\label{sect:optical_data}

The basis for our data reduction is the {\sc GaBoDS/Theli} pipeline
\citep[][with additional features described in
  \citealt{sch13}]{esd05}, which is optimized for processing
multi-chip mosaic-camera data to produce weak-lensing quality final
images. The bulk of our data, especially the lensing band exposures,
are from SuprimeCam.  Below we describe adaptations of the {\sc
  GaBoDS/Theli} pipeline to the SuprimeCam images analysis. We follow
the terminology of \citet{esd05}, and refer the reader to that work
for more in-depth indiscussions of the standard reduction steps.

To pre-process the MegaPrime datasets, we use the highly automated
pipeline of \citet{ehl09}. For a few clusters, we also include data
gathered with the CFH12K camera at the CFHT, which are processed with
the standard version of the {\sc GaBoDS/Theli} pipeline.

For the data processing description, it is helpful to distinguish
between {\it run}-specific steps, which refer to data grouped
according to when it was observed (e.g. single nights, or in our case,
periods of a few months), and {\it set}-specific processing, which is
applied to all observations of a single field. In
Sect.~\ref{sect:pre-processing} we describe the main aspects of the
run-specific pre-processing; in Sect.~\ref{sect:set-processing} we describe the subsequent set-specific
steps of the data reduction.

\subsection{Pre-processing of SuprimeCam data}
\label{sect:pre-processing}

\begin{table*}
\caption{Summary of the SuprimeCam configurations spanned by our data. Column (1) gives the identifier we use to refer to a configuration; columns (2) and (3) specify the range of dates each configuration spans. Column (4) lists the number of chips that are read out. Column (5) lists the manufacturer (along with the number of CCDs, if the array is mixed).  Column (6) briefly describes the main characteristics / changes from the previous configuration. In column (7) we indicate whether we use a configuration for the lensing shape measurements. The bulk of our data is from configuration 10\_2.}
\label{tab:suprimecamconfigs}
\begin{tabular}{c c c c c c c}
\hline
Configuration & Start Date & End Date & No. of & CCD Types & Comments & Used for \\
Name & & & CCDs &  & & Lensing \\
(1) & (2) & (3) & (4) & (5) & (6) & (7) \\
\hline
8 & 2000-07-28 & 2000-11-21 & 8 & MIT/LL (4), SITe (4) & non-linearity & -- \\
9 & 2000-11-21 & 2001-03-17 & 9 & MIT/LL (5), SITe (4) & non-linearity & -- \\
10\_1 & 2001-03-17 & 2002-08-01 & 10 & MIT/LL & new CCDs & $\surd$ \\
10\_2 & 2002-08-02 & 2008-07-01 & 10 & MIT/LL  & new electronics & $\surd$ \\
10\_3 & 2008-07-21 & 2011-07-02 & 10 & Hamamatsu Photonics & new CCDs; major upgrade & -- \\
\end{tabular}
\end{table*}

SuprimeCam \citep{mks02} is one of the best-suited instruments for
cluster weak-lensing observations. The large aperture of the Subaru
telescope \citep[8.2\,m;][]{ika04}, along with the typically good
seeing at Mauna Kea \citep[median seeing 0.7-0.8 arcsec][]{mks02},
enable robust shape measurements of faint galaxies with modest
exposure times. The field of view of $34\arcmin \times 27\arcmin$ is
well matched for observing massive clusters at $z\gtrsim0.2$.

The SuprimeCam detector is a mosaic of 10 CCDs, with $2048\times4092$
pixels each.  The camera underwent numerous upgrades in the time
period of our observations (2000--2008).
Table~\ref{tab:suprimecamconfigs} gives a brief overview of the
different camera configurations for our data. In particular, we
distinguish between ``early'' data taken before March 27, 2001, before
the mosaic was fully populated, and standard 10-CCD data taken
thereafter.

The early data, corresponding to configurations ``8'' and ``9'', are
hampered by several defects. The CCDs have noticeably non-linear
response -- for photometry measurements we correct for this in the
central six CCDs (see App.~\ref{appendix:early_data}). The CCDs have
several cosmetic defects, and one CCD has a pixel indexing issue,
where the central part of the CCD appears offset by $\sim 0.5\arcsec$
(see App.~\ref{appendix:early_data}). Because of these issues, we do
not use data from these early configurations for shape measurements.

In March 2001, the CCDs were replaced with 10 newer MIT/Lincoln Labs
(LL) CCDs. These CCDs are well suited for weak-lensing purposes, and
were used for the bulk of our observations. The only notable issues
are the limited dynamic range (the response becomes saturated at
$\sim\!35000$ ADUs above the bias level), and the noticable
charge-transfer inefficiency (CTI) and lower quantum efficiency of the
top left chip (``w67c1''); this chip was omitted for lensing
purposes. These CCDs remained the heart of SuprimeCam until July
2008. Because of an electronics upgrade in August 2002, this period is
split into two configurations ``10\_1'' and ``10\_2''. In July 2008,
the CCDs were replaced with Hamamatsu Photonics chips, as prototypes
for HyperSuprimeCam. These have the unusual artefact that the pixels
vary in shape and size; the divisions between pixels are not straight
lines, but are curved (S. Miyazaki, private communication).  These
chips do not suffer from the low saturation level of the previous CCD
generation, but the non-linearity is about 1\% over the full range --
this is noticeable if the flat-field does not have similar counts to
the science data. Because we have only a single night of data from
this ``10\_3'' configuration, we do not use these data for our lensing
analysis.

Each change in camera configuration requires independent data
reduction set-up.  Additionally, we split the data into epochs of a
few months, over which the flat-fields are quite stable (see
discussion in Sect.~\ref{sect:flatfields}). For each epoch, we
download additional bias frames, flat-fields, and empty fields to be
used for the superflat (a flat-field derived from ``empty field''
night-sky observations, Sect.~\ref{sect:flatfields}), from the SMOKA
archive \citep{byi02}. From these we assemble the master calibration
frames for each epoch.

\subsubsection{Overscan, bias, \& dark corrections}

The first data reduction step is the subtraction of the {\it bias},
i.e. the expected counts in an exposure of zero seconds.
The bias level is corrected in two steps: first, the count levels
in the overscan pixels of each frame are measured and
subtracted; second, a master bias frame is created by stacking many
zero-length exposures (after first subtracting the overscan).
This master bias is then subtracted from all frames.

The bias level of SuprimeCam is fairly high ($\sim\!10000$ ADU), but the
overscan regions are relatively small ($\sim$30--80 pixels).
Subtracting the overscan line by line, as done by the {\sc
  GaBoDS/Theli} pipeline, is thus noisy, leading to striped features
in the bias frames. All of our science images are taken with
broad-band filters; thus the absolute sky level (and associated noise)
is typically much greater than noise in the bias frames. Hence, noise
from the overscan correction is not an issue.

If the image of a bright star happens to fall on the read-out edge of
a chip, the overscan is significantly enhanced due to spill-over into
the overscan region. When subtracting the overscan for each pixel row,
this leads to an oversubtraction. The resulting dark trail of the star
is later masked.

To create the master bias, we median-combine the bias frames
taken over several months (the same epoch as for the flat-fields; see
next section). An exception is configuration 8, where the electronics
apparently were adjusted between new moon periods; here each new moon
period is treated separately.

For all science and flat-field frames, the appropriate master bias is
subtracted.  Master dark frames, assembled per camera configuration, are
used to identify hot pixels.

\subsubsection{Flat fields, superflats, fringing}
\label{sect:flatfields}

\begin{figure}
\includegraphics[trim=0.6cm 0.3cm 0.5cm 0.3cm,width=\hsize]{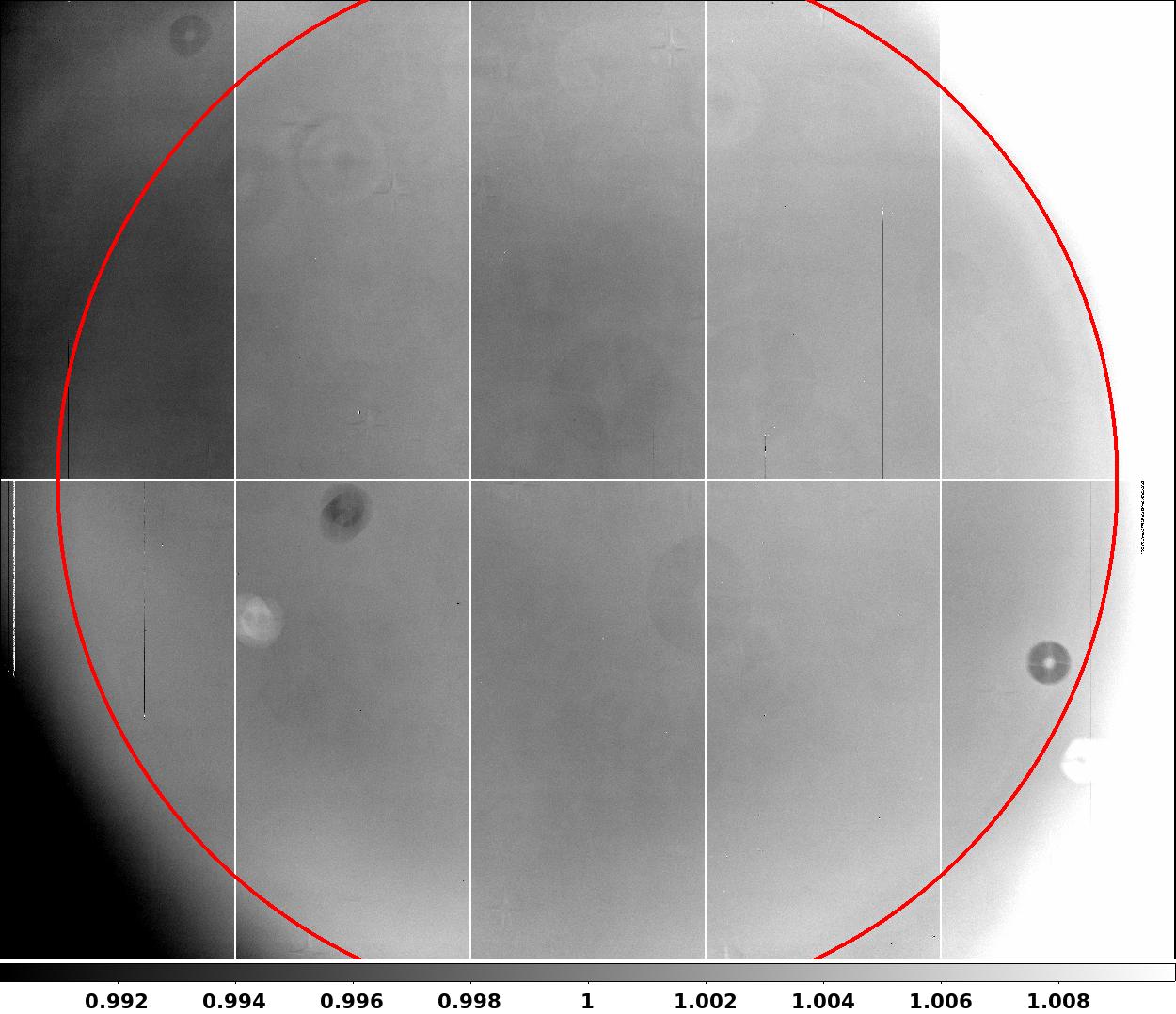}%
\caption{Ratio of two flat-fields in the {\it R}$_{\rm
    C}$ filter, between two epochs (January - August 2004, and January
  - April 2005). Within a radius of 15~arcmin (indicated by the red
  circle), the flat-field pattern is stable to within $\pm1\%$,
  and differs between epochs only through dust grains on the CCD
  window or dewar window, visible as ring-shaped features in this
  flat-field comparison. Beyond the central 15~arcmin, the
  flat-field is considerably less stable.}
\label{fig:flatfields}
\end{figure}

In the second data reduction stage, the counts in science exposures
are corrected for varying pixel sensitivity. Each pixel has an
intrinisic sensitivity; furthermore, the focal plane of any instrument
is inhomogenously illuminated (areas further from the center receive
less light), and the pixel scale varies over the field of view. To
correct for different pixel sensitivities, each science frame is
divided by a {\it flat-field}, an image of a uniformly lit source
(i.e. no spatial variation over the camera's field of view).

We follow a two-stage flat-fielding process, as suggested by
\cite{esd05}. In the first step, a master flat-field constructed from
median-stacking dedicated domeflat or twilight flat observations 
is applied.  To
construct this master flat-field, we downloaded all domeflats and
twilight flats available for the timespan of our observations from the
SMOKA archive.  We investigated the variability of these flat-fields,
and even though we find significant variability in the overall
  illumination pattern between single flats from individual nights, we find
that when averaging over several nights, the flat-fields are very
stable in time within a radius of $\sim\!15$~arcmin from the center
(Fig.~\ref{fig:flatfields}).  Beyond this distance, the field is
strongly vignetted -- the corners of the field receive only about half
as much light as the center.  When averaging flat-fields from many
months, the corners show variability of the order of 5\%. During the
analysis of the point spread function (PSF) (Sect.~\ref{sect:psfcorr})
we find that the PSF can change rapidly across the vignetting
radius. Because of this and the flat-field problems, we mask all
pixels beyond a radius of 15~arcmin.

For the master flat-fields, we average the available flat-field
exposures to epochs of several months each. For each epoch of science
observations, we divide the science images by the appropriate
master flat-field. These flat-fields correct the pixel-to-pixel response
variation, as well as the large-scale illumination (but see below and
Sect.~\ref{sect:scatteredlight}). When both domeflats and twilight
flats are available, we use the one that yields flatter corrected
science frames.

Flat-fields constructed from domeflats and skyflats correct the
pixel-to-pixel response variation, but can leave residual chip-scale
response variations of the order of 3\% \citep{esd05}.  In the second
flat-fielding step, we correct these with a {\it superflat}, which is
constructed from night-time observations of ``empty fields'' -- fields
where the objects are significantly smaller than the chip sizes. Most
of our cluster fields are ``empty'' and can be used in the
construction of the superflat; exceptions are fields with very bright
foreground stars and their reflections within the optical
path. However, multiple exposures of the same field, taken at approximately
the same time, are subject to similar gradients in the sky background,
reflections within the telescope, etc. To mitigate these effects, we
downloaded additional empty field observations, taken within a few
months of our own observations, from the SMOKA archive.

The superflat is constructed by stacking empty field images (each
flat-fielded as described above) where detected objects are removed
\citep[see][for details]{esd05}. The stacked image is then heavily
smoothed and applied as a multiplicative correction to the flat
fielded images.

The superflat also provides the basis for correcting fringing (interference
patterns in thinned CCDs) in the {\it I}$_{\rm C}$, {\it i}$^{+}$, and
{\it z}$^{+}$ exposures. We find that the method of \citet{esd05} works
well on the SuprimeCam data.

\subsubsection{Initial astrometric \& photometric solution}
\label{sect:scamp1}

\begin{figure}
\includegraphics*[width=\hsize]{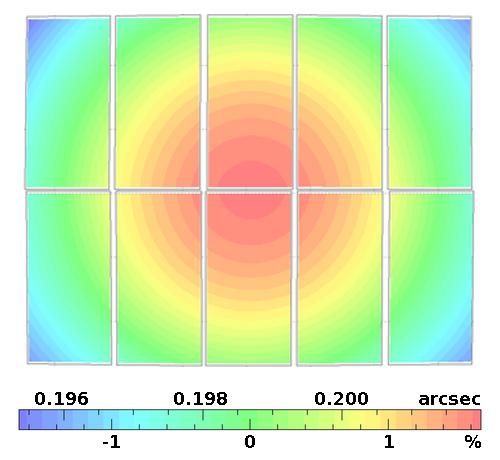}%
\caption{A typical distortion pattern of the SuprimeCam focal plane,
  determined with {\sc SCAMP}. The figure shows the relative positions
  of the CCDs on the sky. The local pixel scale has been color-coded
  in order to illustrate the variation of the mapping between sky
  coordinates and detector coordinates.  The overall variation is
  within $\pm 1.5$\% of the average pixel scale, and its spatial
  variation can be well described by a third-order polynomial.}
\label{fig:scamp_distort}
\end{figure}

We determine an initial astrometric solution for the images in
order to identify objects across exposures for the star-flat
correction (see next section and Paper II). We use {\sc SExtractor}
\citep{bea96} to extract object catalogs for all observations of the same
target field (from all observation nights and filters). For SuprimeCam
images, we include saturated objects in these catalogs since, due to
the low saturation threshold, there is generally very little overlap
in magnitude range between astrometric reference catalogs and the
unsaturated objects in our exposures.

We use {\sc SCAMP} \citep{ber06} to simultaneously find the
astrometric solution for all observations of a given field.  Where
available, the fields are cross-matched to the Sloan Digital Sky
Survey \citep[SDSS DR6;][]{aaa08}, and otherwise to the Two Micron All
Sky Survey \citep[2MASS;][]{scs06}. The robustness of the astrometric
solution is judged from the contrast value of the highest peak in the
cross-correlation of the catalogs (reported by {\sc SCAMP}), and the
distortion patterns of all input configurations. These should have a
regular shape as shown in Fig.~\ref{fig:scamp_distort}; false
solutions produce patterns that deviate strongly from this. To aid
{\sc SCAMP} in finding the correct astrometric solution, we
constructed typical linear astrometric headers for each camera
configuration and rotation (for many of our exposures, the camera was
rotated by $90^{\circ}$ between exposures) from fields with robust
astrometric solutions derived from matching to SDSS. These are
substituted for the original image headers, and greatly expedite the
process of finding the best astrometric solution.

\subsubsection{The star-flat}
\label{sect:scatteredlight}

In the previous steps, we took care to construct a flat-field to
accurately calibrate the response of the camera to illumination from a
uniformly bright sky.  However, for wide-field cameras, this is not a
map of the actual sensitivies of all pixels: On the one hand, the
pixel scale (and therefore the area each pixel subtends on the sky)
can vary by several percent over the field of view
(Fig.~\ref{fig:scamp_distort}), breaking the underlying assumption of
traditional flat-field procedures that each pixel is the same size and
therefore should receive the same amount of light.  Furthermore, light
from a number of sources (the sky itself, the lamps illuminating the
dome-flat screen, etc.) is scattered into the camera
\citep{mac04}. The center of the focal plane is more exposed than the
edges, and so this additional light contribution affects the center of
the image most. The standard flat-field procedure thus
overestimates the sensitivity of pixels at field center (as this is
where the registered counts are highest). Typical gradients are of the
order of 10\% for wide-field cameras such as Wide-Field Imager
\citep[WFI][]{koc03} at the ESO/MPG 2.2m telescope and MegaPrime
\citep{rcg09}. This is equivalent to a variation of the zero-point by
$\sim 0.1$~mag across the field, which is an unacceptable systematic
error for projects requiring precise photometry, such as ours.

The variation in zero-point across the field can be determined from
observations of dense fields of objects with known brightnesses
(e.g. from SDSS), or from series of observations with dither patterns
that are a significant fraction of the field of view.  We use a
combination of these two methods to determine the ``star-flat'', a
correction to the flat-field, for SuprimeCam images, described in detail in
Paper II. We find the magnitude of the correction to be similar to
other wide-field cameras, $\sim\!10$\% across the field.

\subsubsection{Background subtraction}

After the star-flat correction, the sky background in the
images is no longer uniform. We adapt the procedure of \citet{esd05} to
handle non-flat backgrounds robustly by extending it to a true
``two-pass'' method: First, we use the standard {\sc SExtractor}
method to determine the large-scale background, which we then subtract
from the image. In the residual image, we identify objects using a
very low detection threshold. All pixels belonging to objects are
flagged as not-a-number ({\tt NaN}) in the frame to be corrected,
which prevents {\sc SExtractor} from considering them in the
background estimation. We then re-estimate the sky background in these
object-blanked frames, and subtract this background frame from the
original frame.

\subsubsection{Stellar halo subtraction}

\begin{figure*}
\begin{minipage}{0.48\hsize}
\includegraphics[width=\hsize]{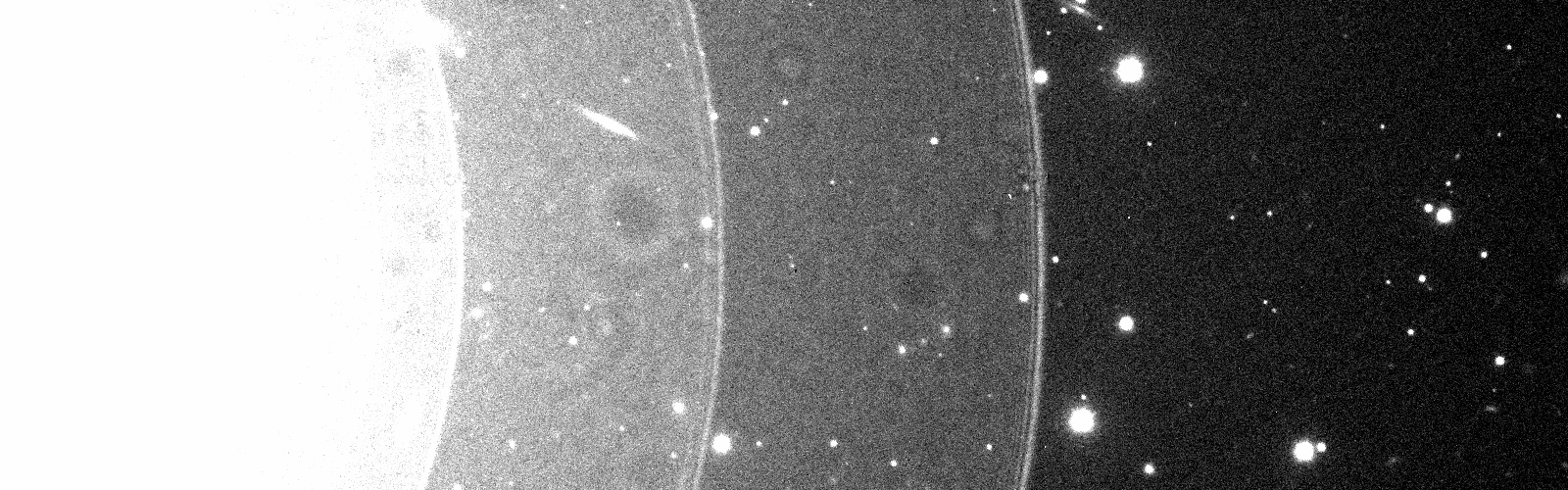}
\includegraphics[width=\hsize]{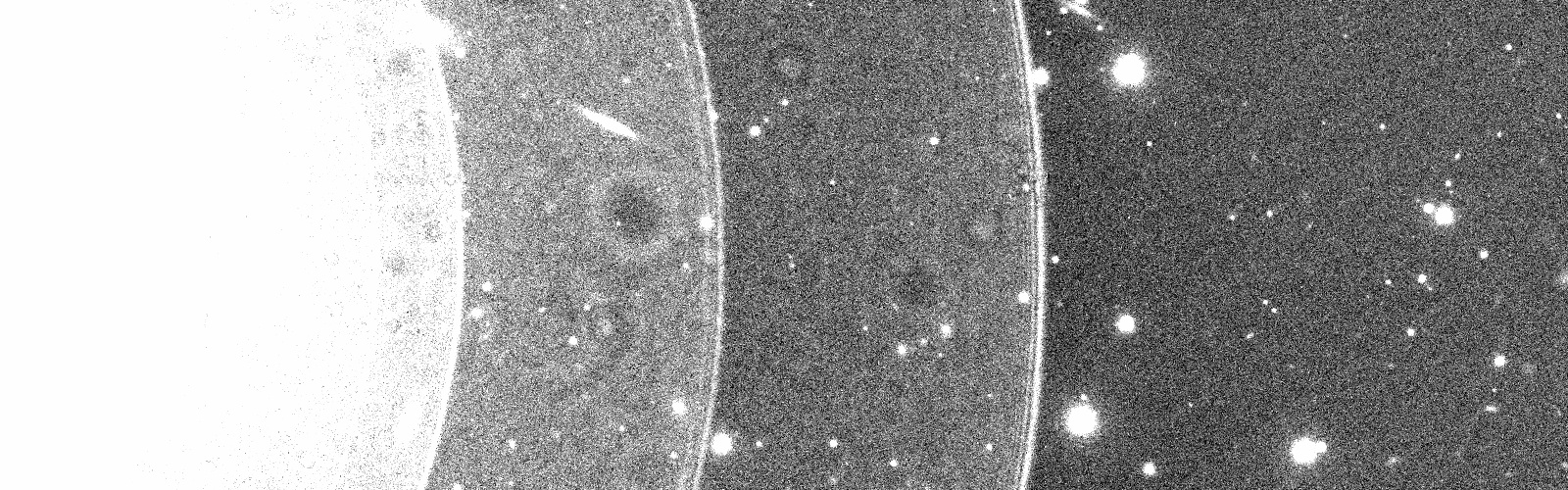}
\includegraphics[width=\hsize]{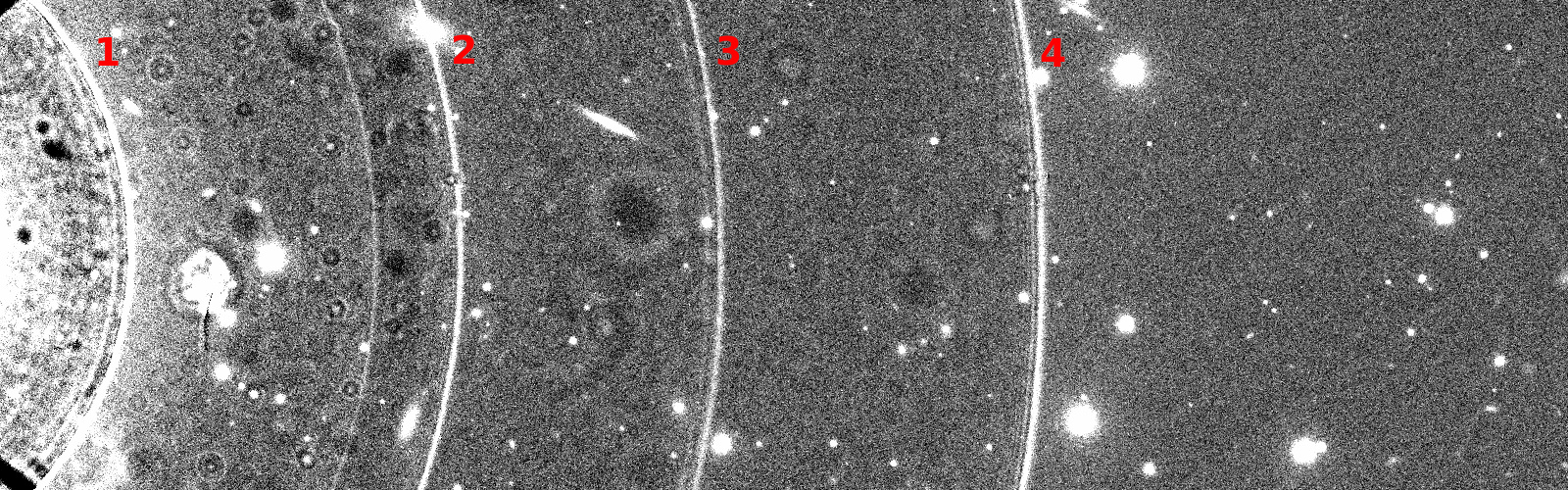}
\end{minipage}\hspace{0.02\hsize}
\begin{minipage}{0.48\hsize}
\includegraphics[width=\hsize]{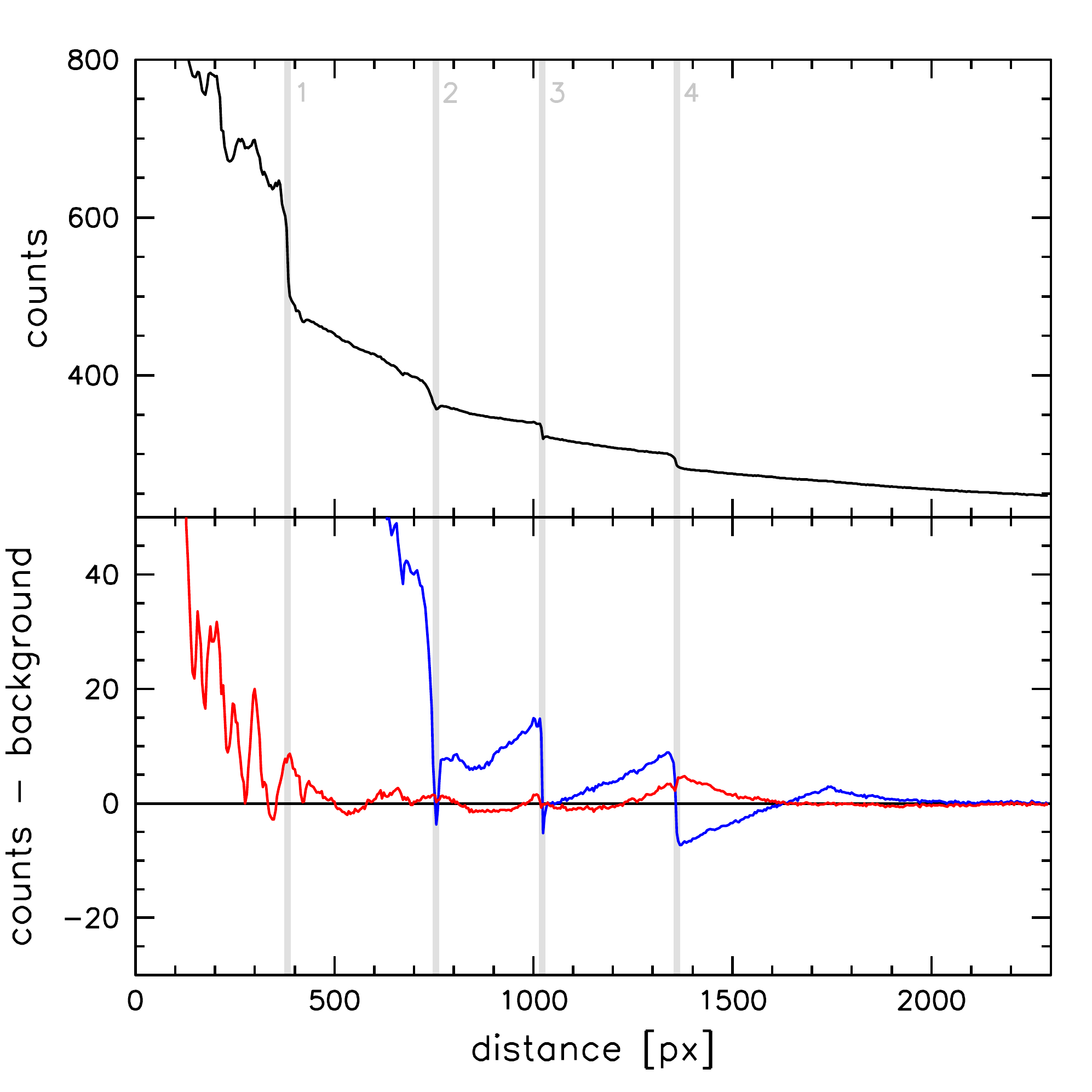}%
\end{minipage}
\caption{Left: Detail of the reflection halos of a bright star. The
  top panel shows an image cutout before background subtraction, with
  a bright star to the left. Roughly centered on the star are four
  reflection halos, whose surface brightness decreases with increasing
  distance from the star. The transitions between the reflection are
  marked by bright rings and a sharp drop in surface brightness. The
  middle panel shows the same image cutout after attempting to
  subtract the background without removing the stellar halo first. The
  remaining background is clearly brighter within each halo edge, and
  darker outside the edges. The bottom panel shows the image after
  subtracting the stellar halo first, and then proceeding with the
  general sky subtraction (the color scale is the same for the middle
  and bottom panels). The sky background is now even across halo
  boundaries. Right: radial profiles of the background counts,
  azimuthally averaged around the same bright star as in the images to
  the left. The top panel shows the background counts before any
  subtraction: note the distinct steps (marked by the grey vertical
  lines), and the nearly linear fall-off between the rings. The rings
  are not precisely concentric; for these profiles, the center of the
  fourth ring was used. The bottom panel shows the radial profiles
  after sky subtraction.  For the blue profile, the standard sky
  subtraction was applied without first removing the stellar halo;
  note the residual saw-tooth pattern extending well beyond the
  original halo. For the red profile, the stellar halo was subtracted
  first, and then the background was subtracted. This technique allows
  us to salvage the area between rings 2 and 4, which would otherwise
  be lost due to the uneven background. A 100 pixel buffer
    (indicated by the grey bars in the right panel) around each ring
  is masked in the lensing image, as well as the area within ring 1.
  For stars as bright as this one, the area within ring 2 is also
  masked, as it is considerably noisier, and further artefacts from
  the reflection halo are visible.}
\label{fig:stellarhalosub}
\end{figure*}

In many of the fields, the brightest stars are surrounded by halos,
which stem from reflections from elements along the optical path, such
as the dewar window and the filter. The largest such halo has a radius
of 4.7\arcmin; i.e. it subtends a noticeable fraction of the
SuprimeCam field.  Within each annulus of reflected light, the
brightness changes only slowly, but the outer edges are marked by
brighter rings.

The smoothing length of the background subtraction described above
would result in sawtooth residuals across these ring boundaries
(Fig.~\ref{fig:stellarhalosub}), rendering a significant area unusable
for reliable object shape measurements (due to non-linear background
variations) -- we would mask out the entire region within
  1500~pixels (5~arcmin) from the star. This would be undesirable and
unnecessary, since the areas within the rings have almost uniform
backgrounds after flat-fielding. We therefore subtract the reflection
halos of the bright stars in the lensing images {\it before} the
background subtraction. This is done with a procedure similar to the
one presented in \citet{shm09}. Preliminary coadded images are
examined for stars with visible halos; once these stars are
identified, the astrometric solution is inverted and used to locate
the stars on the input frames. The position of the halos relative to
the star varies across the field, but can be predicted from the star's
position within the field of view.  The innermost halo (for the
brightest stars, also the second halo; see
Fig.~\ref{fig:stellarhalosub}) and a 100 pixel buffer around each ring
are masked in both background subtraction and later for photometric
and lensing measurements.  A piece-wise linear step function model in
halo-centric radius is then fit to the pixels contained within each
ring, but outside the next smallest ring. This piecewise model is then
subtracted from the original image, scaled to set the background to
the average background level around the largest halo. The result is a
flat image with the stellar core and halo rings masked out (see the
caption of Fig.~\ref{fig:stellarhalosub} for details). We have
  verified that the inclusion of objects thus recovered does not
  cause any systematic shift in the weak lensing mass measurements.

\subsubsection{Weights and masking}
\label{sect:weightmaps}

For each frame, we also create a weight map, following the procedure
in \citet{esd05}. The bases for the weight maps are the normalized
flat-fields. Note that the star-flat correction is {\it not}
applied to the weight maps -- the weights are intended to track the
relative noise across the field, which can be estimated from the
variation in the original flat field.

We also use the weight maps to track bad pixels, by setting their
weight to zero. We use master dark frames and flat-fields to identify hot
and dead pixels, as well as image artefacts such as large dust grains.
As described above, the field area outside a radius of 15 arcmin is
masked due to rapidly varying PSF patterns and instability of the flat
field and scattered-light correction.

In some images, the telescope autoguider blocks light from the upper
$\sim 20\%$ of the top row of chips. The exact position of this shadow
moves from image to image. We automatically detect and mask this
shadow by measuring the background level in the bottom $2/3$ of the
chip and comparing that to the background in the top third. We divide
the upper region into subsets of $500\times80$ pixels.  If the level of
the backgound in one of these subsets is significantly below that of
the lower $2/3$ of the chip, that subset is masked out.

For many of the datasets, the camera was rotated by $90^{\circ}$ for
half of the exposures. This gives us the opportunity to recover sky area
otherwise lost to saturation spikes from bright stars. The count level
of saturated pixels is not constant, however. We automatically mask
saturation spikes by initially identifying extended streaks of highly
elevated counts in pixel columns, and then expanding these masks to
include contiguous elevated regions.

We automatically mask satellite tracks by initially smoothing
the image and then identifying regions on the edge of each
CCD that are elevated above the background level. Satellite trails are
masked when all pixels lying on a straight line connecting any
two of these regions are elevated above the background as well.  

Cosmic rays are detected with the standard neural network of the {\sc
  GaBoDS/Theli pipeline}. Since the edges of cosmic
ray hits are not always included in the mask, we expand each cosmic
ray mask by one pixel in each direction.

The masking steps described above are applied to all images
automatically. In the lensing band, we furthermore mask by hand the
remaining artefacts pertaining to individual exposures (asteroids,
off-axis reflections from bright stars, missed satellite tracks). To
aid this process, we coadd the field both with median stacking and
weighted mean. Comparing these images, image artefacts in individual
frames become readily visible. By inverting the astrometric solution of
the input frames, these artefacts can be efficiently masked on the
original exposures, before the final coaddition.

\begin{figure}
\includegraphics[width=\hsize]{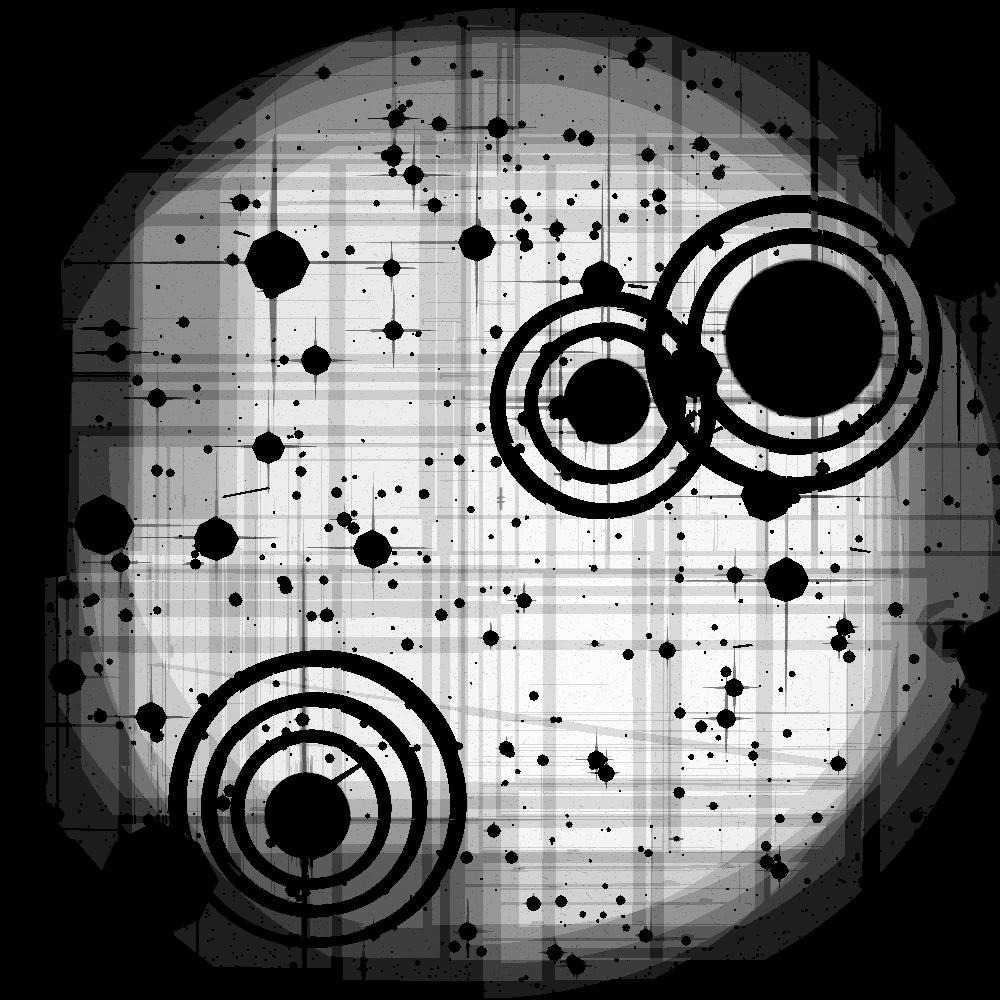}
\caption{An example of a weight map for a coadded lensing image. The
  weight map tracks the total exposure time accumulated at each pixel
  (but scaled for photometric offsets).  A total of eight exposures
  were coadded; the camera was rotated by $90^{\circ}$ after the first
  four. The brightness scales linearly with the weight.  Black areas
  (no weight) have been masked on the final image -- these are mainly
  bright stars, and the rings around the brightest stars.  The uneven
  weight across the image is readily apparent. At the field edges,
  this is largely because the two orientations cover different parts
  of the sky.  For the lensing images, we reject the chip in the top
  left corner because of its noticeable charge transfer inefficiency
  -- this is visible as the low weight regions at the top left and
  bottom left parts of the image. Also discernible as areas of lower
  weight are the chip gaps, saturation spikes from bright stars, and
  masks due to a satellite and image artefacts.} 
\label{fig:weightmask}
\end{figure}

Fig.~\ref{fig:weightmask} shows an example weight map of a coadded
lensing image.

\subsection{Final astrometry, relative photometry, resampling and coaddition}
\label{sect:set-processing}

\begin{figure}
\includegraphics*[width=\hsize]{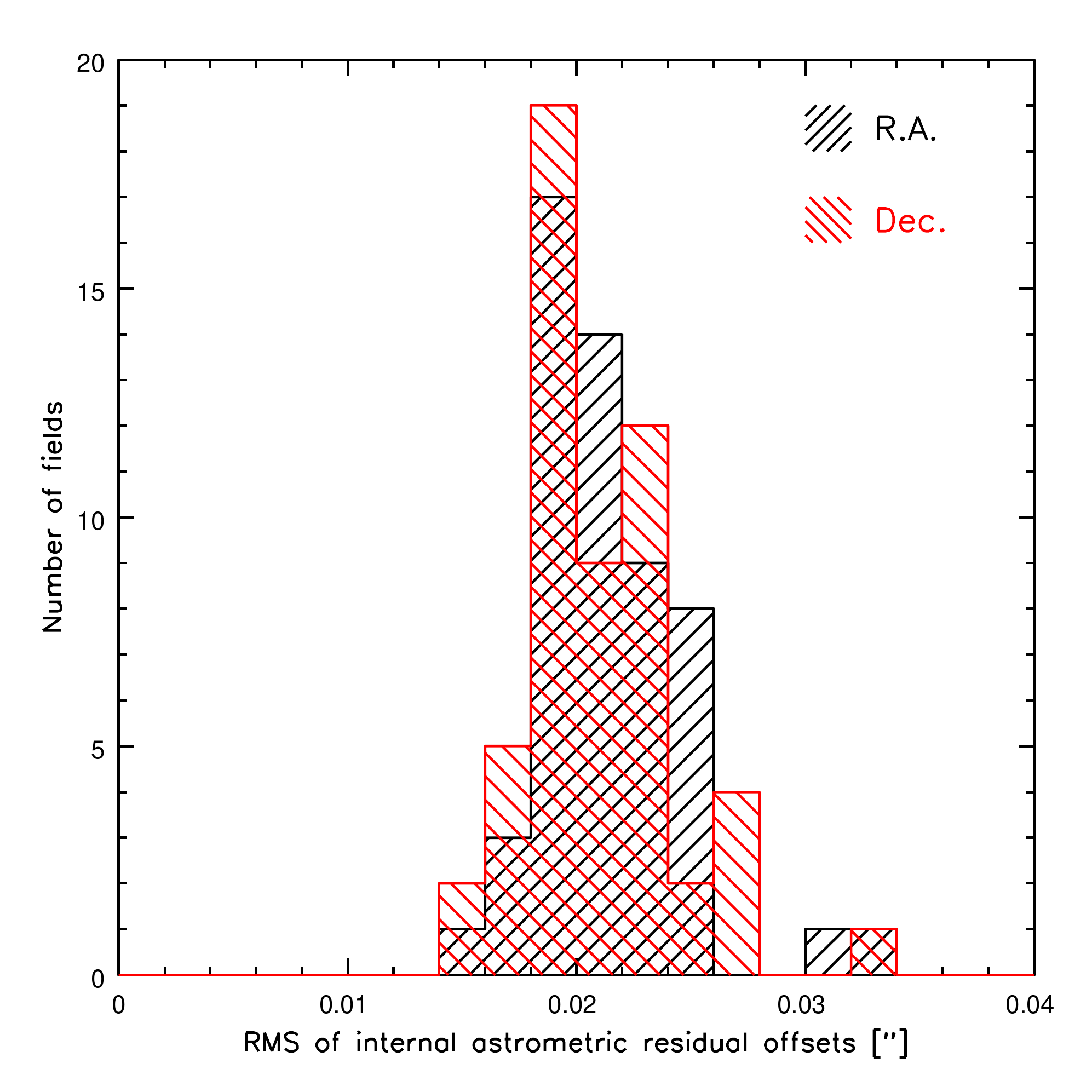}
\caption{Illustration of the astrometric accuracy achieved with {\sc
    SCAMP}. For each cluster field, {\sc SCAMP} computes the RMS of
  all internal astrometric residuals, across all input catalogs. This
  figure shows the distribution of these RMS values for all our fields
  after the second {\sc SCAMP} pass (Sect.~\ref{sect:set-processing}).
  The RMS along the R.A./Dec. axis is shown as a black/red histogram,
  for objects with S/N$>50$.  The remaining astrometric uncertainty is
  only $\sim1/10^{\rm th}$ of the pixel scale.}
\label{fig:scamp_astrerr}
\end{figure}

At this stage, the data reduction switches from run-specific
processing to being set-specific, where {\it set} refers to all
observations of a given (cluster) field.

We run {\sc SCAMP} a second time on all images of a given cluster, for
each image passing the linear solution of the first pass through {\sc
  SCAMP} (Sect.~\ref{sect:scamp1}) as the astrometric starting
point. This time, all saturated objects are removed from the
catalogs. This second pass further improves the internal astrometric
solution -- the positions of objects are now consistent across all
frames to within a fraction of one pixel
(Fig.~\ref{fig:scamp_astrerr}). Such high astrometric precision is
critical for coadding the lensing images, as well as color
measurements across pixel-matched images.  At the same time, {\sc
  SCAMP} determines the relative zero-points of exposures in the same
filter -- this was not possible in the first pass, where saturated
objects were included, and where we had not yet applied the
star-flat. Note that as part of the star-flat correction, we also
determine possible zero-point offsets between the CCDs (see Paper II).

The images (and weight maps) are resampled onto a common
astrometric grid (i.e. common {\tt CRVAL} and {\tt CDELT} WCS
keywords) using {\sc SWarp} \citep{bmr02}. The pixel scale is
homogenized to 0.2~arcsec. We use the {\tt Lanczos3} kernel for
resampling because of its robust signal conservation and noise
properties.  The flux in all images is scaled to an exposure time of
1s. From the resampled images, we create several coadded images:
Firstly, for each filter, we create a median-stacked image from all
available exposures (unless the seeing is worse than 1.5~arcsec, or an
exposure has previously been rejected, e.g. because the sky background
is as high as the saturation limit, or there are bright reflections
over the entire frame).  In the detection band, this will also be the
detection image (see Sect.~\ref{sect:catalogs}), for the other
filters, this image serves primarily visualization purposes.

Due to the heterogeneous nature of our dataset, we choose to measure
object fluxes on individual exposures (see
Sect.~\ref{sect:fluxmeasurements}).  For this purpose, we create one
image for each exposure, pixel-matched to the deep median image;
i.e. we resample the chips from a given exposure onto the common grid.

At this point we select the lensing band(s), where our aim is to
select an image with good seeing, great depth, and uniform observing
conditions. The only filters considered for this purpose are those
with the highest through-puts, i.e. {\it V}$_{\rm J}$, {\it R}$_{\rm
  C}$, {\it I}$_{\rm C}$, {\it i}$^{+}$, and {\it
  i}$^{\star}$. Exposures from SuprimeCam configurations 8, 9, and
10\_3 are not considered. In many fields, the choice is clear, but in some,
we select more than one lensing band. We currently make no effort to
combine shape measurements from different bands. However, these
cluster fields with multiple lensing bands provide an excellent
opportunity for testing our methods across different observations. For
some lensing bands, observations from more than one night are
available. If the seeing / exposure time is markedly different, only
images from the night with the best seeing / greatest depth are
used. If the observations are of similar quality, we use all exposures
for the lensing image, where we create one coadded image per night and
per camera rotation angle, as well as a coadded image from all nights
(see Sect.~\ref{sect:psf_influence_on_mass} for further discussion).
For each lensing image, the input exposures are coadded with a
weighted coaddition (according to the weight maps described in
Sect.~\ref{sect:weightmaps}), which provides the optimal noise
properties under homogeneous conditions. These images are used for
the shape measurements (see Sect.~\ref{sect:shearmeasurements}).

In the coadded lensing image, we mask features that could compromise the shape
measurements. Our images contain large numbers of saturated stars, for
which the PSF wings are visible and need to be masked. The
majority of these can be masked automatically by placing template
masks according to entries in the USNO-B1 catalog \citep[see][for
details]{ehl09}; the rest are masked manually. Large foreground
galaxies are also masked. Our images are deep, and in some of
them, faint Galactic cirrus is visible. We mask the brighest knots of
this nebulosity in the fields that are most affected. We track the
masks in the weight image.

\section{Catalog Creation and Photometry}
\label{sect:catalogs_photom}

\subsection{Object detection}
\label{sect:catalogs}

We detect objects in the median coadded image of the lensing band
using {\sc SExtractor}. Detecting on the median coadd instead of the
lensing image (which is coadded with a weighted coaddition) has the
advantage of avoiding residual spurious artefacts (e.g. asteroid
trails, cosmic ray hits), which may have been missed at the masking
stage. Also, since only the highest-quality exposures were used for
the lensing image, the median image is deeper for some cluster fields,
facilitating object detection. At this stage, we aim to construct
  a catalog that is highly complete and correctly deblends faint
  objects close to one another. This allows us to later test for
  possible biases in the shear measurements from close neighbors (see
  Paper~III). Meeting these requirements requires fairly aggressive {\sc
    SExtractor} settings ({\tt DETECT\_MINAREA=15, 
    DETECT\_THRESH=0.5, ANALYSIS\_THRESH=0.5, 
    DEBLEND\_NTHRESH=64, DEBLEND\_MINCONT=0.00001}), as also
  noted by other authors \citep[e.g.][]{caa07,gfg13}. This catalog
provides our initial master catalog. For the following photometry
measurements, we make heavy use of {\sc SExtractor}'s dual-image mode
to measure photometric properties of the objects in this catalog on
other images (Sect.~\ref{sect:fluxmeasurements}). It will also form
the basis of the object catalogs for which we measure weak-lensing
shape parameters (Sect.~\ref{sect:shapemeassubsect}).  These
  subsequent steps automatically reject any spurious detections,
  both as part of the shape measurement process, as well as with
  requirements on signal-to-noise and object magnitudes in at least
  three filters (see Paper~III for more details).

\subsection{Flux measurements}
\label{sect:fluxmeasurements}

Our dataset is heterogeneous. Some cluster fields have been
observed repeatedly in different seeing conditions and/or different
configurations of SuprimeCam. Furthermore, most of the SuprimeCam
configurations have at least two different types of CCDs in the array,
leading to chip-dependent effective response curves.  To properly
account for these effects, we measure photometric properties from
individual exposures.

Each exposure has been resampled onto an image that is pixel-matched
to the detection image. For photometry, these images are convolved to
a common PSF size using a Gaussian kernel \citep[see][for
details]{ehl09}.  The final seeing is chosen to be the largest seeing
encountered in any single exposure of the cluster in all
filters. However, if the seeing differences are large, the Gaussian
scaling of the PSF is expected to break down. We therefore adopt the
strategy of the EDisCS photometry \citep{wcs05} and limit the maximum
PSF size to be no more than the seeing of the detection image plus
0.3\arcsec. Exposures with seeing worse than this are also included,
but are not convolved.

For photometric redshifts, robust color measurements are essential.
Due to possible color gradients, galaxy colors need to be measured
from flux measurements of the same physical aperture on the galaxy. We
measure fluxes within a fixed aperture of 3\arcsec. This is the same
aperture as was used in the COSMOS survey \citep{caa07}. The COSMOS
survey forms the basis of our reference deep field for computing
cluster masses.  Furthermore, we used the COSMOS photometry and
redshifts to extensively test the quality of our photometric
redshifts, and our algorithm to incorporate individual redshift
probability distributions (see Paper~III).

The flux measurements for each object from the different exposures are
combined according to filter, configuration, and chip type (e.g. {\it
  R}$_{\rm C}$, configuration 8, SITe chip type).  Individual
measurements where any of the object pixels are masked are rejected.
We measure the zero-point offsets between exposures with high
signal-to-noise objects, and adjust the relative zero-points
accordingly; note that the zero-point is no longer a single number for
an exposure, but is specific to the chip type.  The flux measurements
for an object of the same filter, configuration and chip type are
combined with an iterative clipped mean, where measurements are
rejected if they are discrepant by more than $5\sigma$ from the mean
(where $\sigma = \sqrt{\sigma_{\rm meas}^2 + \sigma_{\rm mean}^2}$
accounts for both the error on the measurement $\sigma_{\rm meas}$,
scaled for the presence of correlated noise, and the error on the
mean, $\sigma_{\rm mean}$).  Each galaxy may have multiple
measurements in each filter corresponding to the different
configurations and CCD types with which it was observed. These
  are treated effectively as different filters throughout the analysis
  (although the color terms between different chip types are small).

\subsubsection{Photometric calibration}

For most of our observations, standard star observations are not
available, and even when they are, these exposures are usually too
sparse to allow a robust calibration of the zero-points as a function of
airmass and color terms. A large number (25/\clusterfields) of our
fields fall into the SDSS footprint, allowing us to calibrate the
observations directly. To do so, we identify stars from the lensing
image (where a clean selection of high signal-to-noise, but
unsaturated, stars is essential for the PSF correction; see
Sect.~\ref{sect:psfcorr}). For these stars, we calibrate the absolute
zero-point of a single filter (typically the {\it R}$_{\rm C}$
  band from the 10\_2 configuration, excluding the top left CCD, which
  has a different type from the other nine CCDs) against the SDSS
photometry, taking into account the color terms (due to different
response functions) between the SDSS filters and the filter/chip
  combination that we calibrate here.

For all the MACS clusters that are not in the SDSS field, we took
calibration exposures to allow calibration against SDSS: we took short
(3s) exposures of the cluster field, and a field within SDSS at
similar airmass, back-to-back in time. We can thus calibrate the short
exposure of the SDSS field, and transfer the zero-point to the cluster
exposure. The deeper exposures are then calibrated with respect to the
short cluster exposure.

Of the remaining 11 fields without SDSS calibration, we have MegaPrime
$r^{\prime}$ imaging for six. All MegaPrime imaging is photometrically
calibrated by taking short calibration exposures during photometric
conditions \citep{mac04}. For these fields, we can therefore use the
$r^{\prime}$ band as the absolutely photometrically calibrated band.

This leaves five clusters without SDSS or MegaPrime calibration data.
From the calibrated fields, we find that the extinction-corrected {\it
  R}$_{\rm C}$ zero-point is relatively stable, within $\lesssim
0.1$~mag. Using data from the CFHT SkyProbe monitor\footnote{\tt
  http://www.cfht.hawaii.edu/Instruments/Elixir/skyprobe/}, we
confirmed that the outliers in this distribution correspond to nights
with noticeable extinction due to cloud cover. SkyProbe shows that the
{\it R}$_{\rm C}$ band exposures of the five uncalibrated clusters
were taken on photometric nights; hence we can assign the typical
zero-point to these images.

Photometric redshifts are based mainly on the spectral energy
distribution of object; therefore, the color calibration, i.e. the
zero-point differences between bands, is more critical than the
absolute photometric calibration. In the presence of a spectroscopic
training set for the observed fields, several implementations of
photometric redshift estimators allow one to solve for the relative
zero-points of the different bands. Since we have spectroscopic
redshifts only for a minority of our fields, we instead calibrate the
relative zero-points by matching the colors of stars against the
expected stellar color-color locus \citep[e.g.][]{hsr09}. In Paper II we
describe this procedure in detail.  We find that we can calibrate the
relative zero-points to sufficient precision that our photometric
redshifts have negligible bias.

\section{Shear Measurements}
\label{sect:shearmeasurements}

One of the key ingredients of weak-lensing measurements is robust
shear estimation of faint background galaxies. We use the KSB method
developed by \citet{ksb95}, \citet{luk97}, and \citet{hfk98}, with
modifications by \citet{evb01}. We choose the KSB method because it
has been extensively tested \citep{hvb06,mhb07} and, despite its
simplicity, performs robustly on a variety of data.

\subsection{KSB in a nutshell}
\label{sect:ksbsummary}

Here we provide a brief summary of the KSB algorithm. For a more
in-depth review, see e.g. \cite{bas01}.

In the KSB algorithm, the complex ellipticity $\bm{e}$ for each object
is estimated from the second moments $Q_{ij}$ of the object's light
distribution $I(\vec{\theta})$:
\be
e = e_1 + ie_2 = \frac{Q_{11} - Q_{22} + 2 i Q_{12}}
{Q_{11} + Q_{22}} \quad ,
\ee
where
\be
Q_{ij} = \int \dif^2 \theta\:
  I(\vec{\theta}) \:  W_{r_g}(|\vec{\theta}|)  \: \theta_i \theta_j \quad ,
\label{eq:2nd-moments}
\ee 
and $W_{r_g}$ is a Gaussian weight function of width $r_g$
\citep[we use the {\tt FLUX\_RADIUS} measured by {\sc SExtractor} for
$r_g$; see the discussion in][]{sch08}.  Following \citet{evb01}, we
use the same weight function to define a signal-to-noise ratio of each
object:
\be S/N = \frac{\int \dif^2
  \theta I(\vec{\theta}) W_{r_g}(|\vec{\theta}|)} {\sigma_{\rm sky}
  \sqrt{\int \dif^2 \theta W_{r_g}^2(|\vec{\theta}|) }} \quad ,
\label{eq:lensing-snr}
\ee 
which captures the uncertainty in the shape measurement. This
  shape-specific signal-to-noise measure is different from the flux
  signal-to-noise: for our data, objects with a shape $S/N$ of $\sim
  3$ are highly significant detections with ${\tt FLUX\_ISO} /
  \sigma_{\tt FLUX\_ISO} \sim 30$.

Gravitational lensing has two effects on the observed shapes of
background galaxies: the convergence $\kappa$ scales the image of a
background object isotropically, and the shear $\bfmath{\gamma}$
stretches it anisotropically.  The combined effect is the reduced
shear:
\be
\bm{g} \;=\; \frac{\bfmath{\gamma}}{1-\kappa}\;.
\ee

In the limit of small shear ($g\ll1$), and only small anisotropy of the
telescope's point spread function (PSF), the (seeing-convolved) intrinsic
ellipticity $\bm{\hat{e}}^0$ of an object is transformed to the observed ellipticity
\be
\bm{e} \;=\; \bm{\hat{e}}^0 +  P^{\rm g}\bm{g}
+ P^{\rm sm}\bm{q}^{\star}\; ; \;
P^{\rm g} = P^{\rm sh} - P^{\rm sm}(P^{\star \rm sm})^{-1} P^{\star \rm sh}\;.
\label{eq:ksb}
\ee 
The stellar anisotropy kernel $\bm{q}^{\star}$ describes the anisotropic
component of the PSF; the smear polarizability tensor $P^{\rm sm}$
describes the susceptibility of an object to the PSF anisotropy (and
largely depends on the apparent object size); and the shear
polarizability tensor $P^{\rm sh}$ describes the object response to
the shear. $P^{\rm sm}$ and $P^{\rm sh}$ are measured from an object's
third and fourth order moments. The starred quantities of these
tensors are measured on stars; but note that the weight function must
be adjusted to the object size \citep{hfk98}.

$\bm{q}^{\star}$ is measured from stars (for which the gravitational
shear $\bm{g}$ and intrinsic ellipticity $\bm{\hat{e}}^0$ vanish), so
that the galaxy ellipticities can be corrected for the anisotropy of
the PSF. The reduced shear is then
\be
  \bm{g} = (P^{\rm g})^{-1} (\bm{e}^{\rm aniso} - \bm{e}^{0})\quad ;  \quad
\bm{e}^{\rm aniso} = \bm{e} - P^{\rm sm}\bm{q}^{\star} \quad .
\ee
The source ellipticity $\bm{e}^{0}$ is of course not
known. KSB instead returns
\be
\hat{\bm{g}} = (P^{\rm g})^{-1} \bm{e}^{\rm aniso} \quad .
\ee

Because galaxies are randomly oriented (at least to the precision
required for cluster weak lensing), the average ellipticity of an
unlensed population of galaxies vanishes: $\ave{\bm{e}^{0}} = 0 =
\ave{(P^{\rm g})^{-1}\bm{e}^{0}}$. Hence, the justification for KSB is
that the expectation value $\ave{\hat{\bm{g}}}$ is an estimate of
$\bm{g}$.

Since the trace-free part of the $P^{\rm g}$ tensor is much smaller
than the trace, we follow \citet{evb01} and make the approximations
\be
(P^{\star \rm sm})^{-1} P^{\star \rm sh} \rightarrow \frac{{\rm Tr}[P^{\star \rm sh}]}{{\rm Tr}[P^{\star \rm sm}]}
=: T^{\star}
\quad ; \quad
(P^{\rm g})^{-1} \rightarrow \frac{2}{{\rm Tr}[P^{\rm g}]}
\ee
which also reduces sensitivity to noise \citep{hvb06}.

\subsection{Shape measurements and star selection}
\label{sect:shapemeassubsect}

\begin{figure*}
\includegraphics[width=0.49\hsize]{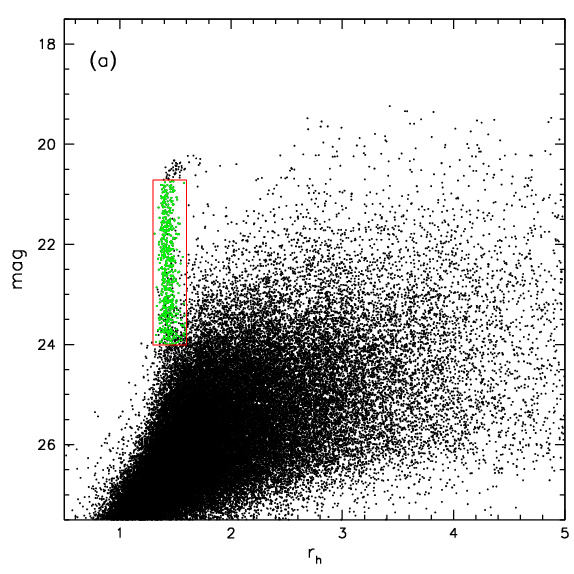}\hspace{0.01\hsize}%
\includegraphics[width=0.49\hsize]{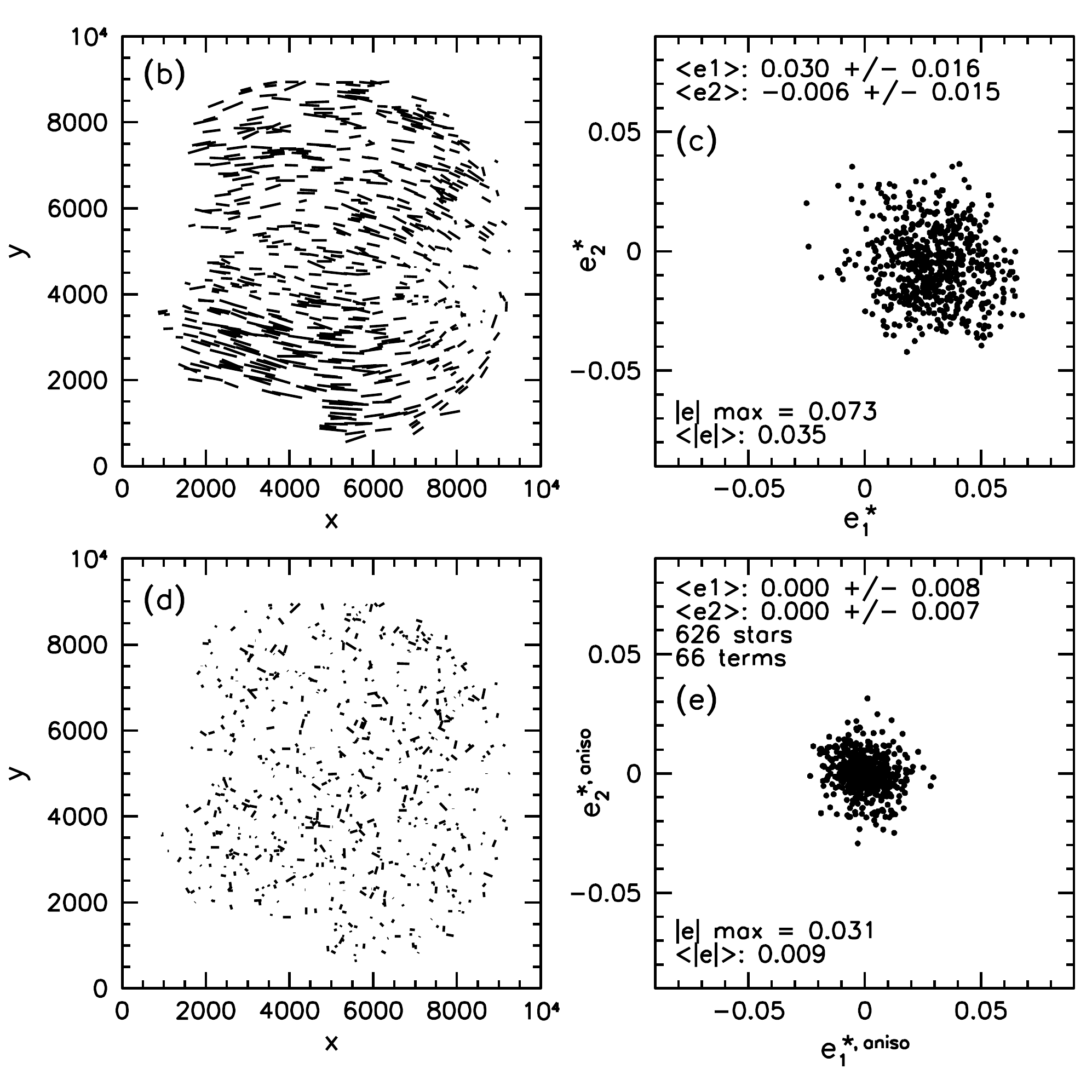}
\caption{Illustration of the star selection and PSF anisotropy
  correction, shown for the \bbullet$\;$ {\it V}$_{\rm J}$ field.
  Panel (a) shows the magnitude - radius diagram for all objects in
  the catalog. The stellar sequence is clearly visible at $r_h \sim
  1.4$px. The preselection of stars is shown as the red box. Stars
  that are not rejected as significant outliers in the initial
  second-order polynomial fit are shown in green. In the figures on
  the right, we illustrate the correction itself. Panel (b) shows the
  uncorrected stellar ellipticity pattern: at the position of each
  star, the measured ellipticity is indicated as a line with a length
  proportional to $\left|e^{\star}\right|$, with orientation $\phi =
  0.5 \arctan(e_2^{\star}/e_1^{\star})$ . Panel (c) shows the
  distribution of $e_1^{\star}$ vs. $e_2^{\star}$ values. Panel (d)
  shows the residual ellipticities after correcting the PSF pattern
  with an eighth-order polynomial (see Fig.~\ref{fig:psf2}); the
  distribution of corrected $e_1^{\star}$ and $e_2^{\star}$ values is
  shown in panel (e). Note that for the SuprimeCam data, the size of
  the coadded output image ($33{\rm arcmin}\times33{\rm arcmin}$) is
  larger than the area covered by the input frames -- because we mask
  pixels more than 15~arcmin from each exposure center, the
  non-zero-weight part of the output image appears roughly
  circular. The ``missing corners'' visible in panels (b) and (d) are
  due to the top-left chip ``w67c1'', which has noticeable CTI and
  lower QE than the other chips, and therefore is rejected.}
\label{fig:psf}
\end{figure*}

For the shape measurements, we first run {\sc SExtractor} in
dual-image mode, with the median coadded image as the detection image, and
the image coadded for lensing (coadded with a weighted average) as the
second, ``measurement'', image. This step mainly serves the purpose of
obtaining the {\sc SExtractor} {\tt FLUX\_RADIUS}($=r_g$) measurement
for each object on the actual lensing image, while retaining the same
object selection and identification as for the photometric catalogs.

For objects with $0.5 \le r_g \le 10$ we measure the ellipticities as
described above with the code {\sc analyseldac} \citep{evb01}.  Larger
objects are unlikely to be background galaxies; while smaller objects
are predominantly spurious detections.

{\sc analyseldac} also provides a more robust measure of the
half-light radius, $r_h$. We select stars for the PSF correction in a
diagram of magnitude vs. $r_h$, where stars with sufficient
signal-to-noise, but which are not saturated, form a well-defined
sequence (Fig.~\ref{fig:psf}). The star selection is refined by fitting
the PSF anisotropy across the field with a second-order polynomial,
and rejecting 5$\sigma$ outliers (for the actual PSF correction we use
a higher-order polynomial; see next section). The number of stars
varies considerably in the sample, from 300 to 3000 per field. With
the roughly circular field of view of radius $\sim 15$~arcmin, this
corresponds to $0.4 - 4 {\rm \; stars \; arcmin}^{-2}$, with the
typical number density being $\sim 1 {\rm \; star \; arcmin}^{-2}$.
The stars selected here also form the basis of the relative
photometric calibration between bands via the stellar color-color
locus (see Paper II); the color-color diagrams confirm that the
stellar sample selected here is fairly clean.

\subsection{PSF anisotropy correction}
\label{sect:psfcorr}

Correcting for the PSF anisotropy, $\bm{q}^{\star} = (P^{\star \rm
  sm})^{-1} \bm{e}^{\star}$, is essential, as it can mimick shear.
However, the PSF can only be measured at discrete locations in the
image plane, namely at the positions of suitable stars.  We measure
$\bm{q}^{\star}$ for the selected stars (see above), and fit a
polynomial function to the spatial variation in both components
(Fig.~\ref{fig:psf}).

An important question here is whether the PSF is stable across chip
boundaries -- this is the case if the CCDs are sufficiently
coplanar. If the CCDs are mounted at different heights, the focus
position and hence PSF shape changes abruptly across the CCD
boundary. On a single exposure, or coadded images with dither patterns
of the size of the gaps between chips, this could be accounted for by
fitting the PSF variation for each CCD separately. For our data,
however, the dither patterns are significantly larger, and for most
fields, the camera has been rotated by 90$^{\circ}$ between exposures.
We have tested for ``jumps'' of the SuprimeCam PSF in images with
excellent seeing and a large density of stars (see
App.~\ref{sect:planarity}) and find that in configurations 10\_1 and
10\_2 (used for shape measurements), the CCDs are remarkably coplanar
-- there are no measurable discrete PSF jumps across chip
boundaries. (This is not the case in early data, another reason to
disregard those data for lensing purposes.) Hence, the PSF pattern can
be corrected across the full field of view, without the need to correct
on a chip-by-chip basis.

\begin{figure*}
\includegraphics[trim=0.3cm 1.3cm 0.5cm 2.5cm,width=0.52\hsize]{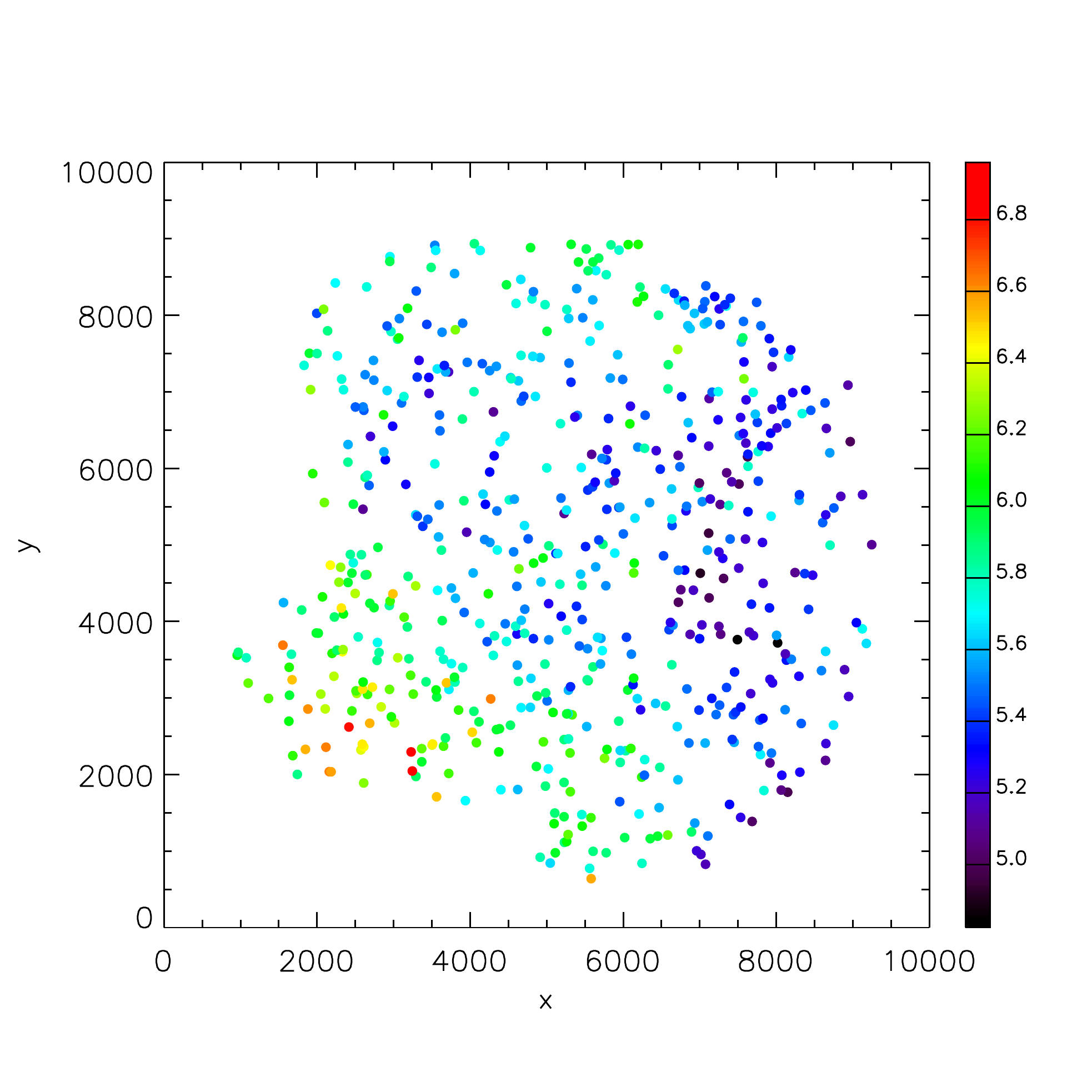}\hspace{0.03\hsize}%
\includegraphics[width=0.45\hsize]{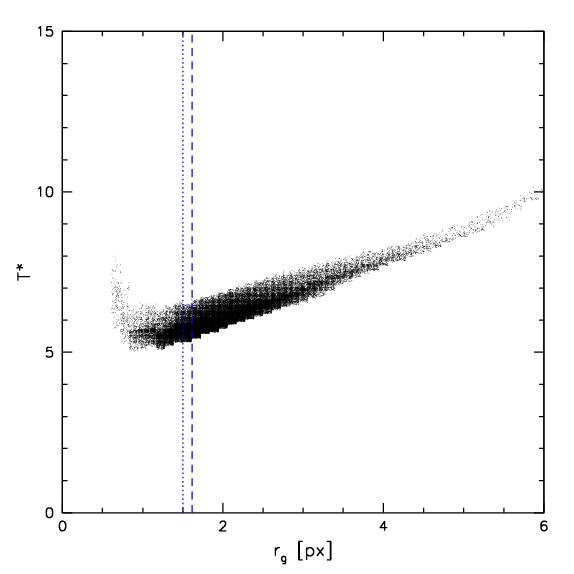}
\caption{Illustration of the PSF isotropy correction as a function of
  position and object size, for the \bbullet$\;$ {\it V}$_{\rm J}$
  field.  The left panel shows the variation of $T^{\star} = {\rm
    Tr}[P^{\star \rm sh}]/{\rm Tr}[P^{\star \rm sm}]$ across the field
  of view. At the position of each star, we indicate $T^{\star}$,
  measured with a weight function of width $r_g=0.6$arcsec. Note the
  20\% variation across the image, which we fit with a second-order
  polynomial. The right panel shows $T^{\star}$ as a function of object
  size, evaluated at each object position with the appropriate weight
  function. The spread in $T^{\star}$ at a given $r_g$ reflects the
  spatial variation shown on the left. We impose a minimum size
  criterion of $r_g>1.5 {\rm px}$ (illustrated by the dotted blue
  line), to avoid the upturn at smaller scales. For comparison, the
  median $r_g$ of stars is indicated by the dashed blue line. }
\label{fig:isocorr}
\end{figure*}

The PSF of SuprimeCam (and MegaPrime) can vary considerably over the
field of view, even in single exposures. We fit the entire field with
a single polynomial, but find that usually a high order polynomial is
required (from fourth order up to a limit of tenth order). Other authors
instead divide the field into subsets and fit these with second order
polynomials, but since this creates discontinuities in the PSF, we
prefer the single, higher-order polynomial.

We developed a number of criteria to judge the quality of the PSF
correction, and to choose the minimum polynomial order required to
achieve a good fit. This process is described in
App.~\ref{sect:psfquality}.

Note that we calculate $\bm{q}^{\star}$ using the weight function of
each star (i.e. $r_g = r_g^{\star}$). As \citet{hfk98} have argued,
all quantities in Eq.~\ref{eq:ksb} should be measured with the same
weight function as the object (galaxy) to be corrected. However, if
the anisotropy of the PSF does not vary with isophote level (which is
a good approximation for many ground-based instruments),
$\bm{q}^{\star}$ is independent of the width of the weight
function. Measuring it with the stellar weight function automatically
reduces the noise in this measurement, making the PSF measurement more
robust. We find no systematic shift in shear measurements when
measuring $\bm{q}^{\star}$ with the galaxy weight function, consistent
with the results of \citet{hvb06}.

\subsection{PSF isotropy correction}
\label{sect:psfisocorr}

The isotropic part of the PSF (expressed as the $P^{\rm g}$ tensor),
circularizes object shapes. If inadequately corrected, this can lead
to a dilution of the shear measurement. For the calculation of $P^{\rm
  g}$, $T^{\star} = {\rm Tr}[P^{\star \rm sh}]/{\rm Tr}[P^{\star \rm
    sm}]$ needs to be measured from stars. This quantity is sensitive
to the size of the weight function and therefore must be measured with
the weight function appropriate for the object to be corrected, and
within the same aperture used for the object. Furthermore, $T^{\star}$
can vary spatially, as the size of the PSF can vary within the field
of view. In well-focused exposures, the PSF tends to be smaller at the
center of the field of view than towards the edges. If this is left
unaccounted for, it can lead to systematic biases in the radial shear
profile, and thus the cluster mass measurement. We measure
  $T^{\star}$ at discrete values $r_g^b$ of the weight function size
  over the range $0.33 \le r_g^b \le 10$, in 0.33 pixel increments.
  For each weight function scale, we fit the spatial variation of
$T^{\star}$ with a second-order polynomial across the images, which
suffices to capture the variation (Fig.~\ref{fig:isocorr}).  For
  each object, we then assign $T^{\star}$ according to the fit for
  $r_g^b$ closest to the object size $r_g$. The trend of $T^{\star}$
  with object size is linear for objects larger than the PSF, but
  shows an upturn at $r_g \lesssim 1.5$~px (right panel of
  Fig.~\ref{fig:isocorr}). Since this upturn is likely an artefact, we
  reject objects with $r_g<1.5$~px. (Note that in the lensing
  analysis, we apply an additional, stricter size criterion based on
  $r_h$; see Paper~III.)

\subsection{Coaddition and PSF correction -- influence on cluster mass measurements?}
\label{sect:psf_influence_on_mass}

Apart from spatial variation, the SuprimeCam PSF is also temporally
variable.  Both the telescope and camera contribute to the anisotropy
of the PSF; i.e. in some fields, rotating the camera by $90^{\circ}$
causes stellar ellipticities to reverse sign (in sky coordinates), and
sometimes the PSF pattern remains largely intact through
rotation. Since for a significant fraction of our data, the camera was
rotated between exposures, we must ask whether the PSF of a coadded
image can still be adequately corrected. The goal of our project is to
measure unbiased cluster masses, and so we evaluate this issue by
testing whether masses measured from coadded images are biased. We
perform this test on the ``worst-case'' fields, where several sets
with very different PSF patterns have been coadded. These are fields
with excellent (but still adequately sampled) seeing ($\sim
0.5\arcsec$), with camera rotation between exposures, and with
exposures often taken on more than one night (note that on all but one
field, the size of the seeing disk is comparable between nights due to
our lensing image selection process). Within a given field, night, and
rotation, the PSF pattern is relatively stable. We therefore coadd
images from these subsets (i.e. for a given subset, only exposures
from the same night and with the same camera rotation angle are used),
and compare masses measured on these images to the image coadded from
all subsets (the mass measurement is described in Paper III).

\begin{figure}
\includegraphics[width=\hsize]{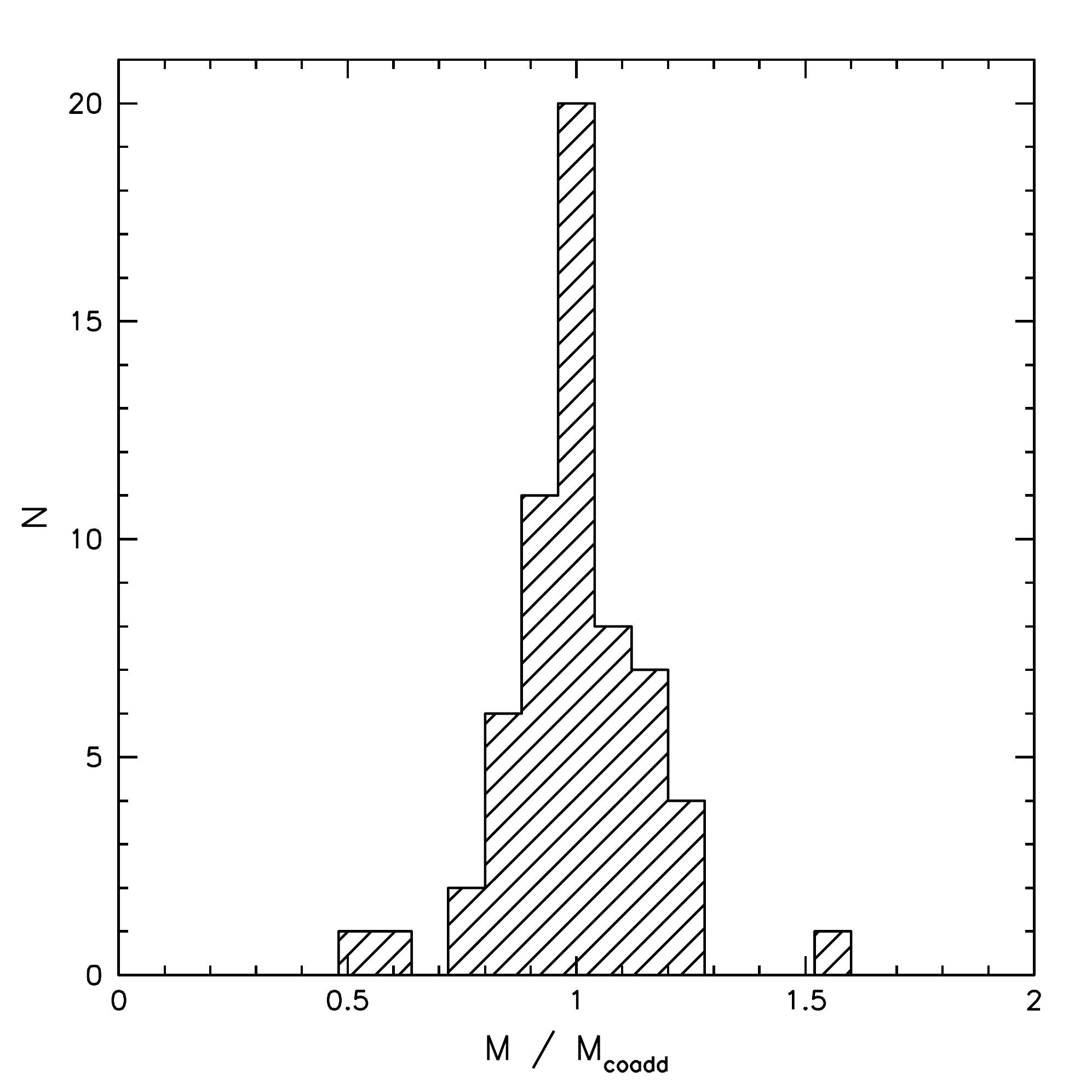}
\caption{The distribution of ratios between masses measured from
  lensing images coadded from single nights and rotations and masses
  measured from the full coadded lensing image (using all available
  nights and rotations). The masses were measured with the
  ``color-cut'' method; see Paper III. No bias is introduced into the
  mass measurements by using full coadded lensing images. The scatter
  in mass ratios is largely due to limited number statistics in some
  images; e.g. the ``outliers'' with values of 0.5 and 1.5 were
  measured from less than 2000 galaxies.}
\label{fig:compare_masses_coadds}
\end{figure}

The distribution of ratios between masses measured on subsets and on
the fully coadded image is shown in
Fig.~\ref{fig:compare_masses_coadds}.  For most comparisons, mass
measurements from different coadded images agree very well: the
distribution clearly peaks at 1, with a median ratio of 0.993,
indicating that using the full coadded image does not lead to biased
cluster mass measurements on average. 

We furthermore investigate the influence of the PSF correction on the
mass measurements. For this purpose, we compare masses determined with
shear measurements corrected only with a second order polynomial to
those using the polynomial order determined according to the process
described in App.~\ref{sect:psfquality} (almost all sixth, eighth, or
tenth order). The change in measured mass when the PSF is not
adequately corrected is best described by an offset of $\sim - 5
\times 10^{13}M_{\odot}$, measured at $2.5\sigma$ significance. For
clusters in the mass range of our sample, this corresponds to mass
underestimates of the order of 1--10\%, illustrating the requirement
of an adequate PSF correction for cluster mass measurements.

To test whether the mass measurements are robust against changes in
the details of how the PSF correction polynomial is determined, we
also compare mass measurements if the order of the polynomial is
decreased or increased by two orders. There is no significant mass
shift. The PSF correction criteria developed in
App.~\ref{sect:psfquality} therefore are sufficient for our purpose.

\subsection{STEP calibration}
\label{sect:step}

\begin{figure*}
\includegraphics[width=0.48\hsize]{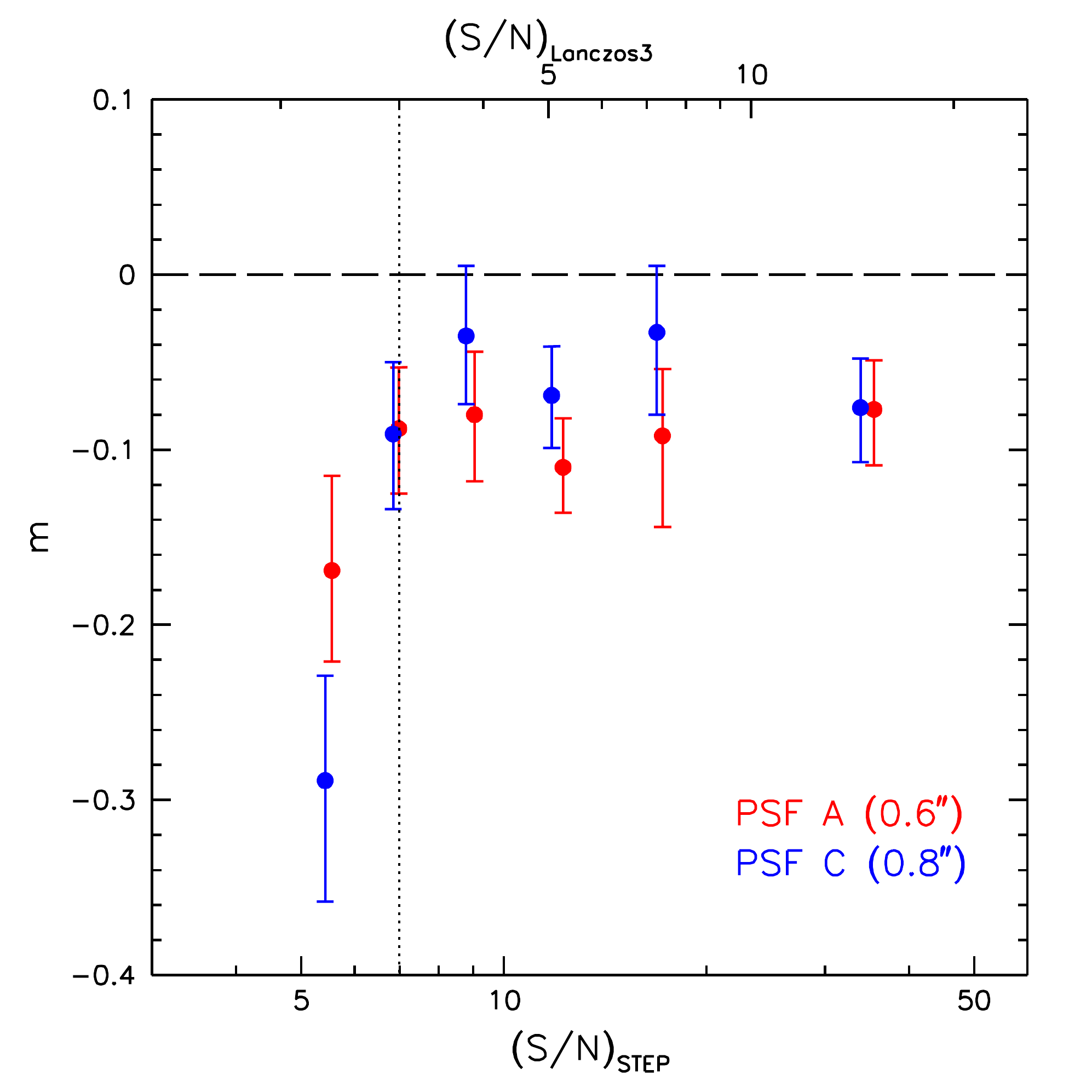}
\includegraphics[width=0.48\hsize]{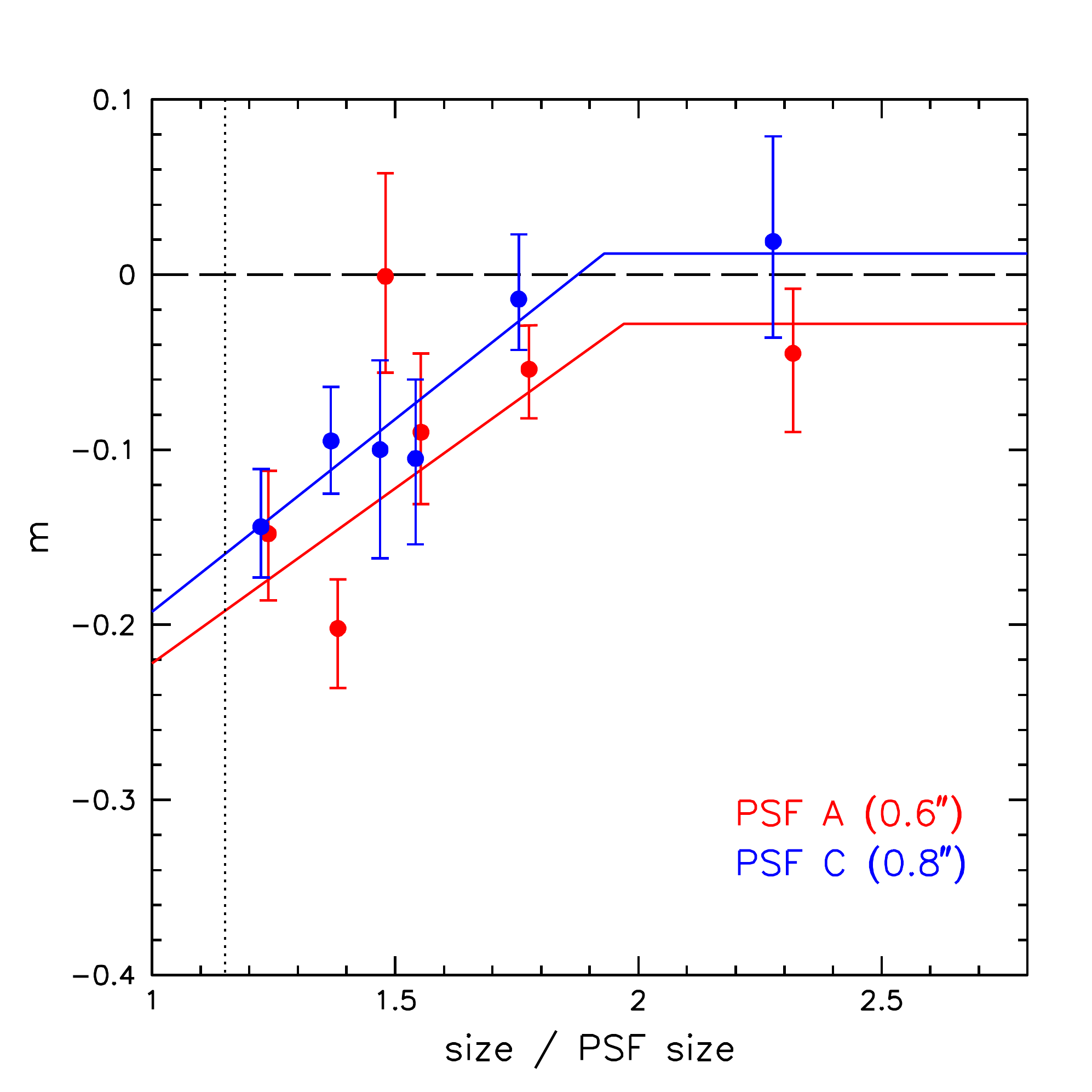}
\caption{Results of our calibration of the shear measurement bias from
  the STEP simulations. Shown is the multiplicative shear bias $m$
  (i.e. a value of $-0.1$ means the estimated shear is 90\% of the
  true shear, since the additive bias is small and consistent with
  zero), as a function of $S/N$ (left panel) and object size (right
  panel). For $(S/N)_{\rm STEP}\gtrsim 7$, the bias is approximately
  constant, $m \sim -0.09$ for PSF A (0.6\arcsec), and $m \sim -0.06$
  for PSF C (0.8\arcsecf). Below this threshold, the magnitude of the
  bias is significantly larger. However, because of the strong
  correlated noise in the STEP2 images, $(S/N)_{\rm STEP}$
  overestimates the true signal-to-noise ratio. Our images are less
  susceptible to correlated noise (due to choosing the {\sc Lanczos3}
  kernel for resampling; see text for details). We find that the
  $(S/N)_{\rm STEP}\gtrsim 7$ threshold approximately corresponds to
  $(S/N)_{\tt Lanczos3}\geq3$ for our images, and hence impose this
  criterion for objects entering the shear analysis. The right panel
  shows the shear bias for objects with $(S/N)_{\rm STEP}\gtrsim 7$ as
  a function of object size, in units of the PSF size (measured as the
  median $r_h$ of stars selected for the PSF correction). There is a
  notable trend with size, in that the magnitude of the bias is larger
  for smaller objects. To correct for this size-dependence, we fit a
  piecewise linear function to the unbinned data, constrained to be a
  constant value for large objects. We find no statistically
  significant difference between corrections for the two shear
  components.  The dotted line indicates the minimum size criterion
  used to reject point sources.}
\label{fig:shapecorr}
\end{figure*}

A crucial element of the analysis is to calibrate the shear
measurement bias inherent to KSB methods, which typically
underestimate the shear \citep{evb01}. Any underestimate of the shear
will result in a direct underestimate of the cluster mass.
Fortunately, the cosmic shear community has led efforts to provide
calibration datasets for shear measurement methods with the Shear
TEsting Programme \citep[STEP,][]{hvb06,mhb07}. We use the simulations
from the STEP 2 project \citep{mhb07} to calibrate the estimator
$\hat{\bm g}$ to the true input shear ${\bm g}$ as a function of the
$S/N$ and size of each galaxy. In Paper III we describe the
characterization of the full probability distribution of $p(\hat{\bm
  g} | \bm g)$; here we summarize the results applicable when
considering simple averaging of the shear estimators.

\subsubsection{Shear measurement bias as a function of S/N}

Fig.~\ref{fig:shapecorr} shows the results for the average
multiplicative shear bias, determined from the STEP data, as a
function of the signal-to-noise ratio measured in the STEP images,
$(S/N)_{\rm STEP}$. The bias is highly dependent on $(S/N)_{\rm
  STEP}$, in the sense that it is consistent with a constant value
above $(S/N)_{\rm STEP}\gtrsim7$, and increases significantly (in
magnitude) for objects with lower $(S/N)_{\rm STEP}$. We find a
slightly smaller correction for PSF model ``C'' with 0.8\arcsec\,
seeing, than for PSF ``A'' with 0.6\arcsec, possibly because the PSF
is better sampled.  Of the PSF models tested in \citet{mhb07}, these
are the most appropriate; since we discard the edges of the field, the
highly elliptical PSF ``D'' and ``E'' are less applicable.

This behavior, with an approximately constant shear bias above a
  threshold signal-to-noise, and large negative bias below the
  threshold, is very similar to other KSB implementations
  \citep{mhb07}. In particular, it is very similar to the ``TS''
  implementation of \citet{ses07} tested in STEP-2. In \citet{hss09},
  those authors also show the shear bias as a function of $S/N$. Taking
  into account the re-scaling of $S/N$ due to correlated noise (see
  below) and the fact that TS apply a constant shear scaling of 1.08,
  the results are in excellent agreement \citep[as would be expected,
    as both methods are based on the same implementation of KSB+ as
    described in][]{evb01}. Using all the PSF models
  tested in STEP-2, these authors detect and subsequently correct for
  a slight $S/N$-dependence also for large $S/N$ \citep{shj10}. In the
  two PSF models we tested, we do not find evidence for such a trend
  with $S/N$, but acknowledge that we might be lacking enough
  statistics to do so.

Fig.~\ref{fig:shapecorr} illustrates strikingly the need for accurate
$S/N$ estimates for each object. By requiring $(S/N)_{\rm STEP}>7$, we
can ensure that we utilize only objects in the regime where the shear
measurement bias is robust, and does not depend sensitively on the
signal-to-noise ratio.

\subsubsection{Accounting for correlated noise}
\label{sect:corr_noise}

A complication with applying the STEP calibration to the actual images
is that the measured noise properties (and thus $S/N$) are sensitive
to correlated noise. Signal-to-noise in images with correlated noise
is overestimated if the estimation procedure does not explicitly
account for its presence. Corrections to account for the effects of
correlated noise have been made to address correlated noise in
photometry measurements \citep{cmd00,masci09} and in shape
measurements \citep{ses07}.

In the actual data, the correlated noise stems from the resampling
process, but the choice of the {\tt Lanczos3} kernel minimizes the
amount of correlation introduced. In the STEP2 images, correlated
noise was artifically introduced by smoothing with a gaussian kernel.

The effects of correlated noise are most pronounced when only a few
pixels are involved in a measurement, and is asymptotic to a constant
correction at a large number of pixels. To quantify the effects of
correlated noise on $S/N$, we created artificial images with Gaussian
noise approximating the noise properties of our images, and 10000
``objects'' (for simplicity, we use normal distributions of a fixed
width) at equal spacing in the images. We then resampled these images
using the same kernel as our actual images, as well as applying a
Gaussian smoothing kernel, as was done for the STEP2 images.

By comparing the distribution of measured $S/N$ values for the shape
measurement procedure for the two resampled images, we are able to
determine the scaling factor between the signal to noise measure on
our images and the STEP2 images. To test for aperture size effects, we
repeat this procedure for objects with FWHM of 2 pixels to 18
pixels. There is only a weak dependence of the scaling on galaxy size
over the range of galaxies accepted into our analysis. We estimate the
scaling of $S/N$ between our images to the STEP images to be
$\approx2.3$. The threshold in shear bias that we see in the STEP
images at $(S/N)_{\rm STEP}\sim7$ thus corresponds to $S/N\sim3$ for
our data. By requiring $S/N\ge3$, we therefore robustly select only
objects for which the average shear calibration bias is not a strong
function of $S/N$. 

The significance $\nu$ that is used in the {\sc imcat}
  implementation of KSB is larger than the $S/N$ measure by a factor
  of 3.5 \citep{evb01}; our $S/N$ cut therefore corresponds to $\nu
  \gtrsim 11.5$. Compared to other weak-lensing studies, this
  $S/N$-cut is relatively conservative \citep[see e.g. Table A1
    of][]{hvb06}. It is also worth keeping in mind that the
  lensing-$S/N$ is lower than the detection significance (see
  Sect.~\ref{sect:ksbsummary}).


\subsubsection{Assigning the correct S/N ratio}

A further practical consideration is necessary to assign the correct
$S/N$ for each object, because {\sc analyseldac} assumes a
constant sky noise level when calculating $S/N$
(Eq.~\ref{eq:lensing-snr}). For our images, with large dither
patterns, rotation between exposures, and ample masking, this is
certainly not the case, as illustrated by a typical weight map shown
in Fig.~\ref{fig:weightmask}.

We correct for this by scaling the reported $S/N$ according to the
local weight. The sky noise used to calculate $S/N$ from
Eq.~\ref{eq:lensing-snr} is the average RMS of the sky background as
measured by {\sc SExtractor} on the lensing image. The relative sky
noise scales with exposure time as $\sigma_{\rm sky}\propto
1/\sqrt{t_{\rm exp}}$ (recall that our images are normalized to a 1s
exposure time). The weight map tracks the effective exposure time per
pixel; therefore we can recover the actual $S/N$ by scaling the
reported value by the square root of the ratio between the average
non-zero weight and the local weight. By comparing with measurements
made on smaller image cut-outs with constant noise, we have verified
that this recovers the true $S/N$ to within a few percent, enough
precision to identify galaxies above the threshold $S/N$ value.

\subsubsection{Shear measurement bias as a function of object size}

In the regime where the average shear measurement bias does not depend
on $(S/N)_{\rm STEP}$, we test for dependence on the size of the
object (in units of PSF size, Fig.~\ref{fig:shapecorr}). We find that
the shear underestimate is worst for objects just larger than the PSF,
and is smallest for well-sampled objects. This is expected and
consistent with other KSB implementations \citep{mhb07}. We therefore
express the correction to be applied to the shear measurement as a
function of object size ($r_h$ as returned by {\sc analyseldac},
scaled by the size of the PSF, defined as the median $r_h$ of stars
selected for the PSF correction). Fig.~\ref{fig:shapecorr} illustrates
the best-fit correction; in Paper III this process is described in
more detail, including how the uncertainties in the shape correction
are propagated to the mass measurements.

\section{Cluster Images and Maps of Shear, Optical Light, and X-ray Emission}
\label{sect:cluster_maps}

\begin{figure*}
\includegraphics[width=0.46\hsize]{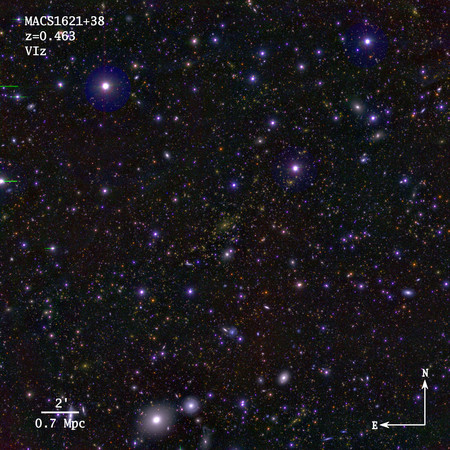}
\includegraphics[width=0.46\hsize]{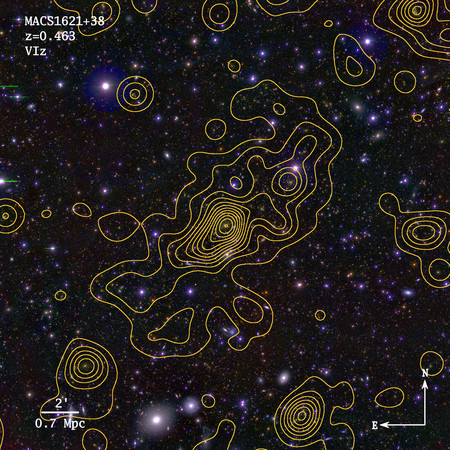}
\includegraphics[width=0.46\hsize]{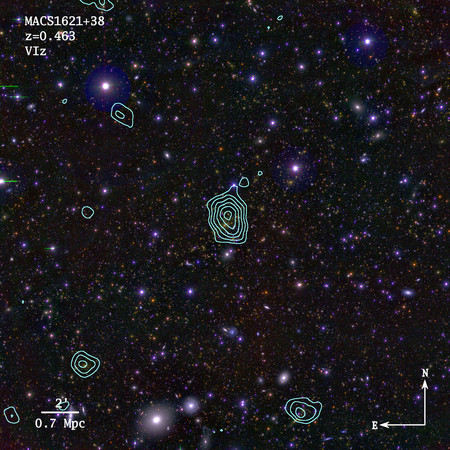}
\includegraphics[width=0.46\hsize]{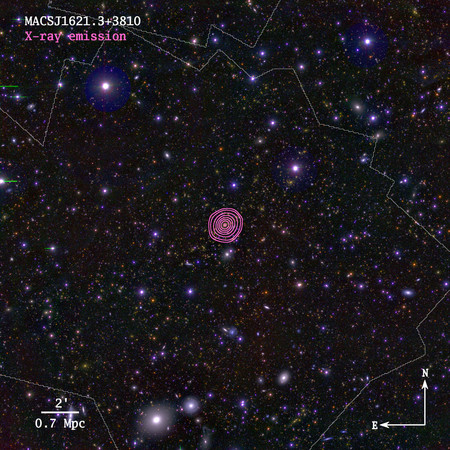}
\begin{minipage}{0.48\hsize}
\vspace{0.3cm}
\includegraphics[width=0.93\hsize]{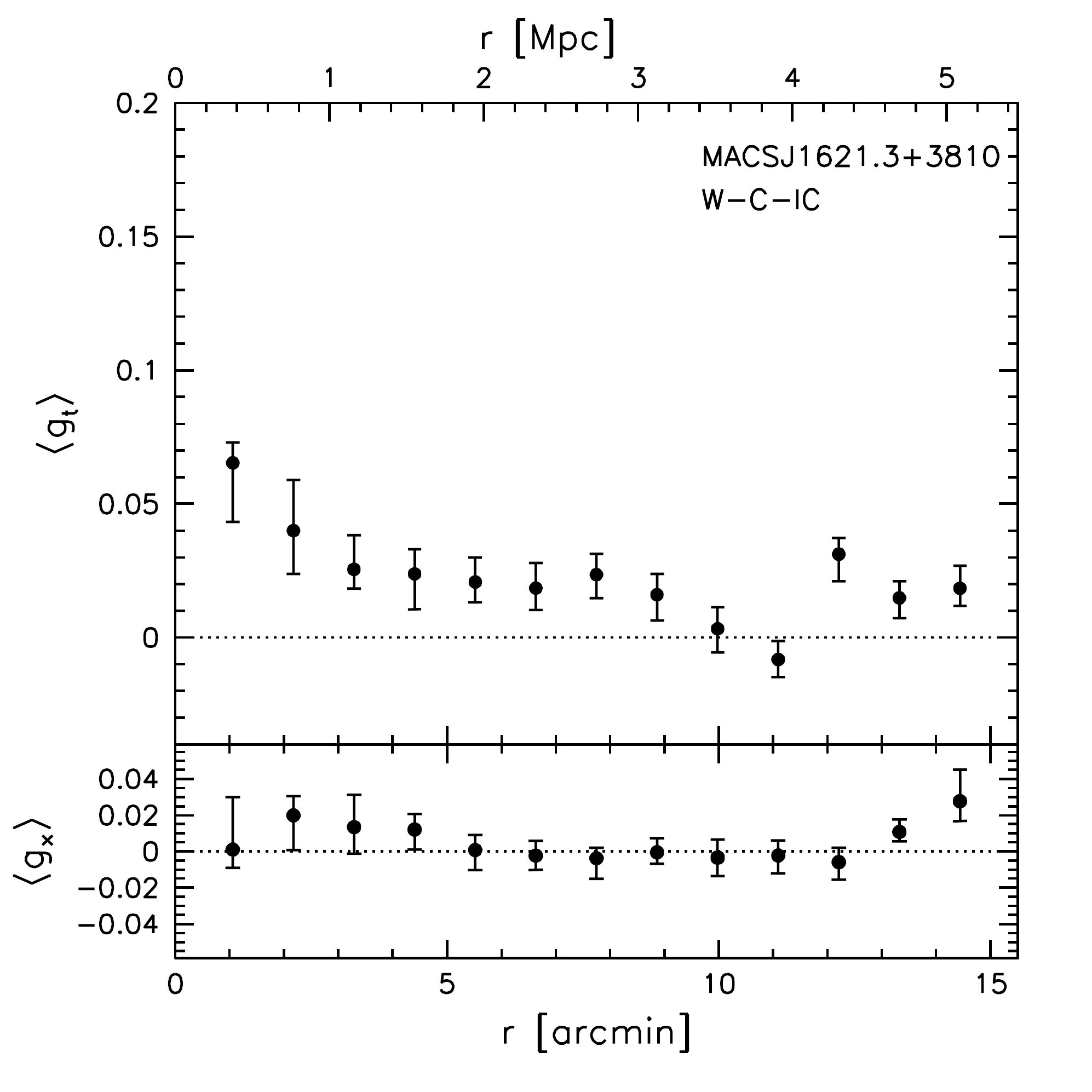}%
\end{minipage}%
\begin{minipage}{0.51\hsize}%
  \caption{The cluster MACSJ1621.3+3810 ($z=0.463$). Each panel above
    shows the $24\arcmin \times 24\arcmin$ optical image composed of
    the SuprimeCam {\it V}$_{\rm J}${\it I}$_{\rm C}${\it z}$^{+}$
    observations. The yellow contours in the top right panel indicate
    the distribution of galaxies on the cluster red sequence, smoothed
    with a gaussian of 3\arcmin width. The blue contours in the bottom
    left panel illustrate the aperture mass map, starting at
    $2.5\sigma$ and increasing by $0.5\sigma$ increments,
    reconstructed from the {\it R}$_{\rm C}$ lensing image. The outer
    radius of the ${M_{\rm ap}}$ filter function corresponds to
    1.5~Mpc at the cluster redshift. In the bottom right panel, the
    pink contours indicate the X-ray emission. The white, thin contour
    illustrates the edge of the Chandra image (merged from four
    exposures); the flux contours are spaced on a square root
    scale. MACSJ1621.3+3810 is in the dynamically relaxed cluster
    sample of A08, though not in the cosmology sample of M10. Despite
    its relative high redshift and low X-ray flux, the
    multi-wavelength analysis reveals a wealth of information. The
    cluster is embedded in a large filament, running from southeast to
    northwest in the image. In an extension of the filament,
    (projected) 4~Mpc to the southeast of MACSJ1621.3+3810, a
    secondary, less massive cluster is seen in both the red sequence
    map and the lensing map. Another secondary cluster, possibly along
    a weaker filament, is located 4~Mpc to the south-southwest. (The
    third such cluster, in the northwest image corner, is detected in
    the {\it I}$_{\rm C}$ band lensing image.)  The figure on the left
    shows the profile of the average tangential and radial shear (top
    and bottom panels, respectively) measured with respect to the
    X-ray centroid, which is at the center of the image. A coherent
    tangential shear signal is detected out to $\sim$3~Mpc.}
\label{fig:MACS1621+38}
\end{minipage}
\end{figure*}

High-quality lensing catalogs represent one of the main ingredients
for cluster mass measurements, which are fully described in Paper III.
With the lensing catalog at hand, however, the two-dimensional shear
field in each cluster field can also be reconstructed to identify mass
overdensities. Together with optical and X-ray images, these provide a
multi-wavelength view of each cluster field.  For each cluster, we
present such a multi-wavelength view in a field of
$24\arcmin\times24\arcmin$, consisting of a three-color image,
contours of the mass distribution as recovered from the shear field,
the surrounding large-scale structure as traced by the light from red
sequence galaxies, and the cluster X-ray emission. MACSJ1621.3+3810 is
shown as an example in Fig.~\ref{fig:MACS1621+38}, the maps for the
other clusters are presented in Appendix~\ref{appendix:clustermaps}.

\subsection{Aperture mass maps and shear profiles}
The lensing maps are calculated with the aperture mass statistic
\citep[$M_{\rm ap}$;][]{sch96}. The aperture mass has the advantages
that it is insensitive to the mass-sheet degeneracy, which is a
concern for clusters that subtend a significant part of the
field of view, and that it can be calculated from shear estimates in a
finite region. We use the filter and weight function advocated by
\citet{ses04} and \citet{hes05}, which follows the expected shear
profile of an NFW profile in order to maximize the signal of real
halos. The scale of the filter is chosen such that the outer radial
limit corresponds to 1.5~Mpc at the cluster redshift. Following
\citet{hes05}, we fix the second free parameter of the filter function
to $x_c=0.15$ .  For these reconstructions, bright galaxies and
galaxies on the red sequence are excluded from the lensing analysis.
The $M_{\rm ap}$ contours shown in Fig.~\ref{fig:MACS1621+38} and
App.~\ref{appendix:clustermaps} are signal-to-noise contours, defined
as in \citet{ses04}. 

For each cluster we furthermore show the azimuthally averaged
tangential and radial shear profiles.

\subsection{Light maps}

To display the distribution of cluster galaxies, and the surrounding
large-scale structure, we identify the red sequence galaxies at the
cluster redshift.  The photometry catalog that we use in the lensing
analysis has been optimized to be highly complete for faint background
galaxies; however, this leads to ``shredding'' of bright, large
galaxies due to excessive deblending. For studying the population of
cluster galaxies, the requirements on the photometry catalog are
different: bright galaxies must be robustly identified and measured,
whereas completeness at the faintest magnitudes can be
compromised. Hence, we create a second set of catalogs with different
{\sc SExtractor} settings tuned to the measurements of cluster members
({\tt DETECT\_MINAREA=12, DETECT\_THRESH=1.5, 
    ANALYSIS\_THRESH=1.5, DEBLEND\_NTHRESH=64, 
    DEBLEND\_MINCONT=0.0001,  FILTER\_NAME = gauss\_1.5\_3x3.conv}).

We identify the red sequence with three filters, i.e. in two
color-magnitude diagrams (CMDs). Only galaxies that are on the red
sequence in both CMDs are considered red sequence members (see also
Paper III). Especially for clusters at higher redshifts, where the
contrast of the red sequence to the back-/foreground population is
low, this strategy boosts the purity of the red sequence sample.

If the Brightest Cluster Galaxy (BCG) is not on the red sequence, as
is the case in several cool-core clusters (Sect.~\ref{sect:bcg_pos}),
it is added to the red sequence population.
From the red sequence galaxy sample, we create luminosity-weighted
maps by smoothing with a Gaussian kernel.

\subsection{X-ray emission maps}

Every cluster in our sample has been followed up with the {\it
  Chandra} X-ray observatory. The analysis of these data is described
in M10 and Mantz et al., in prep. For the purpose of these maps, we
adaptively smooth the processed (and if available, merged from several exposures)
images. 
Point sources
are detected as described in \citet{eab13} and masked before
smoothing.

\subsection{Correspondence between optical, X-ray, and lensing structures}

All clusters are detected in the lensing maps with at least $3\sigma$
significance with the 1.5~Mpc $M_{\rm ap}$ filter; the median
significance is $6\sigma$. For clusters with low significance, the
$M_{\rm ap}$ measurement tends to be compromised by masks of bright
stars located close to the cluster center, reducing the number of
available background galaxies and signal-to-noise (see also
Sect.~\ref{sect:wl_xraycent}).  

The correspondence between the lensing-detected peaks and the optical
and X-ray detections is generally very good, as is expected for
massive clusters with high-quality data. The X-ray emission clearly
indicates the most massive structure in the field; in a few fields,
secondary clusters at the same redshift (e.g. in the
MACSJ0911.2$+$1746 field) or at higher redshifts (e.g. in the
MACSJ1115.8$+$0129 field) are also visible in the X-rays. These
secondary clusters tend to be detected in the lensing map as well,
typically as $\sim3-5\sigma$ peaks.

The optical light maps highlight the large-scale structure surrounding
the target cluster. Several clusters appear to be embedded in
filaments of several Mpc length (e.g. MACSJ1115.8$+$0129,
MACSJ2228.5$+$2036, \rxj, MACSJ1621.3$+$3810). In all fields, other
groups and smaller clusters at the same redshift as the main cluster
are visible. A number of these are also detected in the lensing
maps. Even by eye, the large scatter between optical luminosity and
lensing significance (as measured byt the $M_{\rm ap}$ statistic) is
apparent, as is expected due to the measured scatter between lensing
significance and cluster mass \citep{hty04}, and optical richness and
cluster mass \citep{kma07}. Note that, especially for low significance
peaks, the lensing peak can be noticeably offset from the galaxy
distribution (see below).

In a few fields, there are additional lensing peaks of $3-4\sigma$
significance that are not clearly associated with luminous structure
at the cluster redshift. Several of these correspond to clusters at
higher redshift (e.g. the background clusters in the fields of
MACSJ1115.8$+$0129, A697, Zw7215). Others do not correspond to clear
galaxy overdensities (A697, \mos) and may be caused by shape noise or
low-mass projections along the line of sight.  The frequency of
occurrence of such alignment peaks that do not correspond to massive,
virialized halos is consistent with expectations \citep{hty04,dih10}.

For all clusters targeted her, a coherent azimuthally averaged
tangential shear signal is measured to large scales, $3-5$~Mpc.
The radial shear signal is consistent with zero, especially over
the range over which we fit the tangential shear to determine
cluster masses (0.75--3~Mpc).Qualitatively, the inner part of
the shear profile correlates somewhat with the dynamical state of 
the clusters: in on-going mergers, the central shear profile is
flat, or even decreases in the innermost bins. Since some of these
have clearly bimodal mass distributions (e.g. A1758, \bbullet)
on scales of an arcminute, this is to be expected.

\section{Cluster centering}
\label{sect:cluster_centers}

Robust estimates of the positions of cluster centers are essential
to accurate mass measurements, as significant miscentering can lead to an
underestimate of the weak-lensing mass. As long as the miscentering is
not large compared to the scale of the cluster, however, the effect is
expected to be small \citep{rrk11}. The potential bias can furthermore
be mitigated if the inner cluster region is excluded from the
measurement \citep{msb10}.  The choice of cluster center is also
relevant when comparing to simulations, since different halo center
definitions, such as center-of-mass \citep[e.g.][]{lac94}, most bound
particle or potential minimum \citep[e.g.][]{mrm10,hiw10}, or highest
density peak \citep{tkk08,bek11} approximate different observational
definitions.

X-ray centroids generally provide the most robust measure of the
cluster center. The X-ray flux scales with the square of the gas
density, which in turn follows the overall mass distribution (unless
temporarily separated as in the Bullet Cluster) -- this makes the
X-ray centroid a robust estimator of the cluster center. In
dynamically relaxed clusters hosting a cool core, the correspondence
between the bright X-ray core, the BCG, and the cluster center as
indicated by strong lensing has been shown to be excellent
\citep{all98,sks05}.

In this work, we use the X-ray centroid as the cluster center, as
identified from an iterative analysis of the X-ray emission
within 500~kpc, starting from the X-ray flux peak. 

Here we investigate how two other possible measures of cluster
centers, namely the weak-lensing peak and the position of the BCG,
compare to the position of the X-ray centroid.

\subsection{Weak-lensing peak positions}
\label{sect:wl_xraycent}

\begin{figure}
\includegraphics[width=\hsize]{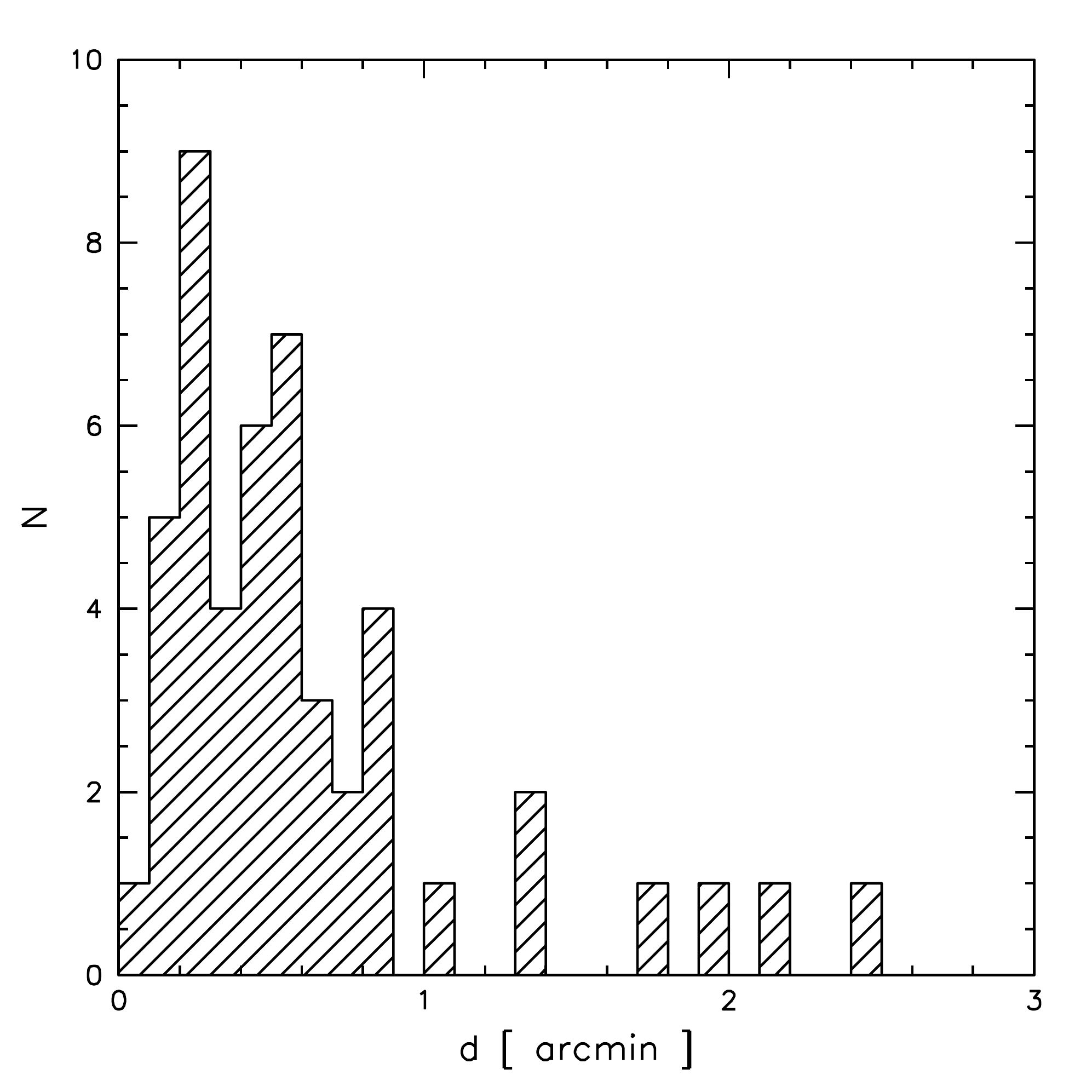}
\caption{Distribution of the distance $d$ between the X-ray centroid
  and the position of the peak of the weak-lensing $M_{\rm ap}$ map,
  measured in arcmin, for the \clusterfields\, clusters in the sample.}
\label{fig:snpeaks-dist}
\end{figure}

Since weak lensing is sensitive to the total cluster mass, one might be
tempted to choose the peak of the weak-lensing mass map as the cluster
center. However, shape noise of the background galaxies and the
smoothing scale inherent to any weak-lensing mass reconstruction cause
a significant dispersion of the reconstructed mass peak with respect
to the true cluster center \citep{dbl11}.  Fig.~\ref{fig:snpeaks-dist}
shows the measured offsets between the X-ray centroid and the peak of
the $M_{\rm ap}$ map for our clusters.  The median of the distribution
is 29~arcsec, and the average is 38~arcsec. For all but seven
clusters, the offset is smaller than 1~arcmin. Most of these seven
clusters have large masks for bright stars close to cluster center,
which affect the determination of the lensing centers. These numbers
are roughly consistent with the analysis of \citet{dbl11}, who cite
offsets between weak-lensing peaks and the cluster centers in N-body
simulations.  Qualitatively, this analysis suggests no significant
offsets between the X-ray centroids and true cluster centers.

\begin{figure*}
\includegraphics[width=0.48\hsize]{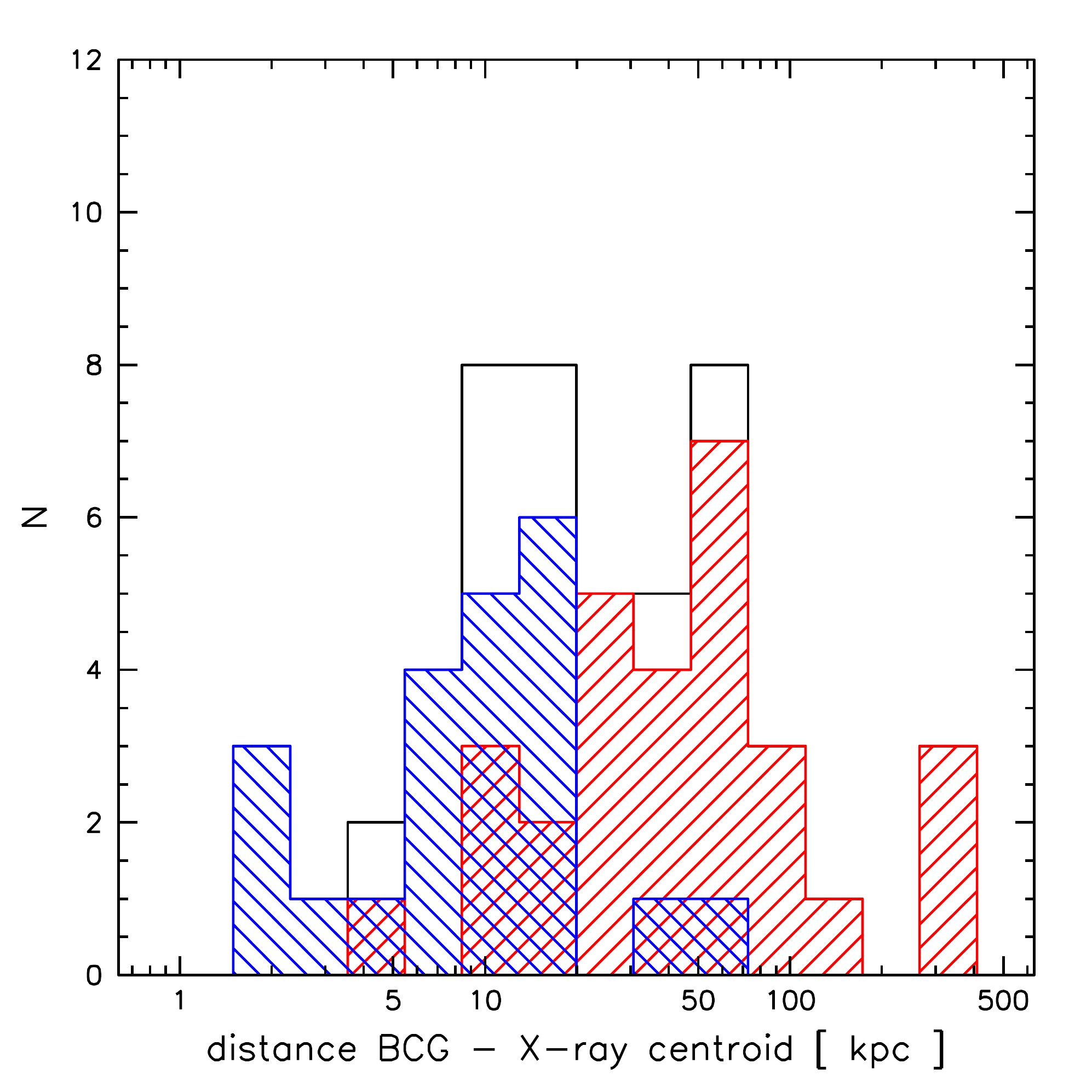}
\includegraphics[width=0.48\hsize]{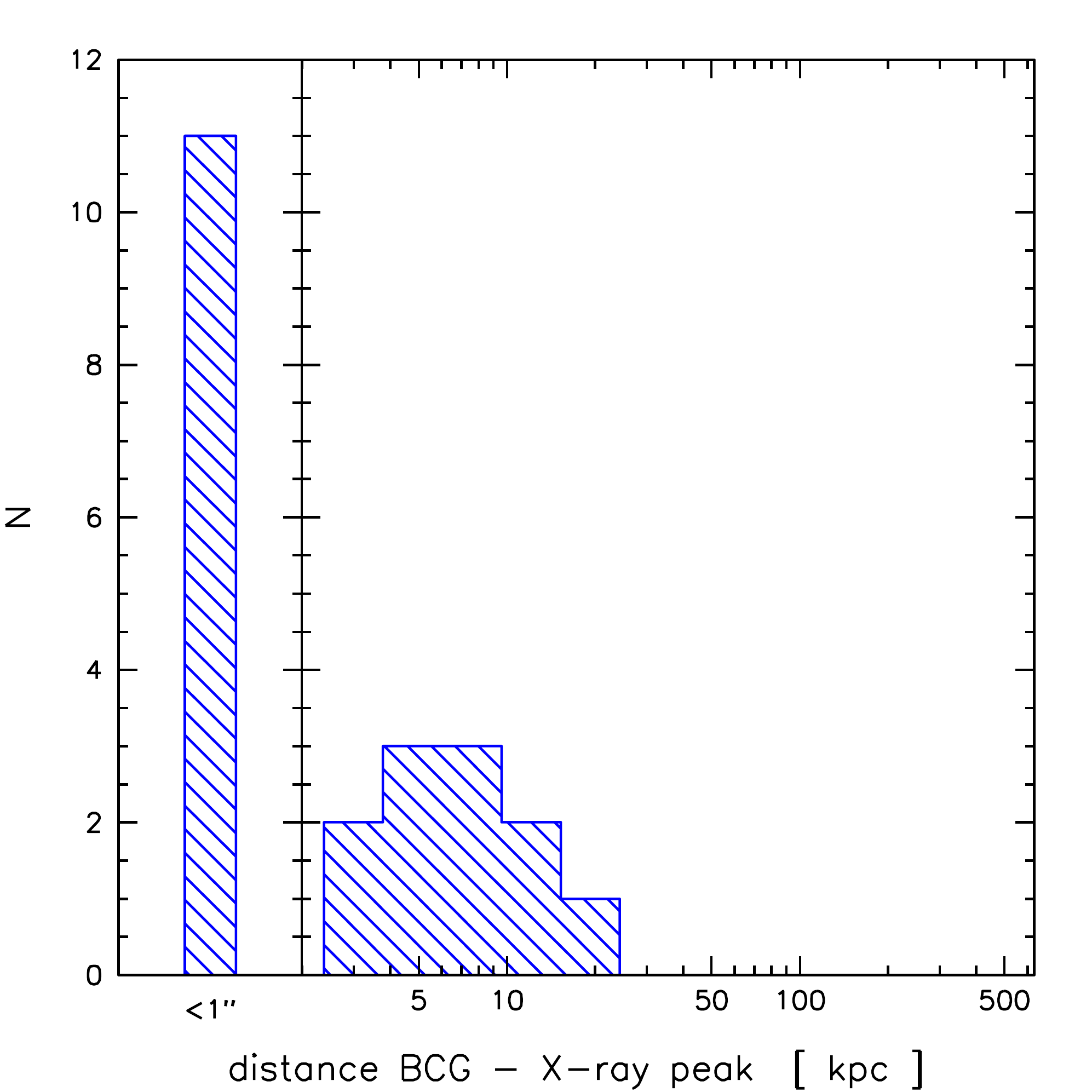}
\caption{Left panel: the distribution of projected offsets between the
  BCG and the X-ray centroid. The black (open) histogram shows the
  total sample; for the blue (hatched downward) and red (hatched
  upward) histogram, the sample has been split into bright-core and
  non-bright-core clusters, respectively. There is a clear correlation
  between BCG -- X-ray offset and dynamical state of the cluster (for
  which bright cores are a good proxy).  Right panel: similar to the
  left panel, but showing the offsets between BCGs and X-ray flux
  peak, for bright-core clusters only. The X-ray flux peak here
  denotes the position of the pixel with the highest flux, after
  binning to 1~arcsec and accounting for X-ray point sources. For half
  the sample, there is no measurable offset between the positions of
  the BCG and X-ray peak.}
\label{fig:bcg_xraycent}
\end{figure*}

\subsection{BCG positions}
\label{sect:bcg_pos}

Cluster studies lacking X-ray data often use the position of the BCG
as the cluster center. Visually, the dominant galaxy of a massive
cluster can usually be unambiguously identified by its brightness, the
presence of an extended stellar halo, and a flock of satellite
galaxies.  In many clusters, the BCG is indeed located near the bottom
of the gravitational well of the cluster
\citep[e.g.][]{all98,sks05,lbk07,blk07}, but this is not always the
case \citep{lim04,sby11}.

For each cluster in our sample, we identify the dominant cluster
galaxy in three steps: The initial BCG candidate is chosen to be the
brightest galaxy on the red sequence within $r_{500, \rm X}$, the
radius within which the average cluster density is 500 times the
critical density as determined from the X-ray data (M10).  However,
for cool-core clusters, the BCG may be bluer than the red sequence. As
a proxy for cool-core clusters, we identify X-ray bright-core clusters
by the criterion that the X-ray luminosity within $0.05 r_{500,\rm X}$
contributes at least 17\% of the total X-ray luminosity within
$r_{500,\rm X}$ \citep{man09}. All clusters from the A08 sample
considered here are identified as bright-core clusters, as well as
three additional clusters.  For these clusters, if there is a brighter
galaxy that is not on the red sequence, but closer to the X-ray
centroid, we update the BCG choice to that galaxy. For 11 out of the
22 bright-core clusters, this is the case. Finally, we visually
inspect the BCG choice, and correct it in four cases. In \mef, the BCG
appears bluer than the red sequence because of the fourth image of a
strongly lensed background galaxy located close to the BCG center
\citep{sel09}; in the other three cases, the BCG is misidentified
because light from the cD halo was not correctly attributed to the
BCG, but rather to superimposed objects (e.g. secondary nuclei,
satellite galaxies). The blue BCGs selected for bright-core clusters
all meet the visual identification criterion; i.e. they are clearly
the galaxy with the largest cD envelope.

\subsubsection{Offsets between BCGs and X-ray centroids}

Fig.~\ref{fig:bcg_xraycent} shows the distribution of offsets between
BCG position and the X-ray centroid.  The offsets are small on a
cluster scale: all but five are less than 100~kpc.  The sample splits
into two populations, one centered at $\sim 10 {\rm kpc}$, and one at
$\sim 50 {\rm kpc}$. When dividing the sample into bright-core and
non-bright-core clusters, it becomes apparent that the two populations
correspond to these two subsamples. The cool-core dichotomy is clearly
linked to the dynamical state of the cluster
\citep[e.g.][]{but96,bpa10}, and largely, the two subsamples
correspond to relaxed / unrelaxed clusters.  We therefore confirm
previous studies that noted the correlation between BCG -- X-ray
offset and dynamical state of the cluster
\citep[e.g.][]{all98,sks05,bhb08,ses09,hmr10,mae12}.

There are two notable exceptions to the typically small distances
between BCG and X-ray offset in bright-core clusters: in \mof$\,$ and
\mosf, the distance is of the order of 50~kpc. \mof$\,$ is a violent
merging cluster, where the X-ray emission is highly asymmetric (von
der Linden et al., in prep).  Although \mosf$\,$ is part of the A08
sample of relaxed clusters, it is one of the few clusters in that
sample where significant substructure in the X-ray emission was noted,
and excised for the hydrostatic equilibrium analysis. The offsets
between the BCG and the X-ray centroid therefore predominantly reflect
asymmetry of the large-scale X-ray emission. 

Fig.~\ref{fig:bcg_xraycent} also shows the measured separations of
BCGs and X-ray flux peaks, defined as the position of the pixel with
the highest flux, after binning to 1~arcsec and accounting for X-ray
point sources, for bright-core clusters (non-bright-core clusters by
definition do not have a pronounced peak). For half the bright-core
clusters, the positions are consistent to within the 1~arcsec
precision with which we measure the X-ray flux peak. The measured
offsets are generally smaller than 10~kpc. Only in clusters where the
cool core is known to have substructure \citep[\mnt;][]{eal11}, or the
X-ray exposure time is short, are the offsets larger.

For non-bright-core clusters, the distribution of BCG -- X-ray
centroid distances extends to larger offsets (median of 49~kpc).  The
clusters where the offset is larger than 100~kpc have clear bimodal
galaxy distributions (\mos, A1758N, \bbullet, \mtt) and/or X-ray
emission with clear substructure (A370).  These clusters also tend to
have more than one dominant galaxy, making the choice of BCG not entirely
unambiguous. Some of these are known to have complex mass
distributions on scales of a few 100~kpc, associated with on-going
merger activity, which clearly violate the assumption of a single,
relaxed halo.  However, even for these clusters, the offsets are
significantly smaller than the cluster size, $r_{500, \rm X} \sim
1.5{\rm Mpc}$.

\subsubsection{Implications for the lensing analysis}

\begin{figure}
\includegraphics[width=\hsize]{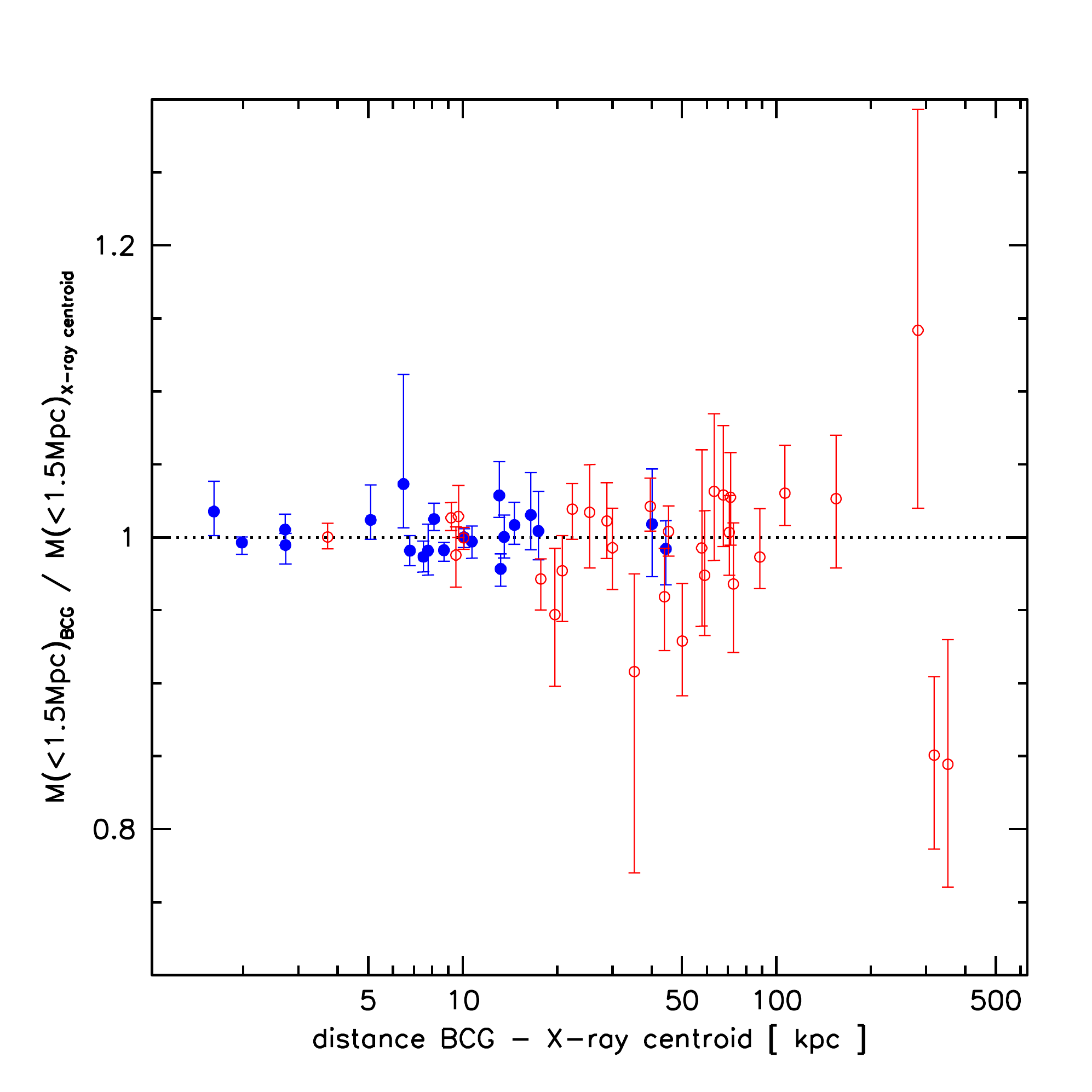}
\caption{The ratios of measured masses when the BCG is chosen as
  cluster center to those centered the X-ray centroid (the adopted
  center in the subsequent analysis). Shown are the median ratios and
  16$^{\rm th}$ and 84$^{\rm th}$ percentiles of bootstrap
  realizations of the source galaxy catalog. Bright-core clusters are
  shown in blue, solid symbols, and non-bright-core clusters as red,
  open symbols. The masses agree exceptionally well especially for
  offsets where the BCG is located within $\lesssim\!100$~kpc of the
  X-ray centroid.}
\label{fig:masses_bcg_xcen}
\end{figure}

For our lensing analysis, we fit the tangential shear profile between
0.75~Mpc and 3~Mpc (see Paper III). The inner cluster regions, on
scales larger than the observed offsets between BCGs and X-ray
centroids, are thereby excluded from the measurement, and we expect
any possible bias from substructure and/or miscentering to be minimal.
We test this assertion explicitly by measuring cluster masses centered
on the BCG, and comparing these to those measured relative to the
X-ray centroid (Fig.~\ref{fig:masses_bcg_xcen}).  The mass
measurements agree exceptionally well: the mean ratio is
$0.999\pm0.002$. For our sample and methodology, miscentering
therefore is not a cause of appreciable systematic uncertainty.
Fig.~\ref{fig:masses_bcg_xcen} shows that for clusters where the BCG
is located within $\lesssim\!100$~kpc of the X-ray centroid, the mass
measurements typically agree within 5\%. For larger offsets found in
bimodal clusters, where the samples of galaxies from which the lensing
mass is determined become increasingly disjoint, the dispersion
appears to become larger. A similar conclusion was reached by
\citet{glb12}, who found from stacking the weak lensing signals of
X-ray identified groups in the COSMOS survey that the most massive
galaxy located close to the X-ray centroid provides a good tracer of
the center-of-mass within $\sim\!$75~kpc.
 
The typically small offsets we find between BCGs and X-ray centers
should be encouraging for SZ cluster surveys, for example, which
select similarly massive clusters, but where the cluster center is
difficult to determine from the survey observations. When X-ray
follow-up observations are not available, our results indicate that
the BCG chosen from high-quality imaging observations is typically a
sufficiently robust indicator for the cluster center. 

For optically selected cluster samples, on the other hand, the
correspondence between our results and centering strategies is less
clear. The clusters in our study are not representative of the
clusters found by optical surveys, which recover large samples of less
massive clusters. \citet{jsw07} find that the probability of choosing
the ``wrong'' BCG decreases with cluster richness, and is
approximately 10\% for the most massive clusters in their sample,
which is in fact reminiscent of the rate of BCG identifications we
manually correct (4/\clusterfields).  The typical offset between
misidentified BCGs and the true cluster center found or expected for
optical surveys \citep{jsw07,rrk11} is on the order of $0.4$~Mpc,
significantly larger than even the BCG -- X-ray centroid offsets we
find for the most extreme bimodal clusters.  However, for large-scale
optical surveys it is not possible to visually vet the choice of
BCG. Automated BCG identification simply based on measured brightness
could fail because of noise in the measurement, lack of sensitivity to
the extended, faint cD halo, oversubtraction of the sky background at
cluster center, etc.  Furthermore, many optical cluster finders use
the red sequence to identify clusters and BCGs; however, in cool-core
clusters, the BCGs are frequently bluer than the red
sequence. Miscentering is expected to be a dominant observational
uncertainty for optical cluster surveys \citep{rrk11}. However, our
results indicate that at least for massive clusters, this can be
improved by better BCG identification -- e.g. through improvements in
photometric measurements, in particular better inclusion of the cD
envelope, as well as allowing the possibility for BCGs to be bluer
than the red sequence.

\section{Summary and Outlook}
\label{sect:summary}

This is the first of a series of papers aimed at measuring accurate
weak-lensing masses for \clusterfields\, of the most X-ray-luminous galaxy clusters
-- the true giants in the observable Universe. The primary goal is to
measure the key mass--observable scaling relations for clusters to
better than 10\% accuracy, a vital prerequisite for current and future
cluster surveys to utilize their full statistical power. To achieve
this goal, we have developed new methods and improved upon existing
ones to measure accurate weak-lensing cluster masses, and have
rigorously quantified the residual sources of systematic uncertainty. 

The cluster sample presented here is the largest to-date for which
weak-lensing masses have been measured with a homogeneous dataset and
methodology. With a redshift range of $0.15 \lesssim z \lesssim 0.7$,
and with half the clusters at $z>0.4$, it extends to higher redshifts
than previous ground-based studies. However, its key distinction is
the emphasis on the minimization and accurate quantification of
residual systematic uncertainties, and the blind nature of the lensing
mass analysis with respect to other mass proxies.

Although the intrinsic scatter of 3D weak-lensing mass measurements is
large ($\sim\!30$\%), cluster sample sizes of $\sim\!50$ bring the
statistical uncertainty on the mean cluster mass (or mean ratio of
weak-lensing mass to other mass proxy) to the 5\% level. Hence,
systematic uncertainties should ideally be controlled to the level of a
few percent in order not to limit weak-lensing mass calibration
efforts.  There are three main sources of systematic uncertainties for
3D mass measurements of individual clusters: the two observational
challenges lie in measuring unbiased estimators of galaxy shapes, and
their redshifts. The third source of systematic uncertainty lies in
relating the measured shear and redshift estimates to the mass of the
cluster.

In this paper, we have laid the basis for the subsequent lensing
analysis, describing a robust data reduction method that aims for both
excellent shape measurements and photometry measurements.  We show
that the shear bias of the KSB method is a strong function of
signal-to-noise ratio, and a function of object size even after low
signal-to-noise objects have been rejected.  Assigning each object the
appropriate shear calibration is critical to this work, in particular
when individual photometric redshift estimates are used. We address
one source of uncertainty for the relation between measured shear and
cluster mass, namely the choice of cluster center. We adopt the
centroid of the X-ray emission as the cluster center, and show that
the location of the dominant cluster galaxy, if correctly identified,
agrees well with the X-ray centroid, with a median projected offset of
only 20~kpc.  Only for the most extreme bimodal cluster mergers, such
as \mos and \bbullet, are the BCG and the X-ray centroid separated by
$\gtrsim$~100~kpc. We find no systematic bias between weak-lensing
mass measurements centered on the BCGs compared to those relative to
the X-ray centroids. For the clusters considered here, and with our
weak-lensing methodology, miscentering therefore is not a source of
systematic uncertainty.

For each cluster, we show optical images and maps of the total mass
distribution measured from weak lensing, cluster structure and
surrounding large-scale structure as traced by red-sequence galaxies,
and the extended X-ray emission. These multi-wavelength maps
illustrate the large-scale structure within which each cluster is
embedded, as well as possible interactions with other mass
concentrations in the field and the presence of foreground and
background structures.

In Paper~II \citep{kla12}, we detail the key methods used for
accurate photometric calibration and determination of photometric
redshifts. We describe how to correct position-dependent
flux-zero-points from repeated observations of the same fields (and
SDSS photometry, when available).  We describe a number of
  improvements to the ``stellar locus method'', allowing us to
  precisely calibrate the relative zero-points between filters
through reference to the narrow intrinsic locus of main sequence stars
in color-color space. We show that with these techniques, we
recover robust photometric redshifts even in the absence of
  calibration data. The quality of the photometric redshifts is
further illustrated with the shear-redshift scaling behind the
clusters.

In Paper~III \citep{alk12}, we proceed to the actual mass
measurements. We develop a novel Bayesian algorithm which utilizes the
full photometric redshift probability distribution, and the full
distribution of KSB shear estimates with respect to the true shear. We
extensively test this method on the COSMOS field and show that, in
terms of the mean recovered cluster mass, the bias of our method is at
most 2\% over the cluster redshift range considered here. We also
measure cluster masses using an improved version of the traditional
``color-cut method'' used in other works, which is applicable to a
larger number of weak-lensing cluster datasets.  We make detailed
estimates of the residual systematic uncertainties of our study,
arriving at a final precision of 7\% on the mean cluster
mass.

In subsequent papers, we will incorporate the weak-lensing mass
measurements into a self-consistent cosmological framework
\citep{mae10} in order to determine improved cosmological constraints
on cosmology and key astrophysical scaling relations.

Accurate and precise absolute calibration of cluster masses will be a critical
requirement for future studies aimed at constraining cosmological
parameters with galaxy clusters.
The methodology introduced in this series of papers should be
straightforwardly adaptable to other projects, utilizing optical
survey data and/or targeted follow-up observations of clusters.

\section*{Acknowledgements}

We thank the anonymous referee for their careful reading of the
manuscript and for comments which helped clarify the text.  We
especially thank them for their quick response when substituting for
the original referee.  We thank Peter Capak for providing updated
COSMOS photometry measurements, and him and Satoshi Miyazaki for
discussions about SuprimeCam data analysis. We thank Matt Becker,
Andrey Kravtsov, Henk Hoekstra, Andisheh Mahdavi, and Tim Schrabback
for helpful discussions on shear measurements and cluster mass
estimates.

This work is supported in part by the U.S. Department of Energy under
contract number DE-AC02-76SF00515. This work was also supported by the
National Science Foundation under Grant No. AST-0807458. MTA and PB
acknowledge the support of NSF grant PHY-0969487. The authors
acknowledge support from programs HST-AR-12654.01-A,
HST-GO-12009.02-A, and HST-GO-11100.02-A provided by NASA through a
grant from the Space Telescope Science Institute, which is operated by
the Association of Universities for Research in Astronomy, Inc., under
NASA contract NAS 5-26555. This work is also supported by the National
Aeronautics and Space Administration through Chandra Award Numbers
TM1-12010X, GO0-11149X, GO9-0141X , and GO8-9119X issued by the
Chandra X-ray Observatory Center, which is operated by the Smithsonian
Astrophysical Observatory for and on behalf of the National
Aeronautics Space Administration under contract NAS8-03060.  DEA
recognizes the support of a Hewlett Foundation Stanford Graduate
Fellowship.

Based in part on data collected at Subaru Telescope (University of
Tokyo) and obtained from the SMOKA, which is operated by the Astronomy
Data Center, National Astronomical Observatory of Japan.  Based on
observations obtained with MegaPrime/MegaCam, a joint project of CFHT
and CEA/DAPNIA, at the Canada-France-Hawaii Telescope (CFHT) which is
operated by the National Research Council (NRC) of Canada, the
Institute National des Sciences de l'Univers of the Centre National de
la Recherche Scientifique of France, and the University of Hawaii.
This research used the facilities of the Canadian Astronomy Data
Centre operated by the National Research Council of Canada with the
support of the Canadian Space Agency.  This research has made use of
the VizieR catalogue access tool, CDS, Strasbourg, France. Funding for
SDSS-III has been provided by the Alfred P. Sloan Foundation, the
Participating Institutions, the National Science Foundation, and the
U.S. Department of Energy Office of Science. The SDSS-III web site is
http://www.sdss3.org/. This research has made use of the NASA/IPAC
Extragalactic Database (NED), which is operated by the Jet Propulsion
Laboratory, Caltech, under contract with NASA.

\bibliography{refs.bib}

\begin{thebibliography}{135}
\expandafter\ifx\csname natexlab\endcsname\relax\def\natexlab#1{#1}\fi

\bibitem[{{Adelman-McCarthy} {et~al.}(2008){Adelman-McCarthy}, {Ag{\"u}eros},
  {Allam}, {Allende Prieto}, {Anderson}, {Anderson}, {Annis}, {Bahcall},
  {Bailer-Jones}, {Baldry}, {Barentine}, {Bassett}, {Becker}, {Beers}, {Bell},
  {Berlind}, {Bernardi}, {Blanton}, {Bochanski}, {Boroski}, {Brinchmann},
  {Brinkmann}, {Brunner}, {Budav{\'a}ri}, {Carliles}, {Carr}, {Castander},
  {Cinabro}, {Cool}, {Covey}, {Csabai}, {Cunha}, {Davenport}, {Dilday}, {Doi},
  {Eisenstein}, {Evans}, {Fan}, {Finkbeiner}, {Friedman}, {Frieman},
  {Fukugita}, {G{\"a}nsicke}, {Gates}, {Gillespie}, {Glazebrook}, {Gray},
  {Grebel}, {Gunn}, {Gurbani}, {Hall}, {Harding}, {Harvanek}, {Hawley},
  {Hayes}, {Heckman}, {Hendry}, {Hindsley}, {Hirata}, {Hogan}, {Hogg}, {Hyde},
  {Ichikawa}, {Ivezi{\'c}}, {Jester}, {Johnson}, {Jorgensen}, {Juri{\'c}},
  {Kent}, {Kessler}, {Kleinman}, {Knapp}, {Kron}, {Krzesinski}, {Kuropatkin},
  {Lamb}, {Lampeitl}, {Lebedeva}, {Lee}, {Leger}, {L{\'e}pine}, {Lima}, {Lin},
  {Long}, {Loomis}, {Loveday}, {Lupton}, {Malanushenko}, {Malanushenko},
  {Mandelbaum}, {Margon}, {Marriner}, {Mart{\'{\i}}nez-Delgado}, {Matsubara},
  {McGehee}, {McKay}, {Meiksin}, {Morrison}, {Munn}, {Nakajima}, {Neilsen},
  {Newberg}, {Nichol}, {Nicinski}, {Nieto-Santisteban}, {Nitta}, {Okamura},
  {Owen}, {Oyaizu}, {Padmanabhan}, {Pan}, {Park}, {Peoples}, {Pier}, {Pope},
  {Purger}, {Raddick}, {Re Fiorentin}, {Richards}, {Richmond}, {Riess}, {Rix},
  {Rockosi}, {Sako}, {Schlegel}, {Schneider}, {Schreiber}, {Schwope}, {Seljak},
  {Sesar}, {Sheldon}, {Shimasaku}, {Sivarani}, {Smith}, {Snedden}, {Steinmetz},
  {Strauss}, {SubbaRao}, {Suto}, {Szalay}, {Szapudi}, {Szkody}, {Tegmark},
  {Thakar}, {Tremonti}, {Tucker}, {Uomoto}, {Vanden Berk}, {Vandenberg},
  {Vidrih}, {Vogeley}, {Voges}, {Vogt}, {Wadadekar}, {Weinberg}, {West},
  {White}, {Wilhite}, {Yanny}, {Yocum}, {York}, {Zehavi}, \& {Zucker}}]{aaa08}
{Adelman-McCarthy}, J.~K., {et~al.} 2008, \apjs, 175, 297

\bibitem[{{Allen}(1998)}]{all98}
{Allen}, S.~W. 1998, \mnras, 296, 392

\bibitem[{{Allen} {et~al.}(2011){Allen}, {Evrard}, \& {Mantz}}]{aem11}
{Allen}, S.~W., {Evrard}, A.~E., \& {Mantz}, A.~B. 2011, \araa, 49, 409

\bibitem[{{Allen} {et~al.}(2008){Allen}, {Rapetti}, {Schmidt}, {Ebeling},
  {Morris}, \& {Fabian}}]{ars08}
{Allen}, S.~W., {Rapetti}, D.~A., {Schmidt}, R.~W., {Ebeling}, H., {Morris},
  R.~G., \& {Fabian}, A.~C. 2008, \mnras, 383, 879

\bibitem[{{Applegate} {et~al.}(2014){Applegate}, {von der Linden}, {Kelly},
  {Allen}, {Allen}, {Burchat}, {Burke}, {Ebeling}, {Mantz}, \&
  {Morris}}]{alk12}
{Applegate}, D.~E., {et~al.} 2014, \mnras, 439, 48

\bibitem[{{Baba} {et~al.}(2002){Baba}, {Yasuda}, {Ichikawa}, {Yagi}, {Iwamoto},
  {Takata}, {Horaguchi}, {Taga}, {Watanabe}, {Okumura}, {Ozawa}, {Yamamoto}, \&
  {Hamabe}}]{byi02}
{Baba}, H., {et~al.} 2002, Report of the National Astronomical Observatory of
  Japan, 6, 23

\bibitem[{{Bacon} {et~al.}(2003){Bacon}, {Massey}, {Refregier}, \&
  {Ellis}}]{bmr03}
{Bacon}, D.~J., {Massey}, R.~J., {Refregier}, A.~R., \& {Ellis}, R.~S. 2003,
  \mnras, 344, 673

\bibitem[{{Bahcall} \& {Fan}(1998)}]{baf98}
{Bahcall}, N.~A. \& {Fan}, X. 1998, \apj, 504, 1

\bibitem[{{Bah{\'e}} {et~al.}(2012){Bah{\'e}}, {McCarthy}, \& {King}}]{bmk11}
{Bah{\'e}}, Y.~M., {McCarthy}, I.~G., \& {King}, L.~J. 2012, \mnras, 421, 1073

\bibitem[{{Bardeau} {et~al.}(2007){Bardeau}, {Soucail}, {Kneib}, {Czoske},
  {Ebeling}, {Hudelot}, {Smail}, \& {Smith}}]{bsk07}
{Bardeau}, S., {Soucail}, G., {Kneib}, J.-P., {Czoske}, O., {Ebeling}, H.,
  {Hudelot}, P., {Smail}, I., \& {Smith}, G.~P. 2007, \aap, 470, 449

\bibitem[{{Bartelmann} \& {Schneider}(2001)}]{bas01}
{Bartelmann}, M. \& {Schneider}, P. 2001, \physrep, 340, 291

\bibitem[{{Becker} \& {Kravtsov}(2011)}]{bek11}
{Becker}, M.~R. \& {Kravtsov}, A.~V. 2011, \apj, 740, 25

\bibitem[{{Benson} {et~al.}(2013){Benson}, {de Haan}, {Dudley}, {Reichardt},
  {Aird}, {Andersson}, {Armstrong}, {Ashby}, {Bautz}, {Bayliss}, {Bazin},
  {Bleem}, {Brodwin}, {Carlstrom}, {Chang}, {Cho}, {Clocchiatti}, {Crawford},
  {Crites}, {Desai}, {Dobbs}, {Foley}, {Forman}, {George}, {Gladders},
  {Gonzalez}, {Halverson}, {Harrington}, {High}, {Holder}, {Holzapfel},
  {Hoover}, {Hrubes}, {Jones}, {Joy}, {Keisler}, {Knox}, {Lee}, {Leitch},
  {Liu}, {Lueker}, {Luong-Van}, {Mantz}, {Marrone}, {McDonald}, {McMahon},
  {Mehl}, {Meyer}, {Mocanu}, {Mohr}, {Montroy}, {Murray}, {Natoli}, {Padin},
  {Plagge}, {Pryke}, {Rest}, {Ruel}, {Ruhl}, {Saliwanchik}, {Saro}, {Sayre},
  {Schaffer}, {Shaw}, {Shirokoff}, {Song}, {Spieler}, {Stalder},
  {Staniszewski}, {Stark}, {Story}, {Stubbs}, {Suhada}, {van Engelen},
  {Vanderlinde}, {Vieira}, {Vikhlinin}, {Williamson}, {Zahn}, \&
  {Zenteno}}]{bhd11}
{Benson}, B.~A., {et~al.} 2013, \apj, 763, 147

\bibitem[{{Bertin}(2006)}]{ber06}
{Bertin}, E. 2006, in Astronomical Society of the Pacific Conference Series,
  Vol. 351, Astronomical Data Analysis Software and Systems XV, ed.
  {C.~Gabriel, C.~Arviset, D.~Ponz, \& S.~Enrique}, 112

\bibitem[{{Bertin} \& {Arnouts}(1996)}]{bea96}
{Bertin}, E. \& {Arnouts}, S. 1996, AJ, 117, 393

\bibitem[{{Bertin} {et~al.}(2002){Bertin}, {Mellier}, {Radovich}, {Missonnier},
  {Didelon}, \& {Morin}}]{bmr02}
{Bertin}, E., {Mellier}, Y., {Radovich}, M., {Missonnier}, G., {Didelon}, P.,
  \& {Morin}, B. 2002, in Astronomical Society of the Pacific Conference
  Series, Vol. 281, Astronomical Data Analysis Software and Systems XI, ed.
  {D.~A.~Bohlender, D.~Durand, \& T.~H.~Handley}, 228

\bibitem[{{Best} {et~al.}(2007){Best}, {von der Linden}, {Kauffmann},
  {Heckman}, \& {Kaiser}}]{blk07}
{Best}, P.~N., {von der Linden}, A., {Kauffmann}, G., {Heckman}, T.~M., \&
  {Kaiser}, C.~R. 2007, \mnras, 379, 894

\bibitem[{{Bildfell} {et~al.}(2008){Bildfell}, {Hoekstra}, {Babul}, \&
  {Mahdavi}}]{bhb08}
{Bildfell}, C., {Hoekstra}, H., {Babul}, A., \& {Mahdavi}, A. 2008, \mnras,
  389, 1637

\bibitem[{{B{\"o}hringer} {et~al.}(2010){B{\"o}hringer}, {Pratt}, {Arnaud},
  {Borgani}, {Croston}, {Ponman}, {Ameglio}, {Temple}, \& {Dolag}}]{bpa10}
{B{\"o}hringer}, H., {et~al.} 2010, \aap, 514, A32

\bibitem[{{B{\"o}hringer} {et~al.}(2004){B{\"o}hringer}, {Schuecker}, {Guzzo},
  {Collins}, {Voges}, {Cruddace}, {Ortiz-Gil}, {Chincarini}, {De Grandi},
  {Edge}, {MacGillivray}, {Neumann}, {Schindler}, \& {Shaver}}]{bsg04}
{B{\"o}hringer}, H., {et~al.} 2004, \aap, 425, 367

\bibitem[{{Borgani} {et~al.}(2001){Borgani}, {Rosati}, {Tozzi}, {Stanford},
  {Eisenhardt}, {Lidman}, {Holden}, {Della Ceca}, {Norman}, \&
  {Squires}}]{brt01}
{Borgani}, S., {et~al.} 2001, \apj, 561, 13

\bibitem[{{Buote} \& {Tsai}(1996)}]{but96}
{Buote}, D.~A. \& {Tsai}, J.~C. 1996, \apj, 458, 27

\bibitem[{{Capak} {et~al.}(2007){Capak}, {Aussel}, {Ajiki}, {McCracken},
  {Mobasher}, {Scoville}, {Shopbell}, {Taniguchi}, {Thompson}, {Tribiano},
  {Sasaki}, {Blain}, {Brusa}, {Carilli}, {Comastri}, {Carollo}, {Cassata},
  {Colbert}, {Ellis}, {Elvis}, {Giavalisco}, {Green}, {Guzzo}, {Hasinger},
  {Ilbert}, {Impey}, {Jahnke}, {Kartaltepe}, {Kneib}, {Koda}, {Koekemoer},
  {Komiyama}, {Leauthaud}, {Le Fevre}, {Lilly}, {Liu}, {Massey}, {Miyazaki},
  {Murayama}, {Nagao}, {Peacock}, {Pickles}, {Porciani}, {Renzini}, {Rhodes},
  {Rich}, {Salvato}, {Sanders}, {Scarlata}, {Schiminovich}, {Schinnerer},
  {Scodeggio}, {Sheth}, {Shioya}, {Tasca}, {Taylor}, {Yan}, \&
  {Zamorani}}]{caa07}
{Capak}, P., {et~al.} 2007, \apjs, 172, 99

\bibitem[{{Casertano} {et~al.}(2000){Casertano}, {de Mello}, {Dickinson},
  {Ferguson}, {Fruchter}, {Gonzalez-Lopezlira}, {Heyer}, {Hook}, {Levay},
  {Lucas}, {Mack}, {Makidon}, {Mutchler}, {Smith}, {Stiavelli}, {Wiggs}, \&
  {Williams}}]{cmd00}
{Casertano}, S., {et~al.} 2000, \aj, 120, 2747

\bibitem[{{Corless} \& {King}(2007)}]{cok07}
{Corless}, V.~L. \& {King}, L.~J. 2007, \mnras, 380, 149

\bibitem[{{Croft} \& {Dailey}(2011)}]{crd11}
{Croft}, R.~A.~C. \& {Dailey}, M. 2011, arXiv:1112.3108

\bibitem[{{Cypriano} {et~al.}(2004){Cypriano}, {Sodr{\'e}}, {Kneib}, \&
  {Campusano}}]{csk04}
{Cypriano}, E.~S., {Sodr{\'e}}, Jr., L., {Kneib}, J.-P., \& {Campusano}, L.~E.
  2004, \apj, 613, 95

\bibitem[{{Dahle}(2006)}]{dah06}
{Dahle}, H. 2006, \apj, 653, 954

\bibitem[{{Dahle} {et~al.}(2002){Dahle}, {Kaiser}, {Irgens}, {Lilje}, \&
  {Maddox}}]{dki02}
{Dahle}, H., {Kaiser}, N., {Irgens}, R.~J., {Lilje}, P.~B., \& {Maddox}, S.~J.
  2002, \apjs, 139, 313

\bibitem[{{Dietrich} {et~al.}(2012){Dietrich}, {B{\"o}hnert}, {Lombardi},
  {Hilbert}, \& {Hartlap}}]{dbl11}
{Dietrich}, J.~P., {B{\"o}hnert}, A., {Lombardi}, M., {Hilbert}, S., \&
  {Hartlap}, J. 2012, \mnras, 419, 3547

\bibitem[{{Dietrich} \& {Hartlap}(2010)}]{dih10}
{Dietrich}, J.~P. \& {Hartlap}, J. 2010, \mnras, 402, 1049

\bibitem[{{Donahue} {et~al.}(1998){Donahue}, {Voit}, {Gioia}, {Lupino},
  {Hughes}, \& {Stocke}}]{dvg98}
{Donahue}, M., {Voit}, G.~M., {Gioia}, I., {Lupino}, G., {Hughes}, J.~P., \&
  {Stocke}, J.~T. 1998, \apj, 502, 550

\bibitem[{{Donovan}(2007)}]{don07}
{Donovan}, D.~A.~K. 2007, PhD thesis, University of Hawai'i at Manoa

\bibitem[{{Ebeling} {et~al.}(2007){Ebeling}, {Barrett}, {Donovan}, {Ma},
  {Edge}, \& {van Speybroeck}}]{ebd07}
{Ebeling}, H., {Barrett}, E., {Donovan}, D., {Ma}, C.-J., {Edge}, A.~C., \&
  {van Speybroeck}, L. 2007, \apjl, 661, L33

\bibitem[{{Ebeling} {et~al.}(2000){Ebeling}, {Edge}, {Allen}, {Crawford},
  {Fabian}, \& {Huchra}}]{eea00}
{Ebeling}, H., {Edge}, A.~C., {Allen}, S.~W., {Crawford}, C.~S., {Fabian},
  A.~C., \& {Huchra}, J.~P. 2000, \mnras, 318, 333

\bibitem[{{Ebeling} {et~al.}(1998){Ebeling}, {Edge}, {Bohringer}, {Allen},
  {Crawford}, {Fabian}, {Voges}, \& {Huchra}}]{eeb98}
{Ebeling}, H., {Edge}, A.~C., {Bohringer}, H., {Allen}, S.~W., {Crawford},
  C.~S., {Fabian}, A.~C., {Voges}, W., \& {Huchra}, J.~P. 1998, \mnras, 301,
  881

\bibitem[{{Ebeling} {et~al.}(2001){Ebeling}, {Edge}, \& {Henry}}]{eeh01}
{Ebeling}, H., {Edge}, A.~C., \& {Henry}, J.~P. 2001, \apj, 553, 668

\bibitem[{{Ebeling} {et~al.}(2010){Ebeling}, {Edge}, {Mantz}, {Barrett},
  {Henry}, {Ma}, \& {van Speybroeck}}]{eem10}
{Ebeling}, H., {Edge}, A.~C., {Mantz}, A., {Barrett}, E., {Henry}, J.~P., {Ma},
  C.~J., \& {van Speybroeck}, L. 2010, \mnras, 407, 83

\bibitem[{{Ehlert} {et~al.}(2013){Ehlert}, {Allen}, {Brandt}, {Xue}, {Luo},
  {von der Linden}, {Mantz}, \& {Morris}}]{eab13}
{Ehlert}, S., {Allen}, S.~W., {Brandt}, W.~N., {Xue}, Y.~Q., {Luo}, B., {von
  der Linden}, A., {Mantz}, A., \& {Morris}, R.~G. 2013, \mnras, 428, 3509

\bibitem[{{Ehlert} {et~al.}(2011){Ehlert}, {Allen}, {von der Linden},
  {Simionescu}, {Werner}, {Taylor}, {Gentile}, {Ebeling}, {Allen}, {Applegate},
  {Dunn}, {Fabian}, {Kelly}, {Million}, {Morris}, {Sanders}, \&
  {Schmidt}}]{eal11}
{Ehlert}, S., {et~al.} 2011, \mnras, 411, 1641

\bibitem[{{Erben} {et~al.}(2009){Erben}, {Hildebrandt}, {Lerchster}, {Hudelot},
  {Benjamin}, {van Waerbeke}, {Schrabback}, {Brimioulle}, {Cordes}, {Dietrich},
  {Holhjem}, {Schirmer}, \& {Schneider}}]{ehl09}
{Erben}, T., {et~al.} 2009, \aap, 493, 1197

\bibitem[{{Erben} {et~al.}(2005){Erben}, {Schirmer}, {Dietrich}, {Cordes},
  {Haberzettl}, {Hetterscheidt}, {Hildebrandt}, {Schmithuesen}, {Schneider},
  {Simon}, {Deul}, {Hook}, {Kaiser}, {Radovich}, {Benoist}, {Nonino}, {Olsen},
  {Prandoni}, {Wichmann}, {Zaggia}, {Bomans}, {Dettmar}, \& {Miralles}}]{esd05}
{Erben}, T., {et~al.} 2005, Astronomische Nachrichten, 326, 432

\bibitem[{{Erben} {et~al.}(2001){Erben}, {Van Waerbeke}, {Bertin}, {Mellier},
  \& {Schneider}}]{evb01}
{Erben}, T., {Van Waerbeke}, L., {Bertin}, E., {Mellier}, Y., \& {Schneider},
  P. 2001, \aap, 366, 717

\bibitem[{{George} {et~al.}(2012){George}, {Leauthaud}, {Bundy}, {Finoguenov},
  {Ma}, {Rykoff}, {Tinker}, {Wechsler}, {Massey}, \& {Mei}}]{glb12}
{George}, M.~R., {et~al.} 2012, arXiv:1205.4262

\bibitem[{{Guo} {et~al.}(2013){Guo}, {Ferguson}, {Giavalisco}, {Barro},
  {Willner}, {Ashby}, {Dahlen}, {Donley}, {Faber}, {Fontana}, {Galametz},
  {Grazian}, {Huang}, {Kocevski}, {Koekemoer}, {Koo}, {McGrath}, {Peth},
  {Salvato}, {Wuyts}, {Castellano}, {Cooray}, {Dickinson}, {Dunlop}, {Fazio},
  {Gardner}, {Gawiser}, {Grogin}, {Hathi}, {Hsu}, {Lee}, {Lucas}, {Mobasher},
  {Nandra}, {Newman}, \& {van der Wel}}]{gfg13}
{Guo}, Y., {et~al.} 2013, \apjs, 207, 24

\bibitem[{{Hamana} {et~al.}(2004){Hamana}, {Takada}, \& {Yoshida}}]{hty04}
{Hamana}, T., {Takada}, M., \& {Yoshida}, N. 2004, \mnras, 350, 893

\bibitem[{{Hartlap} {et~al.}(2009){Hartlap}, {Schrabback}, {Simon}, \&
  {Schneider}}]{hss09}
{Hartlap}, J., {Schrabback}, T., {Simon}, P., \& {Schneider}, P. 2009, \aap,
  504, 689

\bibitem[{{Henry} \& {Arnaud}(1991)}]{hea91}
{Henry}, J.~P. \& {Arnaud}, K.~A. 1991, \apj, 372, 410

\bibitem[{{Hetterscheidt} {et~al.}(2005){Hetterscheidt}, {Erben}, {Schneider},
  {Maoli}, {van Waerbeke}, \& {Mellier}}]{hes05}
{Hetterscheidt}, M., {Erben}, T., {Schneider}, P., {Maoli}, R., {van Waerbeke},
  L., \& {Mellier}, Y. 2005, \aap, 442, 43

\bibitem[{{Hetterscheidt} {et~al.}(2007){Hetterscheidt}, {Simon}, {Schirmer},
  {Hildebrandt}, {Schrabback}, {Erben}, \& {Schneider}}]{hss07}
{Hetterscheidt}, M., {Simon}, P., {Schirmer}, M., {Hildebrandt}, H.,
  {Schrabback}, T., {Erben}, T., \& {Schneider}, P. 2007, \aap, 468, 859

\bibitem[{{Heymans} {et~al.}(2006){Heymans}, {Van Waerbeke}, {Bacon}, {Berge},
  {Bernstein}, {Bertin}, {Bridle}, {Brown}, {Clowe}, {Dahle}, {Erben}, {Gray},
  {Hetterscheidt}, {Hoekstra}, {Hudelot}, {Jarvis}, {Kuijken}, {Margoniner},
  {Massey}, {Mellier}, {Nakajima}, {Refregier}, {Rhodes}, {Schrabback}, \&
  {Wittman}}]{hvb06}
{Heymans}, C., {et~al.} 2006, \mnras, 368, 1323

\bibitem[{{High} {et~al.}(2012){High}, {Hoekstra}, {Leethochawalit}, {de Haan},
  {Abramson}, {Aird}, {Armstrong}, {Ashby}, {Bautz}, {Bayliss}, {Bazin},
  {Benson}, {Bleem}, {Brodwin}, {Carlstrom}, {Chang}, {Cho}, {Clocchiatti},
  {Conroy}, {Crawford}, {Crites}, {Desai}, {Dobbs}, {Dudley}, {Foley},
  {Forman}, {George}, {Gladders}, {Gonzalez}, {Halverson}, {Harrington},
  {Holder}, {Holzapfel}, {Hoover}, {Hrubes}, {Jones}, {Joy}, {Keisler}, {Knox},
  {Lee}, {Leitch}, {Liu}, {Lueker}, {Luong-Van}, {Mantz}, {Marrone},
  {McDonald}, {McMahon}, {Mehl}, {Meyer}, {Mocanu}, {Mohr}, {Montroy},
  {Murray}, {Natoli}, {Nurgaliev}, {Padin}, {Plagge}, {Pryke}, {Reichardt},
  {Rest}, {Ruel}, {Ruhl}, {Saliwanchik}, {Saro}, {Sayre}, {Schaffer}, {Shaw},
  {Schrabback}, {Shirokoff}, {Song}, {Spieler}, {Stalder}, {Staniszewski},
  {Stark}, {Story}, {Stubbs}, {Suhada}, {Tokarz}, {van Engelen}, {Vanderlinde},
  {Vieira}, {Vikhlinin}, {Williamson}, {Zahn}, \& {Zenteno}}]{hhl12}
{High}, F.~W., {et~al.} 2012, arXiv:1205.3103

\bibitem[{{High} {et~al.}(2009){High}, {Stubbs}, {Rest}, {Stalder}, \&
  {Challis}}]{hsr09}
{High}, F.~W., {Stubbs}, C.~W., {Rest}, A., {Stalder}, B., \& {Challis}, P.
  2009, \aj, 138, 110

\bibitem[{{Hilbert} \& {White}(2010)}]{hiw10}
{Hilbert}, S. \& {White}, S.~D.~M. 2010, \mnras, 404, 486

\bibitem[{{Hjorth} {et~al.}(1998){Hjorth}, {Oukbir}, \& {van Kampen}}]{hok98}
{Hjorth}, J., {Oukbir}, J., \& {van Kampen}, E. 1998, \mnras, 298, L1

\bibitem[{{Hoekstra}(2001)}]{hoe01}
{Hoekstra}, H. 2001, \aap, 370, 743

\bibitem[{{Hoekstra}(2003)}]{hoe03}
{Hoekstra}, H. 2003, \mnras, 339, 1155

\bibitem[{{Hoekstra}(2007)}]{hoe07}
{Hoekstra}, H. 2007, \mnras, 379, 317

\bibitem[{{Hoekstra} {et~al.}(2011){Hoekstra}, {Donahue}, {Conselice},
  {McNamara}, \& {Voit}}]{hdc11}
{Hoekstra}, H., {Donahue}, M., {Conselice}, C.~J., {McNamara}, B.~R., \&
  {Voit}, G.~M. 2011, \apj, 726, 48

\bibitem[{{Hoekstra} {et~al.}(1998){Hoekstra}, {Franx}, {Kuijken}, \&
  {Squires}}]{hfk98}
{Hoekstra}, H., {Franx}, M., {Kuijken}, K., \& {Squires}, G. 1998, \apj, 504,
  636

\bibitem[{{Hudson} {et~al.}(2010){Hudson}, {Mittal}, {Reiprich}, {Nulsen},
  {Andernach}, \& {Sarazin}}]{hmr10}
{Hudson}, D.~S., {Mittal}, R., {Reiprich}, T.~H., {Nulsen}, P.~E.~J.,
  {Andernach}, H., \& {Sarazin}, C.~L. 2010, \aap, 513, A37

\bibitem[{{Israel} {et~al.}(2012){Israel}, {Erben}, {Reiprich}, {Vikhlinin},
  {Sarazin}, \& {Schneider}}]{ier11}
{Israel}, H., {Erben}, T., {Reiprich}, T.~H., {Vikhlinin}, A., {Sarazin},
  C.~L., \& {Schneider}, P. 2012, \aap, 546, A79

\bibitem[{{Iye} {et~al.}(2004){Iye}, {Karoji}, {Ando}, {Kaifu}, {Kodaira},
  {Aoki}, {Aoki}, {Chikada}, {Doi}, {Ebizuka}, {Elms}, {Fujihara}, {Furusawa},
  {Fuse}, {Gaessler}, {Harasawa}, {Hayano}, {Hayashi}, {Hayashi}, {Ichikawa},
  {Imanishi}, {Ishida}, {Kamata}, {Kanzawa}, {Kashikawa}, {Kawabata},
  {Kobayashi}, {Komiyama}, {Kosugi}, {Kurakami}, {Letawsky}, {Mikami},
  {Miyashita}, {Miyazaki}, {Mizumoto}, {Morino}, {Motohara}, {Murakawa},
  {Nakagiri}, {Nakamura}, {Nakaya}, {Nariai}, {Nishimura}, {Noguchi},
  {Noguchi}, {Noumaru}, {Ogasawara}, {Ohshima}, {Ohyama}, {Okita}, {Omata},
  {Otsubo}, {Oya}, {Potter}, {Saito}, {Sasaki}, {Sato}, {Scarla}, {Schubert},
  {Sekiguchi}, {Sekiguchi}, {Shelton}, {Simpson}, {Suto}, {Tajitsu}, {Takami},
  {Takata}, {Takato}, {Tamae}, {Tamura}, {Tanaka}, {Terada}, {Torii},
  {Uraguchi}, {Usuda}, {Weber}, {Winegar}, {Yagi}, {Yamada}, {Yamashita},
  {Yamashita}, {Yasuda}, {Yoshida}, \& {Yutani}}]{ika04}
{Iye}, M., {et~al.} 2004, \pasj, 56, 381

\bibitem[{{Jee} {et~al.}(2011){Jee}, {Dawson}, {Hoekstra}, {Perlmutter},
  {Rosati}, {Brodwin}, {Suzuki}, {Koester}, {Postman}, {Lubin}, {Meyers},
  {Stanford}, {Barbary}, {Barrientos}, {Eisenhardt}, {Ford}, {Gilbank},
  {Gladders}, {Gonzalez}, {Harris}, {Huang}, {Lidman}, {Rykoff}, {Rubin}, \&
  {Spadafora}}]{jdh11}
{Jee}, M.~J., {et~al.} 2011, \apj, 737, 59

\bibitem[{{Johnston} {et~al.}(2007){Johnston}, {Sheldon}, {Wechsler}, {Rozo},
  {Koester}, {Frieman}, {McKay}, {Evrard}, {Becker}, \& {Annis}}]{jsw07}
{Johnston}, D.~E., {et~al.} 2007, arXiv:0709.1159

\bibitem[{{Kaiser}(1984)}]{kai84}
{Kaiser}, N. 1984, \apjl, 284, L9

\bibitem[{{Kaiser} {et~al.}(1995){Kaiser}, {Squires}, \& {Broadhurst}}]{ksb95}
{Kaiser}, N., {Squires}, G., \& {Broadhurst}, T. 1995, \apj, 449, 460

\bibitem[{{Kelly} {et~al.}(2014){Kelly}, {von der Linden}, {Applegate},
  {Allen}, {Allen}, {Burchat}, {Burke}, {Ebeling}, {Capak}, {Czoske},
  {Donovan}, {Mantz}, \& {Morris}}]{kla12}
{Kelly}, P.~L., {et~al.} 2014, \mnras, 439, 28

\bibitem[{{Klein} \& {Roodman}(2005)}]{klr05}
{Klein}, J.~R. \& {Roodman}, A. 2005, Annual Review of Nuclear and Particle
  Science, 55, 141

\bibitem[{{Koch} {et~al.}(2003){Koch}, {Odenkirchen}, {Caldwell}, \&
  {Grebel}}]{koc03}
{Koch}, A., {Odenkirchen}, M., {Caldwell}, J.~A.~R., \& {Grebel}, E.~K. 2003,
  Astronomische Nachrichten Supplement, 324, 95

\bibitem[{{Koester} {et~al.}(2007){Koester}, {McKay}, {Annis}, {Wechsler},
  {Evrard}, {Rozo}, {Bleem}, {Sheldon}, \& {Johnston}}]{kma07}
{Koester}, B.~P., {et~al.} 2007, \apj, 660, 221

\bibitem[{{Kravtsov} {et~al.}(2006){Kravtsov}, {Vikhlinin}, \& {Nagai}}]{kvn06}
{Kravtsov}, A.~V., {Vikhlinin}, A., \& {Nagai}, D. 2006, \apj, 650, 128

\bibitem[{{Lacey} \& {Cole}(1994)}]{lac94}
{Lacey}, C. \& {Cole}, S. 1994, \mnras, 271, 676

\bibitem[{{Leauthaud} {et~al.}(2010){Leauthaud}, {Finoguenov}, {Kneib},
  {Taylor}, {Massey}, {Rhodes}, {Ilbert}, {Bundy}, {Tinker}, {George}, {Capak},
  {Koekemoer}, {Johnston}, {Zhang}, {Cappelluti}, {Ellis}, {Elvis}, {Giodini},
  {Heymans}, {Le F{\`e}vre}, {Lilly}, {McCracken}, {Mellier},
  {R{\'e}fr{\'e}gier}, {Salvato}, {Scoville}, {Smoot}, {Tanaka}, {Van
  Waerbeke}, \& {Wolk}}]{lfk10}
{Leauthaud}, A., {et~al.} 2010, \apj, 709, 97

\bibitem[{{Lin} \& {Mohr}(2004)}]{lim04}
{Lin}, Y.-T. \& {Mohr}, J.~J. 2004, \apj, 617, 879

\bibitem[{{Luppino} \& {Kaiser}(1997)}]{luk97}
{Luppino}, G.~A. \& {Kaiser}, N. 1997, \apj, 475, 20

\bibitem[{{Magnier} \& {Cuillandre}(2004)}]{mac04}
{Magnier}, E.~A. \& {Cuillandre}, J.-C. 2004, \pasp, 116, 449

\bibitem[{{Mahdavi} {et~al.}(2008){Mahdavi}, {Hoekstra}, {Babul}, \&
  {Henry}}]{mhb08}
{Mahdavi}, A., {Hoekstra}, H., {Babul}, A., \& {Henry}, J.~P. 2008, \mnras,
  384, 1567

\bibitem[{{Mandelbaum} {et~al.}(2010){Mandelbaum}, {Seljak}, {Baldauf}, \&
  {Smith}}]{msb10}
{Mandelbaum}, R., {Seljak}, U., {Baldauf}, T., \& {Smith}, R.~E. 2010, \mnras,
  405, 2078

\bibitem[{{Mann} \& {Ebeling}(2012)}]{mae12}
{Mann}, A.~W. \& {Ebeling}, H. 2012, \mnras, 420, 2120

\bibitem[{{Mantz} {et~al.}(2010{\natexlab{a}}){Mantz}, $\, {Rapetti}, \&
  {Ebeling}}]{mar10}
{Mantz}, A., $\, \!${Allen}, S.~W., {Rapetti}, D., \& {Ebeling}, H.
  2010{\natexlab{a}}, \mnras, 406, 1759

\bibitem[{{Mantz}(2009)}]{man09}
{Mantz}, A. 2009, PhD thesis, Stanford University

\bibitem[{{Mantz} {et~al.}(2010{\natexlab{b}}){Mantz}, {Allen}, {Ebeling},
  {Rapetti}, \& {Drlica-Wagner}}]{mae10}
{Mantz}, A., {Allen}, S.~W., {Ebeling}, H., {Rapetti}, D., \& {Drlica-Wagner},
  A. 2010{\natexlab{b}}, \mnras, 406, 1773

\bibitem[{{Mantz} {et~al.}(2010{\natexlab{c}}){Mantz}, {Allen}, \&
  {Rapetti}}]{mar10c}
{Mantz}, A., {Allen}, S.~W., \& {Rapetti}, D. 2010{\natexlab{c}}, \mnras, 406,
  1805

\bibitem[{{Marrone} {et~al.}(2012){Marrone}, {Smith}, {Okabe}, {Bonamente},
  {Carlstrom}, {Culverhouse}, {Gralla}, {Greer}, {Hasler}, {Hawkins},
  {Hennessy}, {Joy}, {Lamb}, {Leitch}, {Martino}, {Mazzotta}, {Miller},
  {Mroczkowski}, {Muchovej}, {Plagge}, {Pryke}, {Sanderson}, {Takada}, {Woody},
  \& {Zhang}}]{mso11}
{Marrone}, D.~P., {et~al.} 2012, \apj, 754, 119

\bibitem[{{Masci}(2009)}]{masci09}
{Masci}, F. 2009, {Aperture Photometry Uncertainties assuming Priors and
  Correlated Noise}

\bibitem[{{Massey} {et~al.}(2007){Massey}, {Heymans}, {Berg{\'e}}, {Bernstein},
  {Bridle}, {Clowe}, {Dahle}, {Ellis}, {Erben}, {Hetterscheidt}, {High},
  {Hirata}, {Hoekstra}, {Hudelot}, {Jarvis}, {Johnston}, {Kuijken},
  {Margoniner}, {Mandelbaum}, {Mellier}, {Nakajima}, {Paulin-Henriksson},
  {Peeples}, {Roat}, {Refregier}, {Rhodes}, {Schrabback}, {Schirmer}, {Seljak},
  {Semboloni}, \& {van Waerbeke}}]{mhb07}
{Massey}, R., {et~al.} 2007, \mnras, 376, 13

\bibitem[{{Maughan}(2007)}]{mau07}
{Maughan}, B.~J. 2007, \apj, 668, 772

\bibitem[{{Meneghetti} {et~al.}(2010){Meneghetti}, {Rasia}, {Merten},
  {Bellagamba}, {Ettori}, {Mazzotta}, {Dolag}, \& {Marri}}]{mrm10}
{Meneghetti}, M., {Rasia}, E., {Merten}, J., {Bellagamba}, F., {Ettori}, S.,
  {Mazzotta}, P., {Dolag}, K., \& {Marri}, S. 2010, \aap, 514, A93

\bibitem[{{Miyazaki} {et~al.}(2002){Miyazaki}, {Komiyama}, {Sekiguchi},
  {Okamura}, {Doi}, {Furusawa}, {Hamabe}, {Imi}, {Kimura}, {Nakata}, {Okada},
  {Ouchi}, {Shimasaku}, {Yagi}, \& {Yasuda}}]{mks02}
{Miyazaki}, S., {et~al.} 2002, \pasj, 54, 833

\bibitem[{{Nagai} {et~al.}(2007){Nagai}, {Vikhlinin}, \& {Kravtsov}}]{nvk07}
{Nagai}, D., {Vikhlinin}, A., \& {Kravtsov}, A.~V. 2007, \apj, 655, 98

\bibitem[{{Navarro} {et~al.}(1997){Navarro}, {Frenk}, \& {White}}]{NFW97}
{Navarro}, J.~F., {Frenk}, C.~S., \& {White}, S.~D.~M. 1997, \apj, 490, 493

\bibitem[{{Oguri} \& {Hamana}(2011)}]{ogh11}
{Oguri}, M. \& {Hamana}, T. 2011, \mnras, 414, 1851

\bibitem[{{Okabe} {et~al.}(2010{\natexlab{a}}){Okabe}, {Takada}, {Umetsu},
  {Futamase}, \& {Smith}}]{otu10}
{Okabe}, N., {Takada}, M., {Umetsu}, K., {Futamase}, T., \& {Smith}, G.~P.
  2010{\natexlab{a}}, \pasj, 62, 811

\bibitem[{{Okabe} {et~al.}(2010{\natexlab{b}}){Okabe}, {Zhang}, {Finoguenov},
  {Takada}, {Smith}, {Umetsu}, \& {Futamase}}]{ozf10}
{Okabe}, N., {Zhang}, Y.-Y., {Finoguenov}, A., {Takada}, M., {Smith}, G.~P.,
  {Umetsu}, K., \& {Futamase}, T. 2010{\natexlab{b}}, \apj, 721, 875

\bibitem[{{Pedersen} \& {Dahle}(2007)}]{ped07}
{Pedersen}, K. \& {Dahle}, H. 2007, \apj, 667, 26

\bibitem[{{Perlmutter} {et~al.}(1999){Perlmutter}, {Aldering}, {Goldhaber},
  {Knop}, {Nugent}, {Castro}, {Deustua}, {Fabbro}, {Goobar}, {Groom}, {Hook},
  {Kim}, {Kim}, {Lee}, {Nunes}, {Pain}, {Pennypacker}, {Quimby}, {Lidman},
  {Ellis}, {Irwin}, {McMahon}, {Ruiz-Lapuente}, {Walton}, {Schaefer}, {Boyle},
  {Filippenko}, {Matheson}, {Fruchter}, {Panagia}, {Newberg}, {Couch}, \& {The
  Supernova Cosmology Project}}]{pag99}
{Perlmutter}, S., {et~al.} 1999, \apj, 517, 565

\bibitem[{{Planck Collaboration} {et~al.}(2011){Planck Collaboration}, {Ade},
  {Aghanim}, {Arnaud}, {Ashdown}, {Aumont}, {Baccigalupi}, {Balbi}, {Banday},
  {Barreiro}, \& et~al.}]{pla10}
{Planck Collaboration}, {et~al.} 2011, \aap, 536, A8

\bibitem[{{Postman} {et~al.}(2012){Postman}, {Coe}, {Ben{\'{\i}}tez},
  {Bradley}, {Broadhurst}, {Donahue}, {Ford}, {Graur}, {Graves}, {Jouvel},
  {Koekemoer}, {Lemze}, {Medezinski}, {Molino}, {Moustakas}, {Ogaz}, {Riess},
  {Rodney}, {Rosati}, {Umetsu}, {Zheng}, {Zitrin}, {Bartelmann}, {Bouwens},
  {Czakon}, {Golwala}, {Host}, {Infante}, {Jha}, {Jimenez-Teja}, {Kelson},
  {Lahav}, {Lazkoz}, {Maoz}, {McCully}, {Melchior}, {Meneghetti}, {Merten},
  {Moustakas}, {Nonino}, {Patel}, {Reg{\"o}s}, {Sayers}, {Seitz}, \& {Van der
  Wel}}]{pcb12}
{Postman}, M., {et~al.} 2012, \apjs, 199, 25

\bibitem[{{Predehl} {et~al.}(2010){Predehl}, {Andritschke}, {B{\"o}hringer},
  {Bornemann}, {Br{\"a}uninger}, {Brunner}, {Brusa}, {Burkert}, {Burwitz},
  {Cappelluti}, {Churazov}, {Dennerl}, {Eder}, {Elbs}, {Freyberg}, {Friedrich},
  {F{\"u}rmetz}, {Gaida}, {H{\"a}lker}, {Hartner}, {Hasinger}, {Hermann},
  {Huber}, {Kendziorra}, {von Kienlin}, {Kink}, {Kreykenbohm}, {Lamer},
  {Lapchov}, {Lehmann}, {Meidinger}, {Mican}, {Mohr}, {M{\"u}hlegger},
  {M{\"u}ller}, {Nandra}, {Pavlinsky}, {Pfeffermann}, {Reiprich}, {Robrade},
  {Roh{\'e}}, {Santangelo}, {Sch{\"a}chner}, {Schanz}, {Schmid}, {Schmitt},
  {Schreib}, {Schrey}, {Schwope}, {Steinmetz}, {Str{\"u}der}, {Sunyaev},
  {Tenzer}, {Tiedemann}, {Vongehr}, \& {Wilms}}]{pab10}
{Predehl}, P., {et~al.} 2010, in Society of Photo-Optical Instrumentation
  Engineers (SPIE) Conference Series, Vol. 7732, Society of Photo-Optical
  Instrumentation Engineers (SPIE) Conference Series

\bibitem[{{Rapetti} {et~al.}(2010){Rapetti}, {Allen}, {Mantz}, \&
  {Ebeling}}]{ram10}
{Rapetti}, D., {Allen}, S.~W., {Mantz}, A., \& {Ebeling}, H. 2010, \mnras, 406,
  1796

\bibitem[{{Rapetti} {et~al.}(2013){Rapetti}, {Blake}, {Allen}, {Mantz},
  {Parkinson}, \& {Beutler}}]{rba12}
{Rapetti}, D., {Blake}, C., {Allen}, S.~W., {Mantz}, A., {Parkinson}, D., \&
  {Beutler}, F. 2013, \mnras, 432, 973

\bibitem[{{Rasia} {et~al.}(2012){Rasia}, {Meneghetti}, {Martino}, {Borgani},
  {Bonafede}, {Dolag}, {Ettori}, {Fabjan}, {Giocoli}, {Mazzotta}, {Merten},
  {Radovich}, \& {Tornatore}}]{rmm12}
{Rasia}, E., {et~al.} 2012, New Journal of Physics, 14, 055018

\bibitem[{{Refregier}(2003)}]{ref03}
{Refregier}, A. 2003, \mnras, 338, 35

\bibitem[{{Regnault} {et~al.}(2009){Regnault}, {Conley}, {Guy}, {Sullivan},
  {Cuillandre}, {Astier}, {Balland}, {Basa}, {Carlberg}, {Fouchez}, {Hardin},
  {Hook}, {Howell}, {Pain}, {Perrett}, \& {Pritchet}}]{rcg09}
{Regnault}, N., {et~al.} 2009, \aap, 506, 999

\bibitem[{{Reid} {et~al.}(2010){Reid}, {Verde}, {Jimenez}, \& {Mena}}]{rvj10}
{Reid}, B.~A., {Verde}, L., {Jimenez}, R., \& {Mena}, O. 2010, \jcap, 1, 3

\bibitem[{{Riess} {et~al.}(1998){Riess}, {Filippenko}, {Challis},
  {Clocchiatti}, {Diercks}, {Garnavich}, {Gilliland}, {Hogan}, {Jha},
  {Kirshner}, {Leibundgut}, {Phillips}, {Reiss}, {Schmidt}, {Schommer},
  {Smith}, {Spyromilio}, {Stubbs}, {Suntzeff}, \& {Tonry}}]{rfc98}
{Riess}, A.~G., {et~al.} 1998, \aj, 116, 1009

\bibitem[{{Rozo} {et~al.}(2011){Rozo}, {Rykoff}, {Koester}, {Nord}, {Wu},
  {Evrard}, \& {Wechsler}}]{rrk11}
{Rozo}, E., {Rykoff}, E., {Koester}, B., {Nord}, B., {Wu}, H.-Y., {Evrard}, A.,
  \& {Wechsler}, R. 2011, \apj, 740, 53

\bibitem[{{Rozo} {et~al.}(2010){Rozo}, {Wechsler}, {Rykoff}, {Annis}, {Becker},
  {Evrard}, {Frieman}, {Hansen}, {Hao}, {Johnston}, {Koester}, {McKay},
  {Sheldon}, \& {Weinberg}}]{rwr10}
{Rozo}, E., {et~al.} 2010, \apj, 708, 645

\bibitem[{{Sanderson} {et~al.}(2009){Sanderson}, {Edge}, \& {Smith}}]{ses09}
{Sanderson}, A.~J.~R., {Edge}, A.~C., \& {Smith}, G.~P. 2009, \mnras, 398, 1698

\bibitem[{{Schirmer}(2013)}]{sch13}
{Schirmer}, M. 2013, \apjs, 209, 21

\bibitem[{{Schirmer} {et~al.}(2004){Schirmer}, {Erben}, {Schneider}, {Wolf}, \&
  {Meisenheimer}}]{ses04}
{Schirmer}, M., {Erben}, T., {Schneider}, P., {Wolf}, C., \& {Meisenheimer}, K.
  2004, \aap, 420, 75

\bibitem[{{Schmidt} {et~al.}(2009){Schmidt}, {Vikhlinin}, \& {Hu}}]{svh09}
{Schmidt}, F., {Vikhlinin}, A., \& {Hu}, W. 2009, \prd, 80, 083505

\bibitem[{{Schneider}(1996)}]{sch96}
{Schneider}, P. 1996, \mnras, 283, 837

\bibitem[{{Schrabback}(2008)}]{sch08}
{Schrabback}, T. 2008, PhD thesis,
  http://hss.ulb.uni-bonn.de/2008/1336/1336.htm

\bibitem[{{Schrabback} {et~al.}(2007){Schrabback}, {Erben}, {Simon},
  {Miralles}, {Schneider}, {Heymans}, {Eifler}, {Fosbury}, {Freudling},
  {Hetterscheidt}, {Hildebrandt}, \& {Pirzkal}}]{ses07}
{Schrabback}, T., {et~al.} 2007, \aap, 468, 823

\bibitem[{{Schrabback} {et~al.}(2010){Schrabback}, {Hartlap}, {Joachimi},
  {Kilbinger}, {Simon}, {Benabed}, {Brada{\v c}}, {Eifler}, {Erben},
  {Fassnacht}, {High}, {Hilbert}, {Hildebrandt}, {Hoekstra}, {Kuijken},
  {Marshall}, {Mellier}, {Morganson}, {Schneider}, {Semboloni}, {van Waerbeke},
  \& {Velander}}]{shj10}
{Schrabback}, T., {et~al.} 2010, \aap, 516, A63+

\bibitem[{{Schuecker} {et~al.}(2003){Schuecker}, {B{\"o}hringer}, {Collins}, \&
  {Guzzo}}]{gbc03}
{Schuecker}, P., {B{\"o}hringer}, H., {Collins}, C.~A., \& {Guzzo}, L. 2003,
  \aap, 398, 867

\bibitem[{{Sehgal} {et~al.}(2011){Sehgal}, {Trac}, {Acquaviva}, {Ade},
  {Aguirre}, {Amiri}, {Appel}, {Barrientos}, {Battistelli}, {Bond}, {Brown},
  {Burger}, {Chervenak}, {Das}, {Devlin}, {Dicker}, {Bertrand Doriese},
  {Dunkley}, {D{\"u}nner}, {Essinger-Hileman}, {Fisher}, {Fowler}, {Hajian},
  {Halpern}, {Hasselfield}, {Hern{\'a}ndez-Monteagudo}, {Hilton}, {Hilton},
  {Hincks}, {Hlozek}, {Holtz}, {Huffenberger}, {Hughes}, {Hughes}, {Infante},
  {Irwin}, {Jones}, {Baptiste Juin}, {Klein}, {Kosowsky}, {Lau}, {Limon},
  {Lin}, {Lupton}, {Marriage}, {Marsden}, {Martocci}, {Mauskopf}, {Menanteau},
  {Moodley}, {Moseley}, {Netterfield}, {Niemack}, {Nolta}, {Page}, {Parker},
  {Partridge}, {Reid}, {Sherwin}, {Sievers}, {Spergel}, {Staggs}, {Swetz},
  {Switzer}, {Thornton}, {Tucker}, {Warne}, {Wollack}, \& {Zhao}}]{sta11}
{Sehgal}, N., {et~al.} 2011, \apj, 732, 44

\bibitem[{{Skibba} {et~al.}(2011){Skibba}, {van den Bosch}, {Yang}, {More},
  {Mo}, \& {Fontanot}}]{sby11}
{Skibba}, R.~A., {van den Bosch}, F.~C., {Yang}, X., {More}, S., {Mo}, H., \&
  {Fontanot}, F. 2011, \mnras, 410, 417

\bibitem[{{Skrutskie} {et~al.}(2006){Skrutskie}, {Cutri}, {Stiening},
  {Weinberg}, {Schneider}, {Carpenter}, {Beichman}, {Capps}, {Chester},
  {Elias}, {Huchra}, {Liebert}, {Lonsdale}, {Monet}, {Price}, {Seitzer},
  {Jarrett}, {Kirkpatrick}, {Gizis}, {Howard}, {Evans}, {Fowler}, {Fullmer},
  {Hurt}, {Light}, {Kopan}, {Marsh}, {McCallon}, {Tam}, {Van Dyk}, \&
  {Wheelock}}]{scs06}
{Skrutskie}, M.~F., {et~al.} 2006, \aj, 131, 1163

\bibitem[{{Slater} {et~al.}(2009){Slater}, {Harding}, \& {Mihos}}]{shm09}
{Slater}, C.~T., {Harding}, P., \& {Mihos}, J.~C. 2009, \pasp, 121, 1267

\bibitem[{{Smith} {et~al.}(2009){Smith}, {Ebeling}, {Limousin}, {Kneib},
  {Swinbank}, {Ma}, {Jauzac}, {Richard}, {Jullo}, {Sand}, {Edge}, \&
  {Smail}}]{sel09}
{Smith}, G.~P., {et~al.} 2009, \apjl, 707, L163

\bibitem[{{Smith} {et~al.}(2005){Smith}, {Kneib}, {Smail}, {Mazzotta},
  {Ebeling}, \& {Czoske}}]{sks05}
{Smith}, G.~P., {Kneib}, J.-P., {Smail}, I., {Mazzotta}, P., {Ebeling}, H., \&
  {Czoske}, O. 2005, \mnras, 359, 417

\bibitem[{{The Dark Energy Survey Collaboration}(2005)}]{des05}
{The Dark Energy Survey Collaboration}. 2005, arXiv:astro-ph/0510346

\bibitem[{{Tinker} {et~al.}(2008){Tinker}, {Kravtsov}, {Klypin}, {Abazajian},
  {Warren}, {Yepes}, {Gottl{\"o}ber}, \& {Holz}}]{tkk08}
{Tinker}, J., {Kravtsov}, A.~V., {Klypin}, A., {Abazajian}, K., {Warren}, M.,
  {Yepes}, G., {Gottl{\"o}ber}, S., \& {Holz}, D.~E. 2008, \apj, 688, 709

\bibitem[{{Truemper}(1993)}]{tru93}
{Truemper}, J. 1993, Science, 260, 1769

\bibitem[{{Umetsu} {et~al.}(2012){Umetsu}, {Medezinski}, {Nonino}, {Merten},
  {Zitrin}, {Molino}, {Grillo}, {Carrasco}, {Donahue}, {Mahdavi}, {Coe},
  {Postman}, {Koekemoer}, {Czakon}, {Sayers}, {Mroczkowski}, {Golwala}, {Koch},
  {Lin}, {Molnar}, {Rosati}, {Balestra}, {Mercurio}, {Scodeggio}, {Biviano},
  {Anguita}, {Infante}, {Seidel}, {Sendra}, {Jouvel}, {Host}, {Lemze},
  {Broadhurst}, {Meneghetti}, {Moustakas}, {Bartelmann}, {Ben{\'{\i}}tez},
  {Bouwens}, {Bradley}, {Ford}, {Jim{\'e}nez-Teja}, {Kelson}, {Lahav},
  {Melchior}, {Moustakas}, {Ogaz}, {Seitz}, \& {Zheng}}]{umn12}
{Umetsu}, K., {et~al.} 2012, \apj, 755, 56

\bibitem[{{Vanderlinde} {et~al.}(2010){Vanderlinde}, {Crawford}, {de Haan},
  {Dudley}, {Shaw}, {Ade}, {Aird}, {Benson}, {Bleem}, {Brodwin}, {Carlstrom},
  {Chang}, {Crites}, {Desai}, {Dobbs}, {Foley}, {George}, {Gladders}, {Hall},
  {Halverson}, {High}, {Holder}, {Holzapfel}, {Hrubes}, {Joy}, {Keisler},
  {Knox}, {Lee}, {Leitch}, {Loehr}, {Lueker}, {Marrone}, {McMahon}, {Mehl},
  {Meyer}, {Mohr}, {Montroy}, {Ngeow}, {Padin}, {Plagge}, {Pryke}, {Reichardt},
  {Rest}, {Ruel}, {Ruhl}, {Schaffer}, {Shirokoff}, {Song}, {Spieler},
  {Stalder}, {Staniszewski}, {Stark}, {Stubbs}, {van Engelen}, {Vieira},
  {Williamson}, {Yang}, {Zahn}, \& {Zenteno}}]{vcd10}
{Vanderlinde}, K., {et~al.} 2010, \apj, 722, 1180

\bibitem[{{Vikhlinin} {et~al.}(2009{\natexlab{a}}){Vikhlinin}, {Burenin},
  {Ebeling}, {Forman}, {Hornstrup}, {Jones}, {Kravtsov}, {Murray}, {Nagai},
  {Quintana}, \& {Voevodkin}}]{vbe09}
{Vikhlinin}, A., {et~al.} 2009{\natexlab{a}}, \apj, 692, 1033

\bibitem[{{Vikhlinin} {et~al.}(2009{\natexlab{b}}){Vikhlinin}, {Kravtsov},
  {Burenin}, {Ebeling}, {Forman}, {Hornstrup}, {Jones}, {Murray}, {Nagai},
  {Quintana}, \& {Voevodkin}}]{vkb09}
{Vikhlinin}, A., {et~al.} 2009{\natexlab{b}}, \apj, 692, 1060

\bibitem[{{von der Linden} {et~al.}(2007){von der Linden}, {Best}, {Kauffmann},
  \& {White}}]{lbk07}
{von der Linden}, A., {Best}, P.~N., {Kauffmann}, G., \& {White}, S.~D.~M.
  2007, \mnras, 379, 867

\bibitem[{{White} {et~al.}(2005){White}, {Clowe}, {Simard}, {Rudnick}, {De
  Lucia}, {Arag{\'o}n-Salamanca}, {Bender}, {Best}, {Bremer}, {Charlot},
  {Dalcanton}, {Dantel}, {Desai}, {Fort}, {Halliday}, {Jablonka}, {Kauffmann},
  {Mellier}, {Milvang-Jensen}, {Pell{\'o}}, {Poggianti}, {Poirier},
  {Rottgering}, {Saglia}, {Schneider}, \& {Zaritsky}}]{wcs05}
{White}, S.~D.~M., {et~al.} 2005, \aap, 444, 365

\bibitem[{{Wu} {et~al.}(2010){Wu}, {Rozo}, \& {Wechsler}}]{wrw10}
{Wu}, H.-Y., {Rozo}, E., \& {Wechsler}, R.~H. 2010, \apj, 713, 1207

\bibitem[{{Zhang} {et~al.}(2010){Zhang}, {Okabe}, {Finoguenov}, {Smith},
  {Piffaretti}, {Valdarnini}, {Babul}, {Evrard}, {Mazzotta}, {Sanderson}, \&
  {Marrone}}]{zof10}
{Zhang}, Y.-Y., {et~al.} 2010, \apj, 711, 1033

\end{thebibliography}

\appendix

\section{Early Configuration Processing}
\label{appendix:early_data} 

While most of the data included in this analysis are from
configurations 10\_1 and 10\_2, we also use data from the earlier
configurations 8 and 9.  There are a number of problems associated
with the early data including several cosmetic effects. The largest
problem, however, is the nonlinear response of the CCDs. Since for a
significant fraction of the clusters, {\it R}$_{\rm C}$ and {\it
  I}$_{\rm C}$ images (which are essential for galaxy color
information) were taken in the early configurations, we made an effort
to salvage these data.

\subsection{Non-linear response}

\begin{figure}
\includegraphics[width=\hsize]{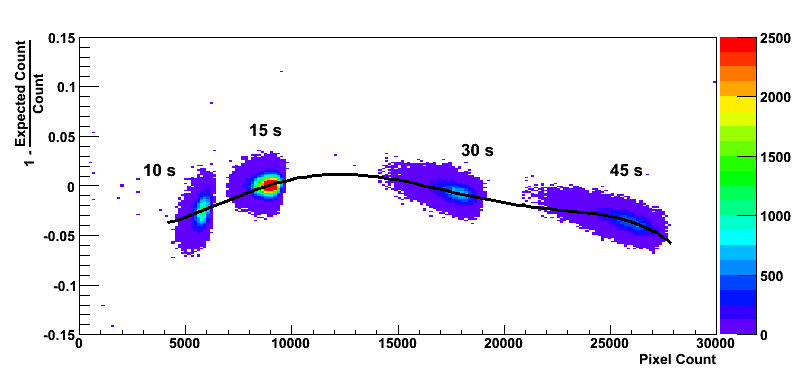}
\caption{The pixel response for an early data MIT/LL chip, normalized
  to the 15-second response, as a function of the pixel count for 10, 15,
  30, and 45 second dome flats. }
\label{fig:nonlin}
\end{figure}

Figure ~\ref{fig:nonlin} shows the detector response to 10, 15, 30 and
45 second domeflats, normalized to the expected response from 15
second exposures. The gain varies by several percent, with a maximum
gain around 10000 pixel counts above the bias level. To correct for
the non-linear response, we fit a polynomial to the data shown and
apply it to the observed pixel counts of science and flat fields,
after overscan and bias subtraction.

We find significant evidence of variable light intensity in the
corners of the focal plane for the series of domeflats used for this
study.  As a result, we cannot derive a correction for the corner
CCDs, and 
we do not use those chips in the early configurations (for
configuration 8, the two left-most chips; for configuration 9, the
chip on the left, and the two right-most chips).

\subsection{Chip defects}

The chips in the early data have a considerable number of dead pixels,
hot pixels, and dirt on the CCDs. Most of these can be
automatically flagged in dark frames and flat-fields, and we mask the
rest by hand.  The two chips in the second column from the left in the array
(``w9c2'' and ``w6c1'') have a $\sim$10\%
brickwall pattern, even in the {\it R}$_{\rm C}$ and {\it I}$_{\rm C}$
bands, but this is entirely removed by flat-fielding.

\begin{figure}
\includegraphics[trim=0.6cm 0.3cm 0.5cm 0.3cm,width=\hsize]{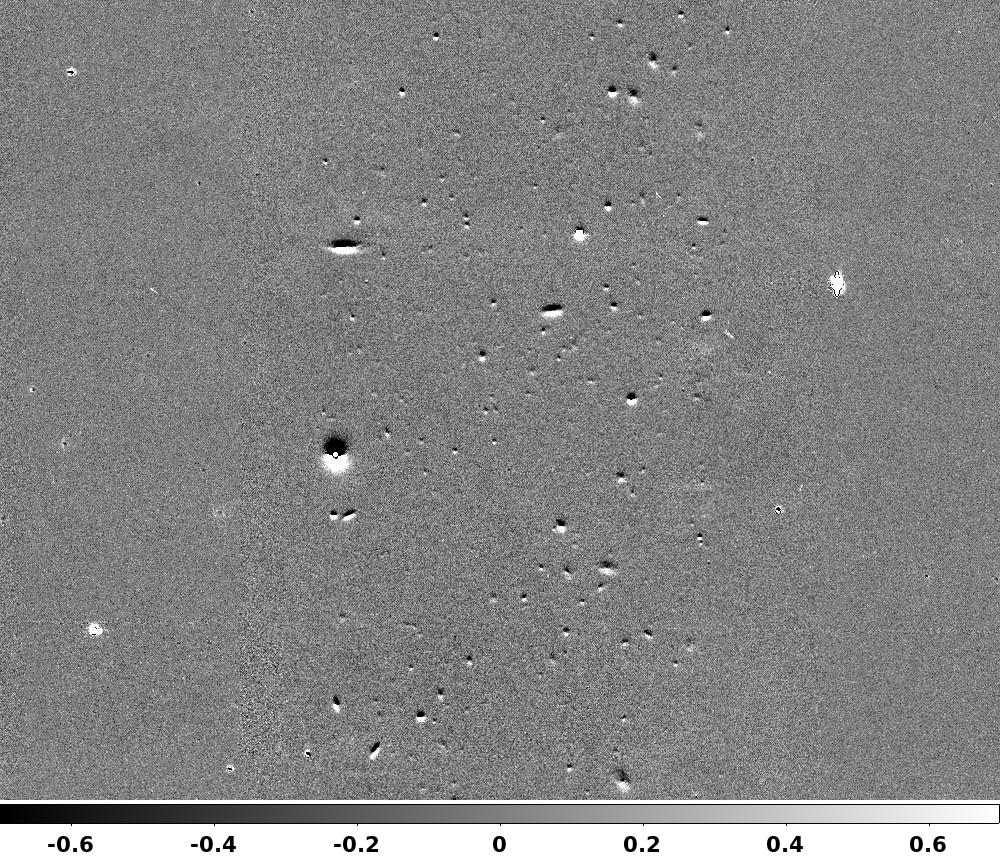}
\caption{Illustration of the pixel indexing error in chip ``w9c2''
  ({\tt DET-ID}=7) of the early configurations. Shown is part of a
  difference image between a single exposure with this chip and the
  median-coadded image of the field. The area shown is $1000\times800$
  pixels, i.e. roughly half the chip-width. On the left and the right
  side of the area shown, the difference image is very smooth (apart
  from saturated stars), indicating good astrometric agreement. In the
  central $\sim 400$ pixels, however, the objects on this chip are
  systematically shifted downwards by $\sim 0.5\arcsec$. This area
  runs about $3/4^{\rm th}$ of the length of the chip, starting from
  the top. The bottom end is marked by a row of hot pixels. To the sides
  it is flanked by $\sim 50$-pixel wide areas of highly correlated
  noise. The width of this whole area appears somewhat variable
  between exposures. In the process of our data reduction, we mask out
  the entire affected area.}  
\label{fig:early_astrom_offset}
\end{figure}

The oddest defect occurs in chip ``w9c2'' (bottom row, second
chip from the left). For $\sim 25\%$ of the chip
area, the $y$-coordinates of the pixels are 2-3 pixels too low, a
defect that must take place during read-out. We found this from
inconsistencies in the astrometric solution between this chip and
other imaging of the same fields. Matching the astrometry to the
remaining chip area, the objects in this strip are displaced by $\sim
0.5\arcsec$ (Fig.~\ref{fig:early_astrom_offset}). We mask this
area.

\section{Verifying the PSF Correction}
\label{appendix:psf}

To determine the PSF anisotropy as a function of position in the image,
we fit a two-dimensional polynomial to the observed ellipticity
components of 
stars (Sect.~\ref{sect:psfcorr}). Fitting over the entire image
(i.e., across chip boundaries) is warranted only if the PSF is continous
across chip boundaries. We describe how we explicitly test for PSF
discontinuity in 
Sect.~\ref{sect:planarity}. In Sect.~\ref{sect:psfquality} we
describe our criteria to choose the order of the polynomials and
validate the resulting PSF correction.

\subsection{Planarity of SuprimeCam}
\label{sect:planarity}


To test for discontinuities in the PSF across chip gaps, we analyze
the observed shapes of stars using a shapelet decomposition
\citep{ref03}.  We extract ``postage stamp'' cutouts of 
regions containing a star in a given exposure, normalize each object
to the same total flux, and 
decompose the flux distribution in terms of the (Cartesian) shapelet
coefficients, 

\begin{equation}
\phi_{n_1n_2} = \int d{\bf x}\; I({\bf x}) \Phi_{n_1n_2}({\bf
  x};\beta), 
\end{equation}
where
\begin{equation}
\Phi_{n_1n_2}({\bf x};\beta)= \cfrac{H_{n_1}({x} / \beta) \: H_{n_2}({y} / \beta)}
{ \left(
  2^{n_1+n_2}\: \pi\: n_1!\:n_2! \:\beta^2\right)^{-1/2}}\:
\exp{\left(-\frac{x^2+y^2}{2\beta^2}\right)}
\end{equation}
and $I({\bf x})$ is the
flux at ${\bf x}$, and $H_n$ are $n$th Hermite polynomials. We find that maximum values of  $n_1,n_2\: = \:5$ (25 
coefficients in total)
provide an accurate measure of the object's
shape without overfitting.

\begin{figure}
\includegraphics[width=\hsize]{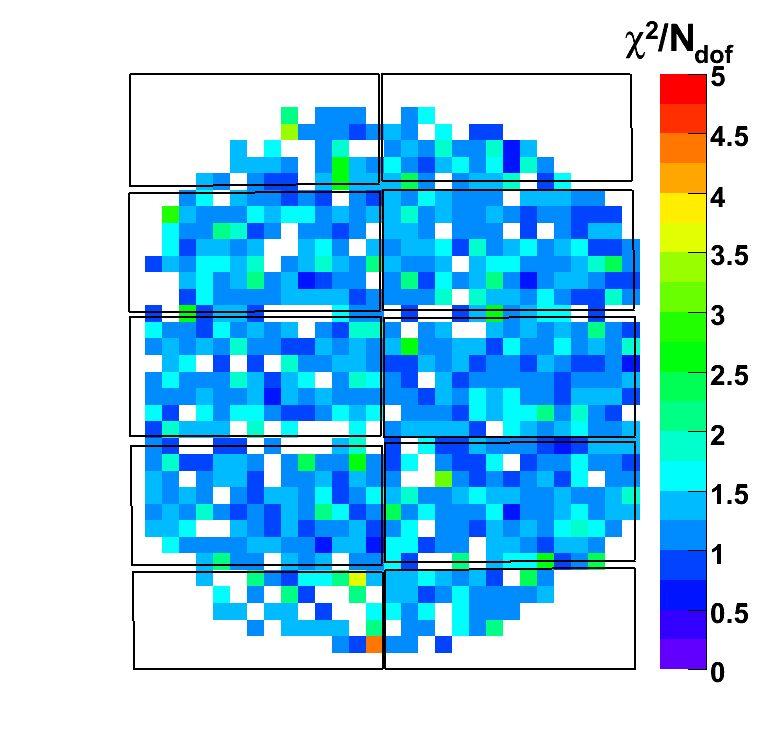}
\caption{A comparison of the shapelet PSF model to the test sample of
  stars across the focal plane.  The color scale shows the value of 
$\langle \chi^2 / N_{\rm dof} \rangle$ in each bin.  We
observe no evidence of PSF discontinuities across chip gaps.}
\label{fig:chimap}
\end{figure}

\begin{figure}
\includegraphics[width=\hsize]{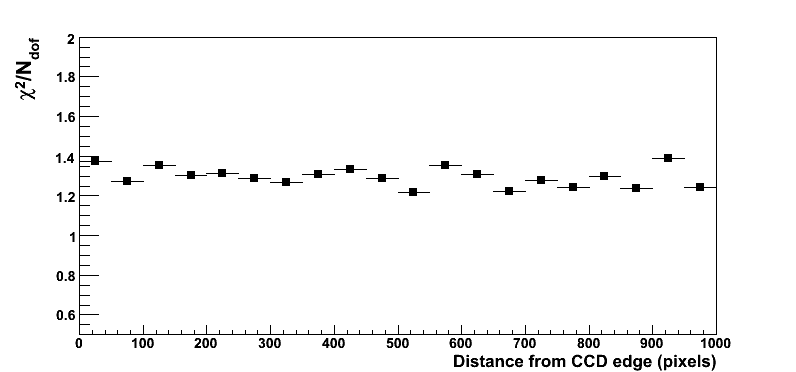}
\caption{A comparison of the shapelet PSF model to the test sample of
  stars as a function of the distance between an object and to the
  nearest chip boundary. There is no evidence of the model being a
  worse description of the PSF near the CCD edges.}
\label{fig:chidist}
\end{figure}

The data used for this analysis are the {\it R}$_{\rm C}$-band images
of MACSJ1931.8$-$2634. The galactic coordinates of MACSJ1931.8$-$2634
are $(l,b)=(12.5669^{\circ},-20.09^{\circ})$; i.e., it is viewed
through the bulge of the Milky Way. Hence the field has a very large
number of stars, and most of these stars are faint enough that they
are not saturated in our exposures - this makes the field an excellent
test case. We randomly separate the $\sim 4200$ stars into equal sized
training and testing samples.  We then decompose the training set, and
for each shapelet component fit a separate third order polynomial
across the field.  This produces a model of the PSF at every position
in the focal plane.  Figure~\ref{fig:chimap} shows the results of a
comparison of the PSF model and the shapes of the stars in the testing
set.  We divide the focal plane into several regions, and in each
region we
define 
\begin{equation}
\langle \chi^{2}_{\nu} \rangle =
\frac{1}{N_{\rm obj}} \sum^{N_{\rm obj}}_{i=1} \frac{1}{N_{\rm pix}} 
\sum^{N_{\rm pix}}_{j=1} \left(\frac{\sum_{n1,n2}\phi^{i}_{n1,n2}
                                \Phi^{i,j}_{n_1,n_2} ({\bf x};\beta) -
  I({\bf x})}{\sigma^{i,j}}\right)^2,
\end{equation}
where $N_{\rm obj}$ is the number of stars in a bin, $N_{\rm pix}$ is
the number of pixels in a postage stamp, $\phi^{i}_{n_1,n_2}$ is the ${n_1,n_2}$
shapelet coefficient of the interpolated PSF at the position of the $i$th star, 
$\Phi^{i,j}_{n_1,n_2}$ is the integral of the shapelet function over pixel $j$  
and $\sigma^{i,j}$ is
the estimated error on the flux in that pixel. Overlayed is a map of
the location of the CCD boundaries.  Any PSF discontinuities would be
visible as jumps or ridges in $\chi^2$ at the chips boundaries.  We
observe no such jumps, and thus no evidence of PSF discontinuities
across CCD gaps.  Figure~\ref{fig:chidist} shows
$\chi^2_{\nu}$ 
as a function of the distance from the center of each object to the closest
CCD edge. 
Again, we observe no evidence of PSF
discontinuities across CCD boundaries.

\subsection{Quality criteria for PSF fits}
\label{sect:psfquality}

For each ellipticity component, we fit a two-dimensional polynomial to
the observed ellipticities of stars: 
$ e_{i}(x,y) = \sum_{j=0}^{n}
\sum_{k=0}^{j} a_{i,jk} x^{k} y^{j-k} $, where $n$ is the highest
order polynomial used in the fit.

We test the goodness-of-fit in several ways.  First, we verify our PSF
interpolation by using a 10-fold cross validation.  We randomly divide
the sample of stars used for the PSF correction into ten groups.  For
each group, we perform a polynomial fit to the stars in the other 9
groups, and calculate the ellipticity residuals for the stars in the
sample that was not used in the fit.  At the end of this process, we
have an estimate of the PSF ellipticity for each star without using
that star in the fit.  We then calculate the standard deviation of
the two ellipticity component residuals, $\sigma_{\Delta e_1}$ and
$\sigma_{\Delta e_2}$.  This technique gives an estimate of the
typical difference between the estimated PSF ellipticity and the true
PSF ellipticity. 
Typical values for $\sigma_{\Delta e_1}$ and $\sigma_{\Delta
e_2}$ are 0.004-0.006 for SuprimeCam images and 0.003-0.004 for
MegaPrime images. 
Images where either component is larger than $0.007$ are excluded from
the lensing analysis.

For cosmic shear studies, other authors have used the uncorrected
stellar ellipticity - galaxy shear correlation as a probe of
unmodelled PSF ellipticity \citep[e.g.][]{bmr03,hss07}. However, while
cosmic shear fields are ideally random pointings, targeted cluster
observations place the cluster center at the center of the
field. Since many PSF patterns, including those of SuprimeCam, display
symmetry around the field center, the cluster shear and stellar
ellipticity are usually correlated. This is clearly visible in the
example PSF pattern shown in Fig.~\ref{fig:psf}, where the measured
ellipticities of the stars are tangential to the field center, much
like the expected cluster shear signal. 

\begin{figure}
\includegraphics[width=\hsize]{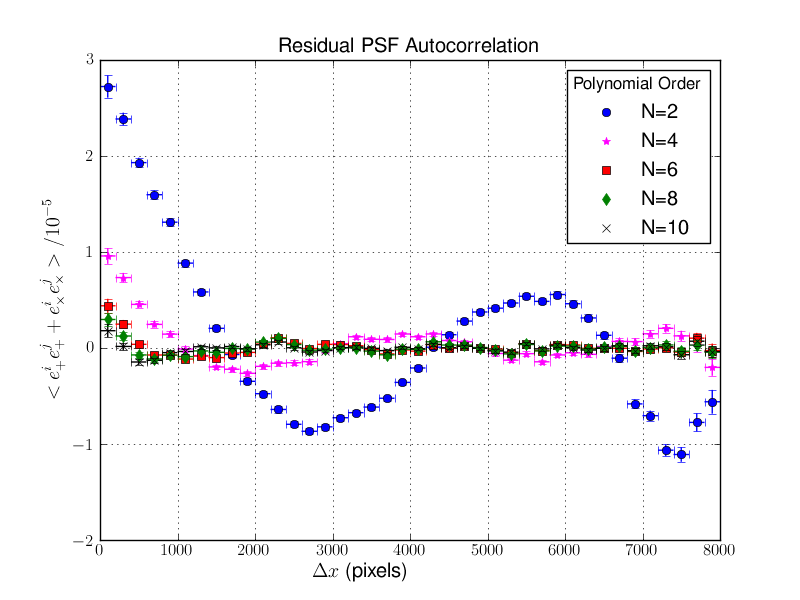}
\caption{The autocorrelation of corrected stellar ellipticites after
  fitting the PSF in the \bbullet {\it V}$_{\rm J}$ field with a
  second, fourth, sixth, eight, and tenth order polynomial. Note the
  large residual (anti-)correlation present after the second order
  fit. Higher order fits suppress this; in this case the eigth order
  polynomial meets all our criteria. }
\label{fig:psf2}
\end{figure}

As another test of the PSF model, we calculate the residual
ellipticity autocorrelation $\langle e^1_+ e^2_+ + e^1_{\times}
e^2_{\times} \rangle$, and require it to be either statistically
consistent with zero, or less than $10^{-5}$ over the range $r_{ac} =
(5 r_{im}^2 / N_{stars})^{1/2}$ and half the size of the image.
Features smaller than that scale are too undersampled to be measured.
A sample residual autocorrelation of stellar ellipticities is shown in
Fig.~\ref{fig:psf2}.

To determine the appropriate maximum polynomial order ($N$) to use, we 
repeat the analysis at every even order
from $N=2$ to $N=10$. (Most of the power comes from the even orders;
the correction is very similar between a given even order and the next
odd order.)
The cross-validation technique provides a natural choice because 
$\sigma_{\Delta e_{1,2}}$ are large when the model 
underfits the data (features exist in the data that are not included in 
the model), and when the order is too large and we overfit (including 
features in the model that are really statistical fluctuations).  
We use the fit that provides the smallest 
$\sigma_{\Delta e_1} + \sigma_{\Delta e_2}$, provided 
that order polynomial passes the autocorrelation
requirements. 
We prefer cross-validation to the $F$-test or the 
likelihood-ratio test because it does not require a per-star ellipticity 
statistical uncertainty, which in this case can be difficult to calculate, 
and it provides an explicit estimate of the interpolation error 
($\sigma_{\Delta e_{1,2}}$).
For about half the fields, we choose an eighth-order
polynomial; the remainder are predominantly sixth- and tenth-order,
with a few fourth-order fits.

\section{Cluster Mass, Light, and Gas Maps}
\label{appendix:clustermaps}

Available in electronic form at
{\tt http://mnras.oxfordjournals.org/lookup/suppl/doi:10.1093/mnras/
stt1945/-/DC1}

\label{lastpage}

\end{document}